\documentclass[letterpaper,american,prl,twocolumn,amsmath,amssymb,superscriptaddress,nofootinbib]{revtex4-2}
\usepackage[T1]{fontenc}
\usepackage{color}
\usepackage{babel}
\usepackage{prettyref}
\usepackage{float}
\usepackage{mathtools}
\usepackage{amsmath}
\usepackage{graphicx}
\usepackage[unicode=true,
 bookmarks=true,bookmarksnumbered=false,bookmarksopen=false,
 breaklinks=false,pdfborder={0 0 1},backref=false,colorlinks=false]
 {hyperref}
\hypersetup{pdftitle={Beyond-mean-field theory for the statistics of neural coordination},
 pdfauthor={Moritz Layer, Moritz Helias, David Dahmen},
 colorlinks=true,linkcolor=black,citecolor=black,urlcolor=black,filecolor=black}

\makeatletter

\pdfpageheight\paperheight
\pdfpagewidth\paperwidth

\providecommand{\tabularnewline}{\\}
\newcommand{\lyxdot}{.}

\usepackage{MnSymbol}  
\usepackage{simplewick}  
\usepackage{prettyref}
\usepackage[latin1]{inputenc}

\newrefformat{eq}{\hyperref[#1]{Eq.~(\ref{#1})}}
\newrefformat{cap}{\hyperref[#1]{Fig.~\ref{#1}}}
\newrefformat{fig}{\hyperref[#1]{Fig.~\ref{#1}}}
\newrefformat{tab}{\hyperref[#1]{Table ~\ref{#1}}}
\newrefformat{sec}{\hyperref[#1]{Sec.~\ref{#1}}}
\newrefformat{subsec}{\hyperref[#1]{Sec.~\ref{#1}}}
\newrefformat{cha}{\hyperref[#1]{Chapter~\ref{#1}}}
\newrefformat{app}{\hyperref[#1]{Appendix~\ref{#1}}}

\newcommand{\T}{\mathrm{T}}

\usepackage{booktabs}
\usepackage[title]{appendix}
\usepackage{aligned-overset}

\makeatother

\begin{document}
\title{Effect of Synaptic Heterogeneity on Neuronal Coordination}
\author{Moritz Layer}
\affiliation{Institute of Neuroscience and Medicine (INM-6) and Institute for Advanced
Simulation (IAS-6) and JARA-Institute Brain Structure-Function Relationships
(INM-10), Jülich Research Centre, Jülich, Germany}
\affiliation{RWTH Aachen University, Aachen, Germany}
\author{Moritz Helias}
\affiliation{Institute of Neuroscience and Medicine (INM-6) and Institute for Advanced
Simulation (IAS-6) and JARA-Institute Brain Structure-Function Relationships
(INM-10), Jülich Research Centre, Jülich, Germany}
\affiliation{Department of Physics, Faculty 1, RWTH Aachen University, Aachen,
Germany}
\author{David Dahmen}
\email{d.dahmen@fz-juelich.de}

\affiliation{Institute of Neuroscience and Medicine (INM-6) and Institute for Advanced
Simulation (IAS-6) and JARA-Institute Brain Structure-Function Relationships
(INM-10), Jülich Research Centre, Jülich, Germany}
\date{\today}
\begin{abstract}
Recent advancements in measurement techniques have resulted in an
increasing amount of data on neural activities recorded in parallel,
revealing largely heterogeneous correlation patterns across neurons.
Yet, the mechanistic origin of this heterogeneity is largely unknown
because existing theoretical approaches linking structure and dynamics
in neural circuits are restricted to population-averaged connectivity
and activity. Here we present a systematic inclusion of heterogeneity
in network connectivity to derive quantitative predictions for neuron-resolved
covariances and their statistics in spiking neural networks. Our study
shows that the heterogeneity in covariances is not a result of variability
in single-neuron firing statistics but stems from the ubiquitously
observed sparsity and variability of connections in brain networks.
Linear-response theory maps these features to the effective connectivity
between neurons, which in turn determines neuronal covariances. Beyond-mean-field
tools reveal that synaptic heterogeneity modulates the variability
of covariances and thus the complexity of neuronal coordination across
many orders of magnitude.
\end{abstract}
\maketitle

\section{Introduction}

Neuronal networks in the brain display largely heterogeneous activity:
common observables such as firing rates \citep{Griffith1966_516,Koch1989_292,Dabrowska21_2632},
coefficients of variation (CVs) \citep{Shinomoto03_2823}, and pair-wise
correlations \citep{Ecker10,Cohen11_811,Dahmen19_13051,dahmen22_e68422}
are widely distributed across neurons. This has important implications
for coding and information processing in the brain, as the coordinated
activity across the enormous number of units in neuronal circuits
is thought to underlie all complex functions \citep{daSilveira13_1,Morenobote14_1410,Stringer19_361,Vyas20_249}.
The causes of heterogeneity in neuronal dynamics are diverse: intrinsic
neuronal properties, external inputs, and the network connectivity
itself are all sources of variability. While these structural and
dynamic heterogeneities can be readily observed with modern experimental
techniques \citep{Jun17_232,Steinmetz2021_6539,Campagnola2022_eabj5861},
understanding their mechanistic relations requires theoretical tools
that are currently still lacking.

In this study, we focus on the effects of connectivity and investigate
the influence of heterogeneity in connections on the activity of networks
of identical neurons receiving homogeneous external input. Previous
work \citep{Roxin11_16217} has shown that a considerable fraction
of the variance, in the distribution of firing rates across neurons
and in the CV of individual neurons' spike trains, in such networks
can already be explained by the distributed number of inputs the neurons
in a network receive. In this study, we go beyond single neuron activities
and focus on the statistics of pair-wise correlations and the related
covariances, which measure how strongly the activities of pairs of
neurons co-fluctuate. Such coordination builds the basis for collective
network activity and function.

\begin{figure*}
\centering{}\includegraphics{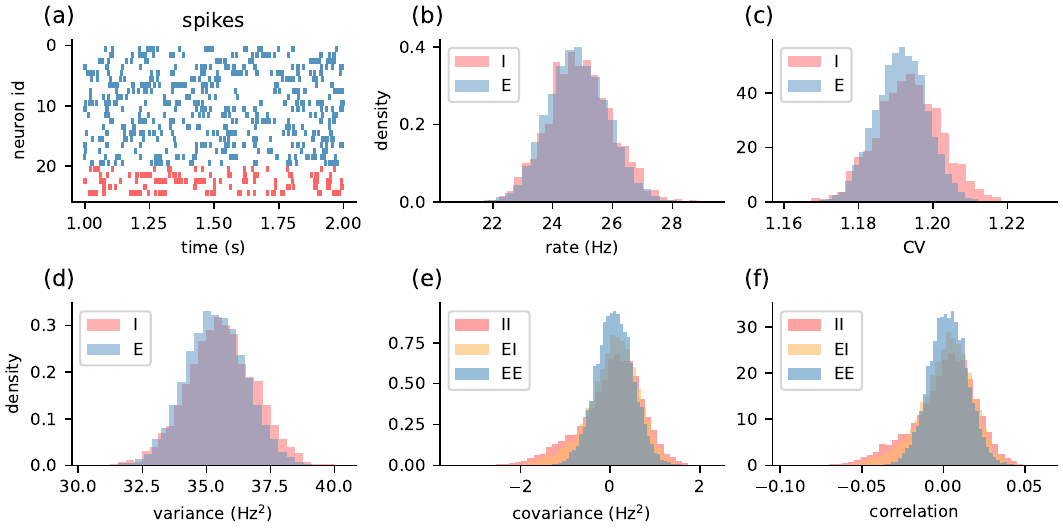}\caption{\label{fig:simulation}Simulation of excitatory-inhibitory (E-I) network
of leaky integrate-and-fire (LIF) neurons. (a) Spike trains of a sample
of 20 excitatory (E, blue) and 5 inhibitory (I, red) neurons. (b)
Distributions of firing rates. (c) Distributions of coefficients of
variation (CVs). (d) Distributions of variances of spike counts measured
in time bins of $1\mathrm{s}$ {[}cf. \prettyref{eq:spike_count_covariances}{]}.
(e) Distributions of spike-count covariances for different pairings
of neurons. (f) Distributions of spike-count correlation coefficients
for different pairings of neurons. For model details and simulation
parameters see Appendix \prettyref{appendix:nest_simulation} for
spectral radius $r=0.49$.}
\end{figure*}

With the exception of small organisms such as Caenorhabditis elegans
\citep{White86_1}, the microconnectome of most biological neuronal
networks is unknown. However, overall connectivity properties and
statistics, like the connection probabilities between different cortical
areas \citep{Markov2014_17,vanAlbada22_201} and cell types \citep{Campagnola2022_eabj5861},
the distance dependence of connections \citep{Schnepel15_3818,Campagnola2022_eabj5861},
or the statistics of synaptic strengths \citep{Sayer90_826,Feldmeyer1999_169,Song05_0507,Lefort09_301,Ikegaya13_293,Loewenstein11_9481,Campagnola2022_eabj5861}
are available nowadays. Hence, rather than a one-to-one relation between
microconnectome and pair-wise covariances \citep{Lindner05_061919,Pernice11_e1002059,Pernice12_031916,Trousdale12_e1002408,Grytskyy13_131,Dahmen16_031024},
a relation between connectivity and covariance on a statistical level
would readily allow the inclusion of this knowledge. To derive such
a relation, common population-level theories \citep{Ginzburg94,Vreeswijk96_1724,Lindner05_061919,Buice10_377,Renart10_587,Tetzlaff12_e1002596,Montbrio15}
cannot be used because they can only describe population-averaged
observables and, in particular, do not capture heterogeneity in covariances
within populations. Here, we instead employ mean-field theory on the
single-neuron level \citep{Trousdale12_e1002408}, which we systematically
compare to network simulations, and we go beyond mean-field theory
by including nontrivial fluctuation terms to obtain the statistics
of covariances between individual neuron pairs\@.

The main difficulty of a single-neuron level approach is that the
predictions of the theory for individual neurons strongly depend on
the specific details of the connectivity. To get a description on
the level of connectivity statistics, we perform a disorder average,
a technique originally developed for spin-glass systems \citep{Sompolinsky82_6860,Helias20_970}
that allows retaining information about the connectivity statistics
while averaging over the realization randomness. As our main results,
we show how to systematically calculate higher moments of neuronal
activity averaged over the disorder in the connectivity using replica
and beyond-mean-field theory, and we use this technique to derive
a relation between the mean and variance of covariances and the mean
and variance of the network connectivity. First results based on a
similar but reduced theoretical approach have already been successfully
applied in the neuroscientific context to infer the dynamical regime
of cortical networks \citep{Dahmen19_13051} and to explain spatial
properties of coordination structures \citep{dahmen22_e68422} and
dimensionality \citep{Dahmen22_365072v3}.

To summarize, we investigate the origin of neuronal coordination structures,
as experimentally observed across various species and cortical areas,
by analyzing covariances in a prototypical network model of cortical
dynamics \citep{Brunel00_183}, namely sparsely connected excitatory
and inhibitory neurons that operate in the balanced state \citep{Vreeswijk96_1724}.
In this model, all neurons have identical parameters and receive homogeneous,
uncorrelated external input. As in biological cortical networks, the
sparsity in the connectivity between neurons \citep{Campagnola2022_eabj5861}
as well as the wide distribution in synaptic amplitudes \citep{Sayer90_826,Feldmeyer1999_169,Song05_0507,Lefort09_301,Ikegaya13_293,Loewenstein11_9481,Campagnola2022_eabj5861}
constitute the source of variability in connections and thereby the
dynamics: Rates, CVs, variances, covariances, and hence correlation
coefficients are all described by distributions with sizable variance
(see \prettyref{fig:simulation}).

The following sections investigate the sources of the variance in
these quantities. \prettyref{sec:Background} introduces mean-field
theory on the single neuron level. In \prettyref{sec:Statistical-description-of},
we derive the main results on how to compute disorder-averaged moments
of neuronal activity, and we calculate explicit expressions for the
mean and variance of covariances. In \prettyref{sec:Discussion},
we discuss our findings and their limitations in the context of the
existing literature.

\section{Background: Linear-response theory of spiking neuronal networks on
a single-neuron level\label{sec:Background}}

To understand the origin of the distribution of covariances, we start
with analyzing a simulated network on a single-neuron level. The
example network throughout this study comprises $8000$ excitatory
(E) and $2000$ inhibitory (I) leaky integrate-and-fire (LIF) neurons
that make connections according to distinct population-specific statistics.
We mimic two abundant features of heterogeneity in connectivity of
brain circuits, that are sparse connections and distributed synaptic
weights. To do so we consider random sparse connectivity $\boldsymbol{J}$,
with connection probability $10\,\%$, giving rise to an excitatory
indegree $K_{\mathrm{\mathrm{E}}}=800$ and an inhibitory indegree
$K_{\mathrm{I}}=200$. To compensate for the imbalance in excitatory
and inhibitory neuron count, we follow the work by \citet{Brunel00_183}
and scale the strengths of existing inhibitory connections with respect
to excitatory ones by a factor $g=-6$ to obtain an asynchronous irregular
dynamic regime. In addition, we distribute synaptic weights of existing
connections according to population-specific normal distributions
$j_{\mathrm{E}}\propto\mathcal{N}\left(j,0.2j\right)$ and $j_{\mathrm{I}}\propto\mathcal{N}\left(gj,0.2j\right)$,
such that the overall heterogeneity in network connectivity is comprised
of the random sparseness of connections and the variable strength
of connections. For more details of the model, see Appendix \prettyref{appendix:nest_simulation}.

\paragraph{Working point}

Given the parameters of the simulated network of leaky integrate-and-fire
neurons, especially the specific realization of the connectivity matrix
$\boldsymbol{J}$, we determine the stationary \emph{working point},
comprising the input statistics $\left(\boldsymbol{\mu},\boldsymbol{\sigma}\right)$
and the firing rates\textbf{ $\boldsymbol{\nu}$}, as done by \citet{Brunel99}
and \citet{Brunel00_183}. To this end, we first neglect correlations
between the neurons and approximate the neurons' inputs as independent
Gaussian white noise processes. In this \emph{diffusion approximation},
the mean input $\mu_{i}$ and input variance $\sigma_{i}^{2}$ of
neuron $i$ are given by
\begin{eqnarray}
\mu_{i} & = & \tau_{m}\left(\sum_{j}J_{ij}\nu_{j}+j\nu_{\mathrm{ext},\mathrm{E}}+gj\nu_{\mathrm{ext},\mathrm{I}}+\frac{I_{\mathrm{ext}}}{C}\right)\,,\label{eq:mean-field_mean}\\
\sigma_{i}^{2} & = & \tau_{m}\left(\sum_{j}J_{ij}^{2}\nu_{j}+j^{2}\nu_{\mathrm{ext},\mathrm{E}}+g^{2}j^{2}\nu_{\mathrm{ext},\mathrm{I}}\right)\,,\label{eq:mean-field_var}
\end{eqnarray}
with membrane time constant $\tau_{\mathrm{m}}$, membrane capacitance
$C$, constant input current $I_{\mathrm{ext}}$, and excitatory and
inhibitory external Poisson noise with rates $\nu_{\mathrm{ext},\mathrm{E}}$
and $\nu_{\mathrm{ext},\mathrm{I}}$ which are fed into the system
via weights $j$ and $gj$, respectively. The firing rates are given
by the Siegert function \citep{Siegert51}
\begin{eqnarray}
\nu_{i} & = & \left\{ \tau_{\mathrm{r}}+\tau_{\mathrm{m}}\sqrt{\pi}\int_{y_{\mathrm{r},i}}^{y_{\mathrm{th},i}}\mathrm{d}s\,f(s)\right\} ^{-1}\,,\label{eq:siegert}\\
f\left(s\right) & = & \mathrm{e}^{s^{2}}\left[1+\mathrm{erf}\left(s\right)\right]\,,\nonumber 
\end{eqnarray}
with refractory period $\tau_{\mathrm{r}}$, and rescaled reset and
threshold voltages
\[
y_{\mathrm{r},i}=\frac{V_{\mathrm{r}}-\mu_{i}}{\sigma_{i}}\,,\quad y_{\mathrm{th},i}=\frac{V_{\mathrm{th}}-\mu_{i}}{\sigma_{i}}\,.
\]
These equations can be solved iteratively in a self-consistent manner.
Given the working point, we can determine the coefficients of variation
using (see Appendix A.1 of Ref. \citep{Brunel00_183}; note that they
use different units)
\begin{equation}
\mathrm{CV}_{i}^{2}=2\pi\left(\tau_{\mathrm{m}}\nu_{i}\right)^{2}\int_{y_{\mathrm{r},i}}^{y_{\mathrm{th},i}}\mathrm{d}x\:\mathrm{e}^{x^{2}}\int_{-\infty}^{x}\mathrm{d}z\,\mathrm{e}^{z^{2}}\left[1+\mathrm{erf}\left(z\right)\right]^{2}\,.\label{eq:cvs}
\end{equation}

\paragraph{Linearization}

The full dynamics of LIF neurons are nonlinear. However, as covariances
measure co-fluctuations of neurons around their working points, we
can study covariances by analyzing linearized dynamics as long as
the fluctuations are sufficiently small. Grytskyy et al. (see Sec.
5 of Ref. \citep{Grytskyy13_131}) show that a network of LIF neurons
can be mapped to a linear rate model with output noise
\begin{equation}
x_{i}\left(t\right)=\int_{-\infty}^{t}h(t-t^{\prime})\sum_{j}W_{ij}\left[x_{j}\left(t^{\prime}-d\right)+\xi_{j}\left(t^{\prime}-d\right)\right]\,\mathrm{d}t^{\prime},\label{eq:OUP}
\end{equation}
with neuronal activity $x_{i}\left(t\right)$, normalized linear response
kernel $h(t)$, synaptic delay $d$, and uncorrelated Gaussian white
noise $\xi_{i}\left(t\right)$, $\left\langle \xi_{i}\right\rangle =0$,
$\left\langle \xi_{i}\left(s\right)\xi_{j}\left(t\right)\right\rangle =D_{ij}\delta\left(s-t\right)$,
with diagonal noise strength matrix $D_{ij}=\delta_{ij}D_{ii}$. The
matrix $\boldsymbol{W}$, referred to as \emph{effective connectivity},
combines the connectivity matrix $\boldsymbol{J}$ with the sensitivity
of neurons to small fluctuations in their input. It is formally given
by the derivative of the stationary firing rate of neuron $i$ {[}\prettyref{eq:siegert}{]}
with respect to the firing rate of neuron $j$ evaluated at the stationary
working point (see Appendix A of Ref. \citep{Helias13_023002})
\begin{equation}
W_{ij}=\frac{\partial\nu_{i}}{\partial\nu_{j}}=\alpha_{i}J_{ij}+\beta_{i}J_{ij}^{2}\,,\label{eq:effective_weights}
\end{equation}
with
\begin{eqnarray*}
\alpha_{i} & = & \sqrt{\pi}\left(\tau_{\mathrm{m}}\nu_{i}\right)^{2}\frac{1}{\sigma_{i}}\left[f\left(y_{\mathrm{th},i}\right)-f\left(y_{\mathrm{r},i}\right)\right]\,,\\
\beta_{i} & = & \sqrt{\pi}\left(\tau_{\mathrm{m}}\nu_{i}\right)^{2}\frac{1}{2\sigma_{i}^{2}}\left[f\left(y_{\mathrm{th},i}\right)y_{\mathrm{th},i}-f\left(y_{\mathrm{r},i}\right)y_{\mathrm{r},i}\right]\,.
\end{eqnarray*}

\paragraph{Spike-count covariances}

In this study we are interested in spike-count covariances in spiking
networks
\begin{equation}
C_{ij}=\frac{1}{T}\left(\left\langle n_{i}n_{j}\right\rangle -\left\langle n_{i}\right\rangle \left\langle n_{j}\right\rangle \right)\,,\label{eq:spike_count_covariances}
\end{equation}
with spike counts $n_{i}$ occurring within bins of size $T$, where
the average, indicated by the brackets $\left\langle \cdot\right\rangle $,
is taken across all bins that can be viewed as trials with different
realizations of the external input. As shown in \citet[Materials and Methods]{Dahmen19_13051},
for stationary processes and large bin sizes spike-count covariances
$C_{ij}$ can be mapped to the time-lag integrated covariances $c_{ij}(\tau)$
between spike trains of neurons $i$ and $j$ (\citealp[see also][]{Tetzlaff08_2133,SheaBrown08_108102},
for more details on definitions of covariances see Appendix \prettyref{appendix:covariances_definitions})
\[
C_{ij}\overset{T\rightarrow\infty}{\rightarrow}\int_{-\infty}^{\infty}c_{ij}\left(\tau\right)\mathrm{d}\tau\,.
\]
In the following the term covariance always refers to $C_{ij}$. Making
use of the Wiener-Khinchin theorem (\prettyref{appendix:Wiener=002013Khinchin-theorem})
allows expressing the time-lag integrated covariances in terms of
the neuronal activities' Fourier components $X_{i}\left(\omega\right)$
at frequency zero 
\begin{equation}
C_{ij}=\left\langle X_{i}\left(0\right)X_{j}\left(0\right)\right\rangle \,,\label{eq:covs_wiener_khinchin}
\end{equation}
which can be evaluated by Fourier transforming \prettyref{eq:OUP},
yielding
\begin{equation}
\boldsymbol{C}=\left(\boldsymbol{1}-\boldsymbol{W}\right)^{-1}\boldsymbol{D}\left(\boldsymbol{1}-\boldsymbol{W}\right)^{-\T}\,.\label{eq:covs_standard_eq}
\end{equation}
For calculating the covariances, we therefore only need the effective
connectivity $\boldsymbol{W}$ and the noise strength $\boldsymbol{D}$.
The correlation coefficients follow as 
\[
\kappa_{ij}=\frac{C_{ij}}{\sqrt{C_{ii}C_{jj}}}\,.
\]

To estimate the noise strength $\boldsymbol{D}$, we assume that the
spike trains are described sufficiently well as renewal processes
for which the variances are given by \citep{Cox66}
\begin{equation}
C_{ii}=\mathrm{CV}_{i}^{2}\nu_{i}\,.\label{eq:variances}
\end{equation}
Using that $\boldsymbol{D}$ is by definition diagonal (see Appendix
\prettyref{appendix:validity_of_theoretical_predictions} for limitations
on exactly matching simulated spike-count covariances with a linear
rate model with uncorrelated white noise input), we can solve \prettyref{eq:covs_standard_eq}
for $\boldsymbol{D}$, which results in
\begin{equation}
D_{ii}=\sum_{j}\left(B^{-1}\right)_{ij}\mathrm{CV}_{j}^{2}\nu_{j}\,,\label{eq:noise_strength_estimate}
\end{equation}
with
\begin{equation}
B_{ij}=\left[\left(\boldsymbol{1}-\boldsymbol{W}\right)^{-1}\right]_{ij}^{2}\,.\label{eq:B}
\end{equation}

\begin{figure*}
\begin{centering}
\includegraphics{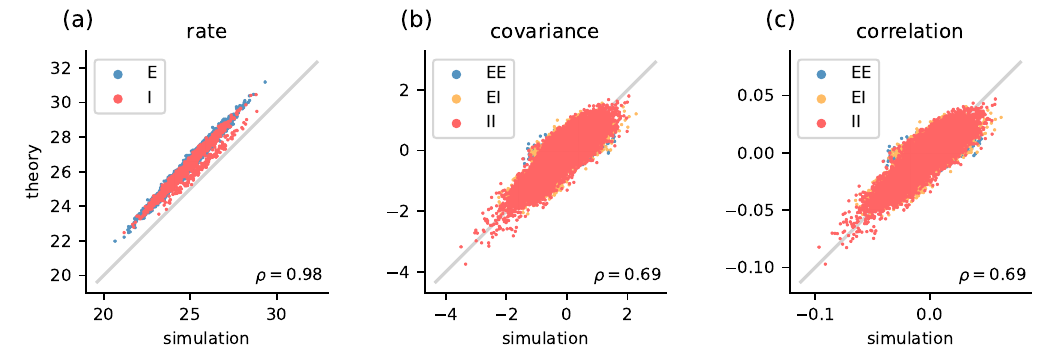}
\par\end{centering}
\centering{}\caption{\label{fig:thy_vs_sim}Simulation results and theoretical estimates
for E-I network of LIF neurons. Here $\rho$ denotes the Pearson correlation
coefficient. (a) Firing rates $\boldsymbol{\nu}$. (b) Covariances
$\boldsymbol{C}$. (c) Correlation coefficients $\boldsymbol{\kappa}$.
For model details and simulation parameters see Appendix \prettyref{appendix:nest_simulation}
for spectral radius $r=0.49$.}
\end{figure*}

The above expressions can be combined to compute theoretical estimates
of the quantities measured in the simulation. To solve the self-consistency
equations for the firing rates and to compute the covariances, we
make use of the Python package NNMT \citep{Layer22_835657}, which
includes optimized implementations of the equations introduced above.
A comparison of theoretical and simulation results is shown in \prettyref{fig:thy_vs_sim}.
For the chosen parameters, simulation and theory correlate strongly,
and the theory appears to capture the primary sources of heterogeneity
in the rates, covariances, and correlation coefficients. Note that
such a good match between theory and simulation cannot be observed
in all parameter regimes of the spiking network; the validity of the
assumptions made and the resulting theoretical estimates depend on
the network state (see Appendix \prettyref{appendix:validity_of_theoretical_predictions}
for further discussion on valid parameter regimes). \prettyref{fig:thy_vs_sim}
also reveals some unexplained variance, particularly pronounced in
the covariances and correlations. This variance is the result of the
finite simulation time and the associated uncertainty in the estimated
covariances. As we show in Appendix \prettyref{appendix:Bias-correction-of},
the covariance estimate bias can be significant and it can only be
corrected for on a statistical level rather than for individual covariances.
Focusing on the statistics of covariances, however, has further advantages:
For realistic network sizes, \prettyref{eq:covs_standard_eq} is a
high-dimensional equation that depends on each and every connection
in the network. Understanding general mechanisms relating network
structure and dynamics is therefore difficult. The covariance statistics
instead summarize the most important aspects of covariances and, for
large neuron populations, can be assumed to be self-averaging \citep{Fischer91,Hertz16_033001,Helias20_970},
which makes them less dependent on connectivity details. Second, \prettyref{eq:covs_standard_eq}
cannot be used for inference based on experimentally measured parameters
because as of yet it is not possible to determine the effective connectivity
or covariances of all neurons in a network. And last, as stated above,
we demonstrate that covariance statistics are more robust measures
than single-neuron covariances, both with respect to finite measurements
as well as to the assumptions made in the derivation above.

\section{Statistical description of covariances\label{sec:Statistical-description-of}}

Expression \eqref{eq:covs_standard_eq} reveals that the statistics
of the covariances $\boldsymbol{C}$, in particular their heterogeneity,
is determined by the statistics and heterogeneity of the effective
connectivity matrix $\boldsymbol{W}$ and the external noise strength
$\boldsymbol{D}$. Our aim here is to derive a description of the
cross-covariance statistics in terms of the statistics of $\boldsymbol{W}$
and $\boldsymbol{D}$. To this end, we derive analytical expressions
for the mean and the variance of the time-lag integrated cross-covariances
averaged over the heterogeneities of the system.

To do this, simply averaging \prettyref{eq:covs_standard_eq} is not
feasible due to $\boldsymbol{W}$ appearing in the inverse matrix
$\left(\boldsymbol{1}-\boldsymbol{W}\right)^{-1}$. Performing an
average over a random connectivity is, however, a well known problem
in the theory of disordered systems \citep{Fischer91,Sompolinsky88_259,Sommers88,Helias20_970},
where it is handled on the level of generating functions. To proceed
analogously, we start with \prettyref{eq:covs_wiener_khinchin}, which
expresses the covariances in terms of the moments of the dynamic variables'
Fourier components at frequency zero. This allows us to write the
covariances in terms of the moment-generating function $Z\left(\boldsymbol{J}\right)$
of the zero-frequency Fourier components $X_{i}$ of the dynamical
equation \eqref{eq:OUP} (see Appendix \prettyref{appendix:Derivation-of-moment}
for more details):
\[
C_{ij}=\left\langle X_{i}X_{j}\right\rangle =\left.\frac{\mathrm{\partial}}{\partial J_{i}}\frac{\mathrm{\partial}}{\partial J_{j}}Z\left(\boldsymbol{J}\right)\right|_{\boldsymbol{J}=0}\,,
\]
with
\begin{eqnarray}
Z\left(\boldsymbol{J}\right) & = & \frac{\widetilde{Z}\left(\boldsymbol{J}\right)}{\widetilde{Z}\left(\boldsymbol{0}\right)}\label{eq:mgf}\\
 & = & \left|\det\left(\boldsymbol{1}-\boldsymbol{W}\right)\right|\int\mathcal{D}\boldsymbol{X}\int\mathcal{D}\widetilde{\boldsymbol{X}}\nonumber \\
 &  & \times\exp\left[\widetilde{\boldsymbol{X}}^{\T}\left(\boldsymbol{1}-\boldsymbol{W}\right)\boldsymbol{X}+\frac{1}{2}\widetilde{\boldsymbol{X}}^{\T}\boldsymbol{D}\widetilde{\boldsymbol{X}}+\boldsymbol{J}^{\T}\boldsymbol{X}\right]\,,\nonumber 
\end{eqnarray}
and $\widetilde{Z}\left(\boldsymbol{0}\right)=\left|\det\left(1-\boldsymbol{W}\right)\right|^{-1}$
the nontrivial normalization of the unnormalized moment-generating
function $\widetilde{Z}\left(\boldsymbol{J}\right)$. Here, $\widetilde{\boldsymbol{X}}$
are auxiliary variables that can be used to calculate the response
function $\left\langle X_{i}\widetilde{X}_{j}\right\rangle $ of neuron
$i$ to a perturbation of neuron $j$ by introducing additional sources
$\widetilde{\boldsymbol{J}}$ in the moment-generating function (see
Appendix \prettyref{appendix:Derivation-of-moment}). Equation \eqref{eq:mgf}
shows that calculating the disorder-average of the covariances boils
down to calculating the disorder-average of the moment-generating
function. In the following two sections, we use this approach to calculate
the mean of the cross-covariances $\left\langle \boldsymbol{C}\right\rangle _{\boldsymbol{W},\boldsymbol{D}}$
and subsequently the variance of the cross-covariances $\left\langle \delta\boldsymbol{C}^{2}\right\rangle _{\boldsymbol{W},\boldsymbol{D}}$,
where $\left\langle \cdot\right\rangle _{\boldsymbol{W},\boldsymbol{D}}$
refers to the average over the randomness in $\boldsymbol{W}$ and
$\boldsymbol{D}$.

\subsection{Mean of cross-covariances\label{subsec:Mean-of-cross-covariances}}

\begin{figure}
\centering{}\includegraphics{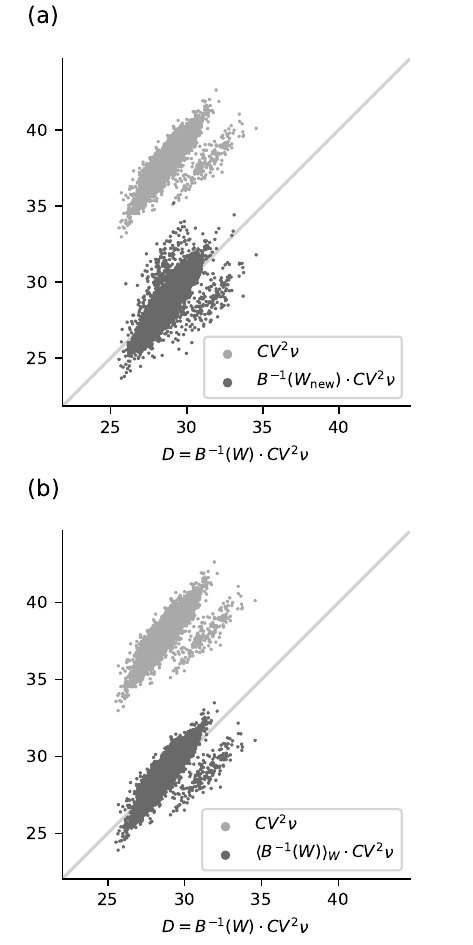}\caption{\label{fig:noise_strength_autocorr}Effect of averaging noise strength
over disorder in $\boldsymbol{W}$ and effect of applying $\boldsymbol{B}^{-1}$
on variances. (a) Noise strength computed using the procedure described
above (\emph{noise strength} in the following) vs. noise strength
computed using a new realization of $\boldsymbol{W}$ (dark gray),
and noise strength vs. variances (light gray). (b) Noise strength
vs. noise strength computed using an average over $100$ realizations
of $\boldsymbol{W}$ (dark gray) and noise strength vs. variances
(light gray). Same excitatory-inhibitory network model as in previous
figures. For model details and simulation parameters see Appendix
\prettyref{appendix:nest_simulation} for spectral radius $r=0.49$.}
\end{figure}

\paragraph{Disorder average}

We begin with the mean cross-covariances, focusing first on the average
over the ensemble of connectivities, indicated in the following by
$\left\langle \cdot\right\rangle _{\boldsymbol{W}}$. In the moment-generating
function {[}\prettyref{eq:mgf}{]}, $\boldsymbol{W}$ occurs linearly
in the exponent of $\widetilde{Z}\left(\boldsymbol{J}\right)$, which
is advantageous for performing the disorder average. However, the
averaging procedure is complicated by two aspects: (1) $\boldsymbol{W}$
contributes to the noise strength $\boldsymbol{D}$ through the variance-rescaling
matrix $\boldsymbol{B}^{-1}$, and (2) the normalization $\widetilde{Z}\left(\boldsymbol{0}\right)$
depends on $\boldsymbol{W}$. However, as illustrated in \prettyref{fig:noise_strength_autocorr},
in practice the first point does not appear to be a problem: \prettyref{fig:noise_strength_autocorr}(a)
indicates that the specifics of $\boldsymbol{D}$ are largely determined
by the details of the variances $\boldsymbol{\mathrm{CV}}^{2}\boldsymbol{\nu}$,
because a different realization of $\boldsymbol{W}$ essentially yields
a similar $\boldsymbol{D}$, and \prettyref{fig:noise_strength_autocorr}(b)
suggests that the effect of the disorder average on $\boldsymbol{D}$
is minimal. For these reasons, we treat $\boldsymbol{D}$ as though
it was independent of the explicit realization of $\boldsymbol{W}$.
To address the second point, an alternative approach based on the
moment-generating functional for the full time-dependent dynamics
(see Appendix \prettyref{appendix:Derivation-of-moment}) could be
utilized. This moment-generating functional has a unit determinant
normalization independent of $\boldsymbol{W}$ \citep{DeDominicis78_4913}.
The disorder average of its frequency space complement, however, introduces
cross-frequency couplings that complicate the further analysis. Here,
instead, we follow \citet{Dahmen19_13051}, and separate the averages
over $\widetilde{Z}\left(\boldsymbol{J}\right)$ and $\widetilde{Z}\left(\boldsymbol{0}\right)$,
\begin{eqnarray}
\left\langle Z\left(\boldsymbol{J}\right)\right\rangle _{\boldsymbol{W}} & =\left\langle \frac{\widetilde{Z}\left(\boldsymbol{J}\right)}{\widetilde{Z}\left(\boldsymbol{0}\right)}\right\rangle _{\boldsymbol{W}}\approx & \frac{\left\langle \widetilde{Z}\left(\boldsymbol{J}\right)\right\rangle _{\boldsymbol{W}}}{\left\langle \widetilde{Z}\left(\boldsymbol{0}\right)\right\rangle _{\boldsymbol{W}}}\,,\label{eq:mgf_normalization}
\end{eqnarray}
as we find that this factorization approach does yield accurate results.
This leaves us with the task of calculating $\left\langle \widetilde{Z}\left(\boldsymbol{J}\right)\right\rangle _{\boldsymbol{W}}$.\footnote{Note that a systematic approach to this factorization approximation
would be to employ the replica trick $\widetilde{Z}\left(\boldsymbol{0}\right)^{-1}=\lim_{n\rightarrow0}\widetilde{Z}\left(\boldsymbol{0}\right)^{n-1}$
and jointly average $\left\langle \widetilde{Z}\left(\boldsymbol{J}\right)\widetilde{Z}\left(\boldsymbol{0}\right)^{n-1}\right\rangle _{\boldsymbol{W}}$
in the limit $n\rightarrow0$, or to average the joint moment-generating
functional for all time points or frequencies that by construction
has a trivial normalization $\widetilde{Z}\left(\boldsymbol{0}\right)=1$
(see \prettyref{appendix:Derivation-of-moment}).}

The disorder average only affects the coupling term and can be expressed
using the moment-generating functions $\phi_{ij}$ of $W_{ij}$,
\begin{align*}
\left\langle \exp\left(-\widetilde{\boldsymbol{X}}^{\mathrm{T}}\boldsymbol{W}\boldsymbol{X}\right)\right\rangle _{\boldsymbol{W}} & =\left\langle \prod_{i,j}\exp\left(-W_{ij}\widetilde{X}_{i}X_{j}\right)\right\rangle _{\boldsymbol{W}}\\
 & =\prod_{i,j}\phi_{ij}\left(-\widetilde{X}_{i}X_{j}\right)\,,
\end{align*}
for independently drawn connections $W_{ij}\sim p_{ij}\left(W_{ij}\right)$.
The moment-generating function can be written in terms of a cumulant
expansion $\phi_{ij}\left(X\right)=\exp\left(\sum_{k=1}^{\infty}\kappa_{k,ij}X^{k}/k!\right)$,
with $k$-th cumulants $\kappa_{k,ij}\,$. For fixed connection probability,
the number of inputs to a neuron scales with the network size $N$.
To keep the input and its fluctuations finite when increasing the
network size, we require synaptic weights to scale with $1/\sqrt{N}$
\citep{Vreeswijk96_1724,Vreeswijk98_1321}, such that the cumulant
expansion is an expansion in $1/\sqrt{N}$. A truncation at the second
cumulant ($\propto N^{-1}$) maps $\boldsymbol{W}$ to a Gaussian
connectivity with distribution $\mathcal{N}\left(\boldsymbol{M},\boldsymbol{\Delta}/N\right)$,
such that
\begin{eqnarray}
\left\langle \widetilde{Z}\left(\boldsymbol{J}\right)\right\rangle _{W} & = & \int\mathcal{D}\boldsymbol{X}\int\mathcal{D}\widetilde{\boldsymbol{X}}\label{eq:cumulant_expanded_mgf}\\
 &  & \times\exp\left[S_{0}\left(\boldsymbol{X},\widetilde{\boldsymbol{X}}\right)+\boldsymbol{J}^{\mathrm{T}}\boldsymbol{X}\right]\nonumber \\
 &  & \times\exp\left[\frac{1}{2N}\sum_{i,j}\Delta_{ij}\widetilde{X}_{i}\widetilde{X}_{i}X_{j}X_{j}\right]\,,
\end{eqnarray}
with 
\[
S_{0}\left(\boldsymbol{X},\widetilde{\boldsymbol{X}}\right)=\widetilde{\boldsymbol{X}}^{\mathrm{T}}\left(\boldsymbol{1}-\boldsymbol{M}\right)\boldsymbol{X}+\frac{1}{2}\widetilde{\boldsymbol{X}}^{\mathrm{T}}\boldsymbol{D}\widetilde{\boldsymbol{X}}\,,
\]
and mean connection weights $M_{ij}=\mathcal{O}\left(1/\sqrt{N}\right)$
as well as variances $\Delta_{ij}=\mathcal{O}(1)$. The higher-order
cumulants are suppressed by the large network size and therefore do
not significantly affect the dynamics of the network. As we show below,
our theory with the Gaussian approximation of the connectivity therefore
faithfully recovers the correlation statistics in the sparse excitatory-inhibitory
spiking network with population-specific connection statistics.

\paragraph{Auxiliary field formulation}

To deal with the four-point coupling term in \prettyref{eq:cumulant_expanded_mgf},
we define auxiliary variables $Q_{i}:=\frac{1}{N}\sum_{j}\Delta_{ij}X_{j}X_{j}$,
which we formally introduce by inserting an identity in the form of
a Fourier-transformed delta distribution:
\begin{eqnarray*}
1 & = & \prod_{i}\int\mathrm{d}Q_{i}\,\delta\left(\frac{1}{N}\sum_{j}\Delta_{ij}X_{j}X_{j}-Q_{i}\right)\\
 & = & \prod_{i}N\int_{-\infty}^{\infty}\mathrm{d}Q_{i}\int_{-\mathrm{i}\infty}^{\mathrm{i}\infty}\frac{\mathrm{d}\widetilde{Q}_{i}}{2\pi\mathrm{i}}\,\\
 &  & \times\exp\left[\widetilde{Q}_{i}\left(\sum_{j}\Delta_{ij}X_{j}X_{j}-NQ_{i}\right)\right]\,.
\end{eqnarray*}
The auxiliary variables $\widetilde{Q}_{i}$ are introduced to express
the delta distribution as an integral. This leads to
\begin{eqnarray}
\left\langle \widetilde{Z}\left(\boldsymbol{J}\right)\right\rangle _{\boldsymbol{W}} & = & \int\mathcal{D}\boldsymbol{Q}\int\mathcal{D}\widetilde{\boldsymbol{Q}}\,\exp\left(-N\boldsymbol{Q}^{T}\widetilde{\boldsymbol{Q}}\right)\label{eq:interaction_with_aux_fields}\\
 & \times & \underbrace{\int\mathcal{D}\boldsymbol{X}\int\mathcal{D}\widetilde{\boldsymbol{X}}\,\exp\left[S_{\boldsymbol{Q},\widetilde{\boldsymbol{Q}}}\left(\boldsymbol{X},\widetilde{\boldsymbol{X}}\right)+\boldsymbol{J}^{\mathrm{T}}\boldsymbol{X}\right]}_{=:\widetilde{Z}_{\boldsymbol{Q},\widetilde{\boldsymbol{Q}}}\left(\boldsymbol{J}\right)}\,,\nonumber \\
S_{\boldsymbol{Q},\widetilde{\boldsymbol{Q}}}\left(\boldsymbol{X},\widetilde{\boldsymbol{X}}\right) & = & \widetilde{\boldsymbol{X}}^{\mathrm{T}}\left(\boldsymbol{1}-\boldsymbol{M}\right)\boldsymbol{X}\nonumber \\
 & + & \frac{1}{2}\widetilde{\boldsymbol{X}}^{\mathrm{T}}\left[\boldsymbol{D}+\mathrm{diag}\left(\boldsymbol{Q}\right)\right]\widetilde{\boldsymbol{X}}\nonumber \\
 & + & \boldsymbol{X}^{\T}\mathrm{diag}\left(\widetilde{\boldsymbol{Q}}^{\T}\boldsymbol{\Delta}\right)\boldsymbol{X}\,.\label{eq:action_s_q(x)}
\end{eqnarray}
Here $\mathrm{diag}\left(\boldsymbol{Q}\right)_{ij}$ refers to a
diagonal matrix with diagonal elements $Q_{i}$. As the action $S_{\boldsymbol{Q},\widetilde{\boldsymbol{Q}}}\left(\boldsymbol{X},\widetilde{\boldsymbol{X}}\right)$
at fixed auxiliary variables describes an auxiliary free theory, \prettyref{eq:interaction_with_aux_fields}
describes the activity of linear rate neurons in a network with disorder-averaged
connectivity $\boldsymbol{M}$ that interact with fluctuating external
variables $\boldsymbol{Q}$ and $\widetilde{\boldsymbol{Q}}$. Inserting
\prettyref{eq:interaction_with_aux_fields} into \prettyref{eq:mgf_normalization}
yields
\begin{eqnarray*}
\left\langle Z\left(\boldsymbol{J}\right)\right\rangle _{\boldsymbol{W}} & = & \int\mathcal{D}\boldsymbol{Q}\int\mathcal{D}\widetilde{\boldsymbol{Q}}\,\frac{\widetilde{Z}_{\boldsymbol{Q},\widetilde{\boldsymbol{Q}}}\left(\boldsymbol{J}\right)}{\widetilde{Z}_{\boldsymbol{Q},\widetilde{\boldsymbol{Q}}}\left(\boldsymbol{0}\right)}\\
 &  & \times\frac{\exp\left(-N\boldsymbol{Q}^{T}\widetilde{\boldsymbol{Q}}\right)\widetilde{Z}_{\boldsymbol{Q},\widetilde{\boldsymbol{Q}}}\left(\boldsymbol{0}\right)}{\int\mathcal{D}\boldsymbol{P}\int\mathcal{D}\widetilde{\boldsymbol{P}}\,\exp\left(-N\boldsymbol{P}^{T}\widetilde{\boldsymbol{P}}\right)\widetilde{Z}_{\boldsymbol{P},\widetilde{\boldsymbol{P}}}\left(\boldsymbol{0}\right)}\\
 & =: & \int\mathcal{D}\boldsymbol{Q}\int\mathcal{D}\widetilde{\boldsymbol{Q}}\,p\left(\boldsymbol{Q},\widetilde{\boldsymbol{Q}}\right)Z_{\boldsymbol{Q},\widetilde{\boldsymbol{Q}}}\left(\boldsymbol{J}\right)\,,
\end{eqnarray*}
with joint probability distribution
\begin{eqnarray}
p\left(\boldsymbol{Q},\widetilde{\boldsymbol{Q}}\right) & = & \frac{\exp\left(-S\left(\boldsymbol{Q},\widetilde{\boldsymbol{Q}}\right)\right)}{\int\mathcal{D}\boldsymbol{P}\int\mathcal{D}\widetilde{\boldsymbol{P}}\,\exp\left(-S\left(\boldsymbol{P},\widetilde{\boldsymbol{P}}\right)\right)}\,,\nonumber \\
S\left(\boldsymbol{Q},\widetilde{\boldsymbol{Q}}\right) & = & N\boldsymbol{Q}^{T}\widetilde{\boldsymbol{Q}}-\ln\left[\widetilde{Z}_{\boldsymbol{Q},\widetilde{\boldsymbol{Q}}}\left(\boldsymbol{0}\right)\right]\,,\label{eq:q_action}
\end{eqnarray}
and properly normalized moment generating function $Z_{\boldsymbol{Q},\widetilde{\boldsymbol{Q}}}\left(\boldsymbol{J}\right)=\widetilde{Z}_{\boldsymbol{Q},\widetilde{\boldsymbol{Q}}}\left(\boldsymbol{J}\right)/\widetilde{Z}_{\boldsymbol{Q},\widetilde{\boldsymbol{Q}}}\left(\boldsymbol{0}\right)$.
These equations imply that the disorder-average of arbitrary moments
$\left\langle X_{i_{1}}\cdots X_{i_{k}}\right\rangle $ can be calculated
by determining the corresponding moments $\left\langle X_{i_{1}}\cdots X_{i_{k}}\right\rangle _{\boldsymbol{Q},\widetilde{\boldsymbol{Q}}}$
with respect to the auxiliary free theory and averaging them over
the auxiliary variables:
\begin{eqnarray}
\left\langle \left\langle X_{i_{1}}\cdots X_{i_{k}}\right\rangle \right\rangle _{\boldsymbol{W}} & = & \int\mathcal{D}\boldsymbol{Q}\int\mathcal{D}\widetilde{\boldsymbol{Q}}\,\label{eq:disorder_average_mapped_to_q_average}\\
 &  & \times p\left(\boldsymbol{Q},\tilde{\boldsymbol{Q}}\right)\left\langle X_{i_{1}}\cdots X_{i_{k}}\right\rangle _{\boldsymbol{Q},\widetilde{\boldsymbol{Q}}}\,,\nonumber \\
\left\langle X_{i_{1}}\cdots X_{i_{k}}\right\rangle _{\boldsymbol{Q},\widetilde{\boldsymbol{Q}}} & = & \left.\frac{\partial}{\partial J_{i_{1}}}\cdots\frac{\partial}{\partial J_{i_{k}}}Z_{\boldsymbol{Q},\widetilde{\boldsymbol{Q}}}\left(\boldsymbol{J}\right)\right|_{\boldsymbol{J}=\boldsymbol{0}}\,.\nonumber 
\end{eqnarray}

\paragraph{Saddle-point approximation}

Due to the prefactor $N$ in \prettyref{eq:q_action} and the scalar
products in $\widetilde{Z}_{\boldsymbol{Q},\widetilde{\boldsymbol{Q}}}\left(\boldsymbol{0}\right)$
with $N$ contributions, we expect $p\left(\boldsymbol{Q},\widetilde{\boldsymbol{Q}}\right)$
to peak sharply for $N\rightarrow\infty$, such that we can perform
a saddle-point approximation. To lowest order, we expect $p\left(\boldsymbol{Q},\widetilde{\boldsymbol{Q}}\right)\approx\delta\left(\boldsymbol{Q}-\boldsymbol{Q}^{*}\right)\delta\left(\widetilde{\boldsymbol{Q}}-\widetilde{\boldsymbol{Q}}^{*}\right)$,
with the saddle-point $\boldsymbol{Q}^{*},\widetilde{\boldsymbol{Q}}^{*}$
determined by
\[
\left.\frac{\partial}{\partial\widetilde{Q}_{i}}S\left(\boldsymbol{Q},\widetilde{\boldsymbol{Q}}\right)\right|_{\boldsymbol{Q}^{*},\widetilde{\boldsymbol{Q}}^{*}}=0\,,\quad\left.\frac{\partial}{\partial Q_{i}}S\left(\boldsymbol{Q},\widetilde{\boldsymbol{Q}}\right)\right|_{\boldsymbol{Q}^{*},\widetilde{\boldsymbol{Q}}^{*}}=0\,,
\]
which yields

\begin{eqnarray}
Q_{i}^{*} & = & \frac{1}{N}\sum_{j}\Delta_{ij}\left\langle X_{j}X_{j}\right\rangle _{\boldsymbol{Q}^{*},\widetilde{\boldsymbol{Q}}^{*}}\,,\label{eq:saddle-1}\\
\widetilde{Q}_{i}^{*} & = & \frac{1}{2N}\left\langle \widetilde{X}_{i}\widetilde{X}_{i}\right\rangle _{\boldsymbol{Q}^{*},\widetilde{\boldsymbol{Q}}^{*}}\,,\nonumber 
\end{eqnarray}
with second moments evaluated at the saddle point. The moments can
be calculated explicitly by solving the Gaussian integrals (see Appendix
\prettyref{appendix:Saddle-points-and}). Using the shorthand $\boldsymbol{R}\coloneqq\left(\boldsymbol{1}-\boldsymbol{M}\right)^{-1}$,
we find $\left\langle X_{i}X_{j}\right\rangle _{\boldsymbol{Q}^{*},\widetilde{\boldsymbol{Q}}^{*}}=\left\{ \boldsymbol{R}\left[\boldsymbol{D}+\mathrm{diag}\left(\boldsymbol{Q}^{*}\right)\right]\boldsymbol{R}^{\T}\right\} _{ij}$
and $\left\langle \widetilde{X}_{i}\widetilde{X}_{i}\right\rangle _{\boldsymbol{Q}^{*},\widetilde{\boldsymbol{Q}}^{*}}=0$,
and solving for the saddle point yields
\begin{eqnarray*}
Q_{i}^{*} & = & \frac{1}{N}\sum_{j,k,l,m}\left(\boldsymbol{1}-\frac{1}{N}\Delta\cdot\boldsymbol{R}^{\circ2}\right)_{ij}^{-1}\Delta_{jk}R_{kl}D_{lm}R_{km}\,,\\
\widetilde{Q}_{i}^{*} & = & 0\,,
\end{eqnarray*}
with $\boldsymbol{R}^{\circ2}\coloneqq\boldsymbol{R}\odot\boldsymbol{R}$
and $\odot$ denoting the element-wise (Hadamard) product.

Finally, making use of the Wiener-Khinchin theorem {[}\prettyref{eq:covs_khinchin}{]}
and inserting the solution of the saddle-point equations into \prettyref{eq:disorder_average_mapped_to_q_average}
yields the mean covariances averaged across the disorder of the connectivity:
\begin{align*}
\left\langle \boldsymbol{C}\right\rangle _{\boldsymbol{W}} & =\left\langle \boldsymbol{X}\boldsymbol{X}^{\T}\right\rangle _{\boldsymbol{Q}^{*},\widetilde{\boldsymbol{Q}}^{*}}\\
 & =\left(\boldsymbol{1}-\boldsymbol{M}\right)^{-1}\left[\boldsymbol{D}+\mathrm{diag}\left(\boldsymbol{Q}^{*}\right)\right]\left(\boldsymbol{1}-\boldsymbol{M}\right)^{-\T}\,.
\end{align*}
Averaging over the disorder in $\boldsymbol{D}$ then yields 
\begin{align}
\left\langle \boldsymbol{C}\right\rangle _{\boldsymbol{W},\boldsymbol{D}}\label{eq:mean_cov-1}\\
=(\boldsymbol{1} & -\boldsymbol{M})^{-1}\left\{ \overline{\boldsymbol{D}}+\mathrm{diag}\left[\boldsymbol{Q}^{*}\left(\boldsymbol{D}=\overline{\boldsymbol{D}}\right)\right]\right\} \left(\boldsymbol{1}-\boldsymbol{M}\right)^{-\T}\,.\nonumber 
\end{align}
Here $\overline{\boldsymbol{D}}$ denotes the disorder-averaged noise
strength {[}cf. \prettyref{fig:noise_strength_autocorr}(b) and discussion
after \prettyref{eq:var_of_cov}{]}. Note that the saddle point $\boldsymbol{Q}^{*}\left(\boldsymbol{D}=\overline{\boldsymbol{D}}\right)$
together with $\overline{\boldsymbol{D}}$ yields an effective noise
strength which shifts average variances and covariances. Importantly,
it is only the heterogeneity in the connectivity $\boldsymbol{W}$
that causes this shift. Average covariances are insensitive to heterogeneity
in the noise strengths $\boldsymbol{D}$; they only depend on the
average $\overline{\boldsymbol{D}}$.

\subsection{Variance of cross-covariances}

\paragraph{Replica method}

Calculating the variances of covariances across the ensemble of possible
network connectivities
\begin{eqnarray}
\left\langle \delta\boldsymbol{C}^{2}\right\rangle _{\boldsymbol{W}} & = & \left\langle \boldsymbol{C}^{\circ2}\right\rangle _{\boldsymbol{W}}-\left\langle \boldsymbol{C}\right\rangle _{\boldsymbol{W}}^{\circ2},\label{eq:disorder_var_cov}
\end{eqnarray}
requires making use of the replica method \citep{ZinnJustin96,Helias20_970}
and deriving an expression for the disorder-averaged moment-generating
function of the replicated system $\left\langle Z\left(\boldsymbol{J}\right)Z\left(\boldsymbol{K}\right)\right\rangle _{\boldsymbol{W}}$,
as this allows calculating disorder averages of arbitrary squared
moments $\left\langle \left\langle X_{i_{1}}\cdots X_{i_{k}}\right\rangle ^{2}\right\rangle _{\boldsymbol{W}}$,
which occur in the first term in \prettyref{eq:disorder_var_cov}.
The procedure is completely analogous to the previous section's derivations.
However, the disorder average now affects the term 
\begin{alignat*}{1}
\left\langle \exp\left(\widetilde{\boldsymbol{X}}^{\mathrm{T}}\boldsymbol{W}\boldsymbol{X}+\widetilde{\boldsymbol{Y}}^{\mathrm{T}}\boldsymbol{W}\boldsymbol{Y}\right)\right\rangle _{\boldsymbol{W}}\\
=\prod_{i,j}\exp\Bigg[\sum_{k=1}^{\infty}\frac{\kappa_{k,ij}}{k!} & \left(\widetilde{X}_{i}X_{j}+\widetilde{Y}_{i}Y_{j}\right){}^{k}\Bigg]\,,
\end{alignat*}
where $\boldsymbol{X}$ and $\boldsymbol{Y}$ refer to the activity
in the first and second replicon, respectively. A cumulant expansion
up to second order introduces --- along four-point couplings separately
in \textbf{$\boldsymbol{X}$} and $\boldsymbol{Y}$ similar to the
one in \prettyref{eq:cumulant_expanded_mgf} --- a replica coupling
term 
\[
\exp\left(\frac{1}{N}\sum_{ij}\Delta_{ij}\widetilde{X}_{i}\widetilde{Y}_{i}X_{j}Y_{j}\right)\,.
\]
To deal with the four-point couplings, we again introduce auxiliary
variables
\begin{eqnarray*}
Q_{XX,i} & = & \frac{1}{N}\sum_{j}\Delta_{ij}X_{j}X_{j}\,,\\
Q_{YY,i} & = & \frac{1}{N}\sum_{j}\Delta_{ij}Y_{j}Y_{j}\,,\\
Q_{XY,i} & = & \frac{1}{N}\sum_{j}\Delta_{ij}X_{j}Y_{j}\,,
\end{eqnarray*}
and obtain a relation similar to \prettyref{eq:disorder_average_mapped_to_q_average},
\begin{alignat}{1}
\left\langle \left\langle X_{i_{1}}\cdots X_{i_{k}}\right\rangle ^{2}\right\rangle _{\boldsymbol{W}} & =\int\mathcal{D}\boldsymbol{Q}\int\mathcal{D}\widetilde{\boldsymbol{Q}}\,\label{eq:disorder_average_mapped_to_q_average_replica}\\
\times p\left(\boldsymbol{Q},\widetilde{\boldsymbol{Q}}\right) & \left\langle X_{i_{1}}\cdots X_{i_{k}}Y_{i_{1}}\cdots Y_{i_{k}}\right\rangle _{\boldsymbol{Q},\widetilde{\boldsymbol{Q}}}\,,\nonumber \\
\left\langle X_{i_{1}}\cdots X_{i_{k}}Y_{i_{1}}\cdots Y_{i_{k}}\right\rangle _{\boldsymbol{Q},\widetilde{\boldsymbol{Q}}}\nonumber \\
=\frac{\partial}{\partial J_{i_{1}}}\cdots\frac{\partial}{\partial J_{i_{k}}} & \frac{\partial}{\partial K_{i_{1}}}\cdots\left.\frac{\partial}{\partial K_{i_{k}}}Z_{\boldsymbol{Q},\widetilde{\boldsymbol{Q}}}\left(\boldsymbol{J},\boldsymbol{K}\right)\right|_{\boldsymbol{J},\boldsymbol{K}=\boldsymbol{0}}\,,\nonumber 
\end{alignat}
but with
\begin{alignat}{1}
p\left(\boldsymbol{Q},\widetilde{\boldsymbol{Q}}\right) & =\frac{\exp\left(-S\left(\boldsymbol{Q},\widetilde{\boldsymbol{Q}}\right)\right)}{\int\mathcal{D}\boldsymbol{P}\int\mathcal{D}\widetilde{\boldsymbol{P}}\,\exp\left(-S\left(\boldsymbol{P},\widetilde{\boldsymbol{P}}\right)\right)}\,,\nonumber \\
\mathcal{S}\left(\boldsymbol{Q},\widetilde{\boldsymbol{Q}}\right) & =N\boldsymbol{Q}_{XX}^{\T}\widetilde{\boldsymbol{Q}}_{XX}+N\boldsymbol{Q}_{XY}^{\T}\widetilde{\boldsymbol{Q}}_{XY}\label{eq:replica_action}\\
 & +N\boldsymbol{Q}_{YY}^{\T}\widetilde{\boldsymbol{Q}}_{YY}-\ln\widetilde{Z}_{\boldsymbol{Q},\widetilde{\boldsymbol{Q}}}\left(\boldsymbol{0},\boldsymbol{0}\right)\,,\nonumber \\
\widetilde{Z}_{\boldsymbol{Q},\widetilde{\boldsymbol{Q}}}\left(\boldsymbol{J},\boldsymbol{K}\right) & =\int\mathcal{D}\boldsymbol{X}\int\mathcal{D}\widetilde{\boldsymbol{X}}\int\mathcal{D}\boldsymbol{Y}\int\mathcal{D}\widetilde{\boldsymbol{Y}}\nonumber \\
\times\exp & \Big[S_{\boldsymbol{Q}_{XX},\widetilde{\boldsymbol{Q}}_{XX}}\left(\boldsymbol{X},\widetilde{\boldsymbol{X}}\right)+S_{\boldsymbol{Q}_{YY},\widetilde{\boldsymbol{Q}}_{YY}}\left(\boldsymbol{Y},\widetilde{\boldsymbol{Y}}\right)\nonumber \\
 & +\widetilde{\boldsymbol{X}}^{\T}\mathrm{diag}\left(\boldsymbol{Q}_{XY}\right)\widetilde{\boldsymbol{Y}}+\boldsymbol{X}^{\T}\mathrm{diag}\left(\widetilde{\boldsymbol{Q}}_{XY}^{\T}\boldsymbol{\Delta}\right)\boldsymbol{Y}\nonumber \\
 & +\boldsymbol{J}^{\T}\boldsymbol{X}+\boldsymbol{K}^{\T}\boldsymbol{Y}\Big]\,,\nonumber \\
Z_{\boldsymbol{Q},\widetilde{\boldsymbol{Q}}}\left(\boldsymbol{J},\boldsymbol{K}\right) & =\widetilde{Z}_{\boldsymbol{Q},\widetilde{\boldsymbol{Q}}}\left(\boldsymbol{J},\boldsymbol{K}\right)/\widetilde{Z}_{\boldsymbol{Q},\widetilde{\boldsymbol{Q}}}\left(\boldsymbol{0},\boldsymbol{0}\right)\,,\nonumber 
\end{alignat}
where $\boldsymbol{Q}$ and $\widetilde{\boldsymbol{Q}}$ are shorthand
notations denoting all auxiliary variables, and $S_{\boldsymbol{Q}_{XX},\widetilde{\boldsymbol{Q}}_{XX}}\left(\boldsymbol{X},\widetilde{\boldsymbol{X}}\right)$
and $S_{\boldsymbol{Q}_{YY},\widetilde{\boldsymbol{Q}}_{YY}}\left(\boldsymbol{Y},\widetilde{\boldsymbol{Y}}\right)$
are given by \prettyref{eq:action_s_q(x)}.

\paragraph{Saddle-point approximation}

As in \prettyref{subsec:Mean-of-cross-covariances}, we approximate
$p\left(\boldsymbol{Q},\widetilde{\boldsymbol{Q}}\right)$ as a delta
function at the saddle-point $\boldsymbol{Q}^{*},\widetilde{\boldsymbol{Q}}^{*}$
(for details see Appendix \prettyref{appendix:Saddle-points-and}),
and with \prettyref{eq:disorder_average_mapped_to_q_average_replica}
to lowest order we get 
\begin{eqnarray}
\left\langle C_{ij}^{2}\right\rangle _{\boldsymbol{W}} & = & \left\langle \left\langle X_{i}X_{j}\right\rangle ^{2}\right\rangle _{\boldsymbol{W}}\nonumber \\
 & = & \int\mathcal{D}\boldsymbol{Q}\int\mathcal{D}\widetilde{\boldsymbol{Q}}\,p\left(\boldsymbol{Q},\widetilde{\boldsymbol{Q}}\right)\left\langle X_{i}X_{j}Y_{i}Y_{j}\right\rangle _{\boldsymbol{Q},\widetilde{\boldsymbol{Q}}}\nonumber \\
 & = & \int\mathcal{D}\boldsymbol{Q}\int\mathcal{D}\widetilde{\boldsymbol{Q}}\,p\left(\boldsymbol{Q},\widetilde{\boldsymbol{Q}}\right)\label{eq:covariance_after_wick}\\
 &  & \times\Big(\left\langle X_{i}X_{j}\right\rangle _{\boldsymbol{Q},\widetilde{\boldsymbol{Q}}}\left\langle Y_{i}Y_{j}\right\rangle _{\boldsymbol{Q},\widetilde{\boldsymbol{Q}}}\nonumber \\
 &  & +\left\langle X_{i}Y_{i}\right\rangle _{\boldsymbol{Q},\widetilde{\boldsymbol{Q}}}\left\langle X_{j}Y_{j}\right\rangle _{\boldsymbol{Q},\widetilde{\boldsymbol{Q}}}\nonumber \\
 &  & +\left\langle X_{i}Y_{j}\right\rangle _{\boldsymbol{Q},\widetilde{\boldsymbol{Q}}}\left\langle X_{j}Y_{i}\right\rangle _{\boldsymbol{Q},\widetilde{\boldsymbol{Q}}}\Big)\nonumber \\
 & \approx & \left\langle X_{i}X_{j}\right\rangle _{\boldsymbol{Q}^{*},\widetilde{\boldsymbol{Q}}^{*}}\left\langle Y_{i}Y_{j}\right\rangle _{\boldsymbol{Q}^{*},\widetilde{\boldsymbol{Q}}^{*}}\nonumber \\
 &  & +\left\langle X_{i}Y_{i}\right\rangle _{\boldsymbol{Q}^{*},\widetilde{\boldsymbol{Q}}^{*}}\left\langle X_{j}Y_{j}\right\rangle _{\boldsymbol{Q}^{*},\widetilde{\boldsymbol{Q}}^{*}}\nonumber \\
 &  & +\left\langle X_{i}Y_{j}\right\rangle _{\boldsymbol{Q}^{*},\widetilde{\boldsymbol{Q}}^{*}}\left\langle X_{j}Y_{i}\right\rangle _{\boldsymbol{Q}^{*},\widetilde{\boldsymbol{Q}}^{*}}\nonumber \\
 & = & \left\langle X_{i}X_{j}\right\rangle _{\boldsymbol{Q}^{*},\widetilde{\boldsymbol{Q}}^{*}}\left\langle Y_{i}Y_{j}\right\rangle _{\boldsymbol{Q}^{*},\widetilde{\boldsymbol{Q}}^{*}}\nonumber \\
 & \approx & \left\langle \left\langle X_{i}X_{j}\right\rangle \right\rangle _{\boldsymbol{W}}^{2}\nonumber \\
 & = & \left\langle C_{ij}\right\rangle _{\boldsymbol{W}}^{2}\,,\label{eq:no_var_cov_for_lowest_order}
\end{eqnarray}
where we used Wick's theorem, which is allowed by the fact that, for
$\boldsymbol{Q}$ and $\widetilde{\boldsymbol{Q}}$ given and fixed,
$\widetilde{Z}_{\boldsymbol{Q},\widetilde{\boldsymbol{Q}}}\left(\boldsymbol{J},\boldsymbol{K}\right)$
describes a Gaussian theory, and the fact that all cross-replica correlators
$\left\langle X_{i}Y_{j}\right\rangle _{\boldsymbol{Q}^{*},\widetilde{\boldsymbol{Q}}^{*}}$
vanish at the saddle point (see Appendix \prettyref{appendix:Saddle-points-and}).

\paragraph{Fluctuations around the saddle-point}

Equation \eqref{eq:no_var_cov_for_lowest_order} implies that the
variance of covariances is zero in the saddle-point approximation,
and we need to account for Gaussian fluctuations of the auxiliary
fields around their saddle points by making a Gaussian approximation
of $p\left(\boldsymbol{Q},\widetilde{\boldsymbol{Q}}\right)$. The
crucial fluctuations are the ones of $\boldsymbol{Q}_{XY}$ and $\widetilde{\boldsymbol{Q}}_{XY}$,
as they can potentially preserve the replica coupling and thus lead
to nonvanishing variance contributions of cross-replica correlators
$\left\langle X_{i}Y_{i}\right\rangle _{\boldsymbol{Q}^{*},\widetilde{\boldsymbol{Q}}^{*}}$.
Away from the saddle points, the correlators in \prettyref{eq:covariance_after_wick}
depend on $\boldsymbol{Q}$ and $\widetilde{\boldsymbol{Q}}$ in a
complicated manner. To render the integrals in \prettyref{eq:covariance_after_wick}
solvable in the Gaussian approximation, we perform a Taylor expansion
of the correlators around the saddle points $\boldsymbol{Q}^{*},\widetilde{\boldsymbol{Q}}^{*}$,
which effectively is an expansion of $\widetilde{Z}_{\boldsymbol{Q},\widetilde{\boldsymbol{Q}}}\left(\boldsymbol{J},\boldsymbol{K}\right)$
(see Appendix \prettyref{appendix:fluctuations_around_saddle-points}
for more details). In the first term of \prettyref{eq:covariance_after_wick},
leading order fluctuations in $\boldsymbol{Q}_{XY}$ and $\widetilde{\boldsymbol{Q}}_{XY}$
depend on correlators with an odd number of variables of each replicon.
Therefore, this term cannot yield a contribution to the variance due
to fluctuations of $\boldsymbol{Q}_{XY}$ and $\widetilde{\boldsymbol{Q}}_{XY}$.
The major replica coupling arises from the second and third terms
in \prettyref{eq:covariance_after_wick}. We note that the third term
contains off-diagonal elements of correlators which are suppressed
by a factor $1/N$ with respect to the diagonal ones. Therefore, we
can neglect this term for cross-covariances as well and only keep
the second term in \prettyref{eq:covariance_after_wick} as the leading
order contribution. For autocovariances the second and third terms
in \prettyref{eq:covariance_after_wick} are the same, yielding an
additional factor $2$. Introducing $\delta\boldsymbol{Q}=\boldsymbol{Q}-\boldsymbol{Q}^{*}$
and defining $\delta\widetilde{\boldsymbol{Q}}$ equivalently, we
obtain
\begin{alignat}{1}
\left\langle X_{i}Y_{i}\right\rangle _{\boldsymbol{Q},\widetilde{\boldsymbol{Q}}} & =\sum_{k}\left\langle X_{i}Y_{i}\widetilde{X}_{k}\widetilde{Y}_{k}\right\rangle _{\boldsymbol{Q}^{*},\widetilde{\boldsymbol{Q}}^{*}}\delta Q_{XY,k}\nonumber \\
+\sum_{k,l} & \Delta_{kl}\left\langle X_{i}Y_{i}X_{l}Y_{l}\right\rangle _{\boldsymbol{Q}^{*},\widetilde{\boldsymbol{Q}}^{*}}\delta\widetilde{Q}_{XY,k}+\mathcal{O}\left(\left|\delta\boldsymbol{Q}\right|^{2},\left|\delta\widetilde{\boldsymbol{Q}}\right|^{2}\right)\nonumber \\
 & =\sum_{k}\left\langle X_{i}\widetilde{X}_{k}\right\rangle _{\boldsymbol{Q}^{*},\widetilde{\boldsymbol{Q}}^{*}}^{2}\delta Q_{XY,k}\nonumber \\
+\sum_{k,l} & \Delta_{kl}\left\langle X_{i}X_{l}\right\rangle _{\boldsymbol{Q}^{*},\widetilde{\boldsymbol{Q}}^{*}}^{2}\delta\widetilde{Q}_{XY,k}+\mathcal{O}\left(\left|\delta\boldsymbol{Q}\right|^{2},\left|\delta\widetilde{\boldsymbol{Q}}\right|^{2}\right)\,,\label{eq:expanded_correlator}
\end{alignat}
where we used that cross-replica correlators vanish at the saddle
point. Inserting the above fluctuation expansion result around $\boldsymbol{Q}_{XY}^{*}$
and $\widetilde{\boldsymbol{Q}}_{XY}^{*}$ into \prettyref{eq:covariance_after_wick}
leads to
\begin{alignat}{1}
\int & \mathcal{D}\boldsymbol{Q}\int\mathcal{D}\widetilde{\boldsymbol{Q}}\,p\left(\boldsymbol{Q},\widetilde{\boldsymbol{Q}}\right)\left\langle X_{i}Y_{i}\right\rangle _{\boldsymbol{Q},\widetilde{\boldsymbol{Q}}}\left\langle X_{j}Y_{j}\right\rangle _{\boldsymbol{Q},\widetilde{\boldsymbol{Q}}}\nonumber \\
= & \sum_{k,l}\left\langle X_{i}\widetilde{X}_{k}\right\rangle _{\boldsymbol{Q}^{*},\widetilde{\boldsymbol{Q}}^{*}}^{2}\left\langle X_{j}\widetilde{X}_{l}\right\rangle _{\boldsymbol{Q}^{*},\widetilde{\boldsymbol{Q}}^{*}}^{2}\left\langle \delta Q_{XY,k}\delta Q_{XY,l}\right\rangle _{\boldsymbol{Q},\widetilde{\boldsymbol{Q}}}\nonumber \\
+ & \sum_{k,l,m}\left\langle X_{i}\widetilde{X}_{k}\right\rangle _{\boldsymbol{Q}^{*},\widetilde{\boldsymbol{Q}}^{*}}^{2}\Delta_{lm}\left\langle X_{j}X_{m}\right\rangle _{\boldsymbol{Q}^{*},\widetilde{\boldsymbol{Q}}^{*}}^{2}\left\langle \delta Q_{XY,k}\delta\widetilde{Q}_{XY,l}\right\rangle _{\boldsymbol{Q},\widetilde{\boldsymbol{Q}}}\nonumber \\
+ & \sum_{k,l,m}\left\langle X_{j}\widetilde{X}_{k}\right\rangle _{\boldsymbol{Q}^{*},\widetilde{\boldsymbol{Q}}^{*}}^{2}\Delta_{lm}\left\langle X_{i}X_{m}\right\rangle _{\boldsymbol{Q}^{*},\widetilde{\boldsymbol{Q}}^{*}}^{2}\left\langle \delta Q_{XY,k}\delta\widetilde{Q}_{XY,l}\right\rangle _{\boldsymbol{Q},\widetilde{\boldsymbol{Q}}}\nonumber \\
+ & \sum_{k,l,m,n}\left\langle X_{i}X_{m}\right\rangle _{\boldsymbol{Q}^{*},\widetilde{\boldsymbol{Q}}^{*}}^{2}\left\langle X_{j}X_{n}\right\rangle _{\boldsymbol{Q}^{*},\widetilde{\boldsymbol{Q}}^{*}}^{2}\Delta_{km}\Delta_{ln}\nonumber \\
 & \quad\times\left\langle \delta\widetilde{Q}_{XY,k}\delta\widetilde{Q}_{XY,l}\right\rangle _{\boldsymbol{Q},\widetilde{\boldsymbol{Q}}}\,.\label{eq:covs_second_term_gaussian_expanded}
\end{alignat}

Next, we consider the Gaussian approximation of $p\left(\boldsymbol{Q},\widetilde{\boldsymbol{Q}}\right)$
with

\[
\mathcal{S}\left(\boldsymbol{Q},\widetilde{\boldsymbol{Q}}\right)=\mathcal{S}\left(\boldsymbol{Q}^{*},\widetilde{\boldsymbol{Q}}^{*}\right)+\frac{1}{2}\left(\delta\boldsymbol{Q}_{XY},\delta\widetilde{\boldsymbol{Q}}_{XY}\right)\boldsymbol{\mathcal{S}}^{(2)}\left(\begin{array}{c}
\delta\boldsymbol{Q}_{XY}\\
\delta\widetilde{\boldsymbol{Q}}_{XY}
\end{array}\right)\,,
\]
where $\boldsymbol{\mathcal{S}}^{(2)}$ contains the second derivatives
with respect to the auxiliary fields
\[
\boldsymbol{\mathcal{S}}^{(2)}=\left(\begin{array}{cc}
\left.\frac{\partial\mathcal{S}\left(\boldsymbol{Q},\widetilde{\boldsymbol{Q}}\right)}{\partial\boldsymbol{Q}_{XY}\partial\boldsymbol{Q}_{XY}}\right|_{\boldsymbol{Q}^{*},\widetilde{\boldsymbol{Q}}^{*}} & \left.\frac{\partial\mathcal{S}\left(\boldsymbol{Q},\widetilde{\boldsymbol{Q}}\right)}{\partial\boldsymbol{Q}_{XY}\partial\widetilde{\boldsymbol{Q}}_{XY}}\right|_{\boldsymbol{Q}^{*},\widetilde{\boldsymbol{Q}}^{*}}\\
\left.\frac{\partial\mathcal{S}\left(\boldsymbol{Q},\widetilde{\boldsymbol{Q}}\right)}{\partial\widetilde{\boldsymbol{Q}}_{XY}\partial\boldsymbol{Q}_{XY}}\right|_{\boldsymbol{Q}^{*},\widetilde{\boldsymbol{Q}}^{*}} & \left.\frac{\partial\mathcal{S}\left(\boldsymbol{Q},\widetilde{\boldsymbol{Q}}\right)}{\partial\widetilde{\boldsymbol{Q}}_{XY}\partial\widetilde{\boldsymbol{Q}}_{XY}}\right|_{\boldsymbol{Q}^{*},\widetilde{\boldsymbol{Q}}^{*}}
\end{array}\right)\,,
\]
which allows evaluating the correlators of the auxiliary fields in
\prettyref{eq:covs_second_term_gaussian_expanded} (see Appendix \prettyref{appendix:Correlators-of-auxiliary}
for details). Inserting the results, to leading order we find (see
Appendix \prettyref{appendix:Disorder-averaged-variance-of} for details)
\begin{eqnarray}
\left\langle C_{ij}^{2}\right\rangle _{\boldsymbol{W}} & = & \left(1+\delta_{ij}\right)\biggl[\left(\boldsymbol{1}-\frac{1}{N}\boldsymbol{R}^{\T\circ2}\boldsymbol{\Delta}\right)^{-1}\left\langle \boldsymbol{X}\boldsymbol{X}^{\T}\right\rangle _{\boldsymbol{Q}^{*},\widetilde{\boldsymbol{Q}}^{*}}^{\circ2}\nonumber \\
 &  & \left(\boldsymbol{1}-\frac{1}{N}\boldsymbol{R}^{\T\circ2}\boldsymbol{\Delta}\right)^{-\T}\biggr]_{ij}\nonumber \\
 &  & -\delta_{ij}\left\langle C_{ij}\right\rangle _{\boldsymbol{W}}^{2}\,.\label{eq:disorder_averaged_second_moment}
\end{eqnarray}
To get the variances rather than the second moments, we subtract the
squared mean covariances $\left\langle C_{ij}\right\rangle _{\boldsymbol{W}}^{2}$.
However, for the setup that we study here the squared mean cross-covariances
are of the order $\mathcal{O}\left(1/N^{2}\right)$ and therefore
negligible. Taking into account that $\boldsymbol{R}=\left(\boldsymbol{1}-\boldsymbol{M}\right)^{-1}\approx\boldsymbol{1}$,
which holds as long as the network is inhibition dominated \footnote{$\boldsymbol{M}$ scales as $\mathcal{O}(N^{-1/2})$ and in inhibition-dominated
networks, eigenvalues of $\boldsymbol{M}$ are far away from the divergence
at $1$.}, we find the following expression for the disorder-averaged variance
of cross-covariances (see Appendix \prettyref{appendix:Disorder-averaged-variance-of}
for full expression)

\begin{eqnarray}
\left\langle \delta\boldsymbol{C}^{2}\right\rangle _{\boldsymbol{W}} & = & \left(\boldsymbol{1}-\boldsymbol{S}\right)^{-1}\left(\boldsymbol{D}+\mathrm{diag}\left[\boldsymbol{Q}^{*}\left(\boldsymbol{D}\right)\right]\right)^{2}\left(\boldsymbol{1}-\boldsymbol{S}\right)^{-\T},\label{eq:var_of_cov}
\end{eqnarray}
where we wrote $\boldsymbol{S}=\boldsymbol{\Delta}/N$.

However, if the noise strength \textbf{$\boldsymbol{D}$} has to be
estimated using \prettyref{eq:noise_strength_estimate}, this expression
is still dependent on the specific realization of $\boldsymbol{W}$,
both implicitly through the estimates of the single-neuron rates and
CVs described in \prettyref{sec:Background} and explicitly through
the matrix $\boldsymbol{B}$ {[}\prettyref{eq:B}{]}. Since the right
hand side of \prettyref{eq:var_of_cov} depends nonlinearly on $\boldsymbol{D}$,
averaging over the statistics of $\boldsymbol{D}$ introduces terms
depending on the heterogeneity of $\boldsymbol{D}$. However, \prettyref{fig:cov_stats_dependence_on_D_heterogeneity}
in the Appendix shows that heterogeneity in $\boldsymbol{D}$ ---
both via the explicit dependence on $\boldsymbol{W}$ and via the
implicit dependence through distributed firing rates and CVs ---
is negligible for the statistics of cross-covariances. This can be
understood by considering the structure of \prettyref{eq:var_of_cov}:
The matrices $\left(\boldsymbol{1}-\boldsymbol{S}\right)^{-1}$ are
multiplied with $\boldsymbol{D}$, such that any heterogeneity in
$\boldsymbol{D}$ is averaged out. An E-I network is an illustrative
example, with $\left(\boldsymbol{1}-\boldsymbol{S}\right)^{-1}=\boldsymbol{1}+\boldsymbol{U}$
with a $2\times2$ block matrix $\boldsymbol{U}$ whose entries are
homogeneous in each population block, such that the matrix product
effectively is an average over $\boldsymbol{D}$.

To obtain an average $\boldsymbol{D}$ that is not depending on a
specific realization of $\boldsymbol{W}$, we follow \prettyref{eq:noise_strength_estimate}
and set 
\begin{equation}
\overline{D}_{ii}=\sum_{j}\left(\boldsymbol{1}-\boldsymbol{S}\right)_{ij}\cdot\overline{\mathrm{CV}_{j}^{2}}\overline{\nu_{j}}\,,\label{eq:noise_strength_estimate_realization_independent}
\end{equation}
which inserted into the disorder-averaged expression for the autocovariances
{[}\prettyref{eq:mean_cov-1}{]} yields the correct autocovariances:
\begin{align}
\left\langle C_{ii}\right\rangle {}_{\boldsymbol{W},\boldsymbol{D}} & =\left[\left(\boldsymbol{1}-\boldsymbol{M}\right)^{-1}\left\{ \overline{\boldsymbol{D}}+\mathrm{diag}\left[\boldsymbol{Q}^{*}\left(\overline{\boldsymbol{D}}\right)\right]\right\} \left(\boldsymbol{1}-\boldsymbol{M}\right)^{-\T}\right]_{ii}\nonumber \\
 & \approx\left\{ \overline{\boldsymbol{D}}+\mathrm{diag}\left[\left(\boldsymbol{1}-\boldsymbol{S}\right)^{-1}\boldsymbol{S}\cdot\mathrm{diag}\left(\overline{\boldsymbol{D}}\right)\right]\right\} _{ii}\label{eq:approx_D}\\
 & =\sum_{j}\left(\boldsymbol{1}-\boldsymbol{S}\right)_{ij}^{-1}\overline{D}_{jj}\nonumber \\
 & =\overline{\mathrm{CV}_{i}^{2}}\overline{\nu_{i}}\,.\nonumber 
\end{align}
Here we used $\left(\boldsymbol{1}-\boldsymbol{M}\right)^{-1}\approx\boldsymbol{1}$
and $\boldsymbol{Q}^{*}\approx\left(\boldsymbol{1}-\boldsymbol{S}\right)^{-1}\boldsymbol{S}\cdot\mathrm{diag}\left(\boldsymbol{D}\right)$.
The realization-independent estimates $\overline{\nu_{i}}$ and $\overline{\mathrm{CV}_{i}^{2}}$
of the rates and CVs, respectively, can be obtained using standard
population-resolved mean-field theory \citep{Brunel99,Brunel00_183},
which only requires knowing the statistics of $\boldsymbol{W}$. A
procedure similar to the one described in \prettyref{sec:Background}
can be used: In the population view, however, the indices $i,j$ no
longer denote single neurons but rather populations of equal neurons.
In \prettyref{eq:mean-field_mean} $J_{ij}$ is replaced by $K_{ij}J_{ij}$
and $J_{ij}^{2}$ in \prettyref{eq:mean-field_var} is replaced by
$K_{ij}J_{ij}^{2}$, where $K_{ij}$ is the indegree from population
$j$ to population $i$, and $J_{ij}$ then is interpreted as the
mean synaptic weight from population $j$ to population $i$.

Replacing $\boldsymbol{D}$ in \prettyref{eq:var_of_cov} by \prettyref{eq:noise_strength_estimate_realization_independent}
yields a fully realization-independent disorder-averaged estimate
of the variance of cross-covariances.

\subsection{Singularities}

Next, we discuss the interpretation of the derived formulas. Thereto,
we need to have a closer look at the effective noise strength $\boldsymbol{D}+\mathrm{diag}\left[\boldsymbol{Q}^{*}\left(\boldsymbol{D}\right)\right]$,
which occurs in both the mean {[}\prettyref{eq:mean_cov-1}{]} and
the variances {[}\prettyref{eq:var_of_cov}{]} of covariances. Using
\prettyref{eq:noise_strength_estimate_realization_independent}, we
find that the impact of heterogeneity on the effective noise cancels:
\begin{align}
\mathrm{diag}\left(\boldsymbol{D}\right)+\boldsymbol{Q}^{*}\left(\boldsymbol{D}\right) & \approx\mathrm{diag}\left(\boldsymbol{D}\right)+\left(\boldsymbol{1}-\boldsymbol{S}\right)^{-1}\boldsymbol{S}\cdot\mathrm{diag}\left(\boldsymbol{D}\right)\nonumber \\
 & =\left(\boldsymbol{1}-\boldsymbol{S}\right)^{-1}\cdot\mathrm{diag}\left(\boldsymbol{D}\right)\label{eq:intermediate1}\\
 & =\left(\boldsymbol{1}-\boldsymbol{S}\right)^{-1}\left(\boldsymbol{1}-\boldsymbol{S}\right)\cdot\boldsymbol{a}\nonumber \\
 & \approx\boldsymbol{a}\:,\nonumber 
\end{align}
where $a_{i}=\overline{\mathrm{CV}_{i}^{2}}\overline{\nu_{i}}$ is
the vector of estimated autocovariances. This is because we specifically
choose the noise strength $\boldsymbol{D}$ such that autocovariances
match those from the spiking networks: As heterogeneity is increased,
external fluctuations get amplified by the factor $\left(\boldsymbol{1}-\boldsymbol{S}\right)^{-1}$
in \prettyref{eq:intermediate1}. To achieve that autocovariances
do not diverge, external inputs need to be scaled down according to
\prettyref{eq:noise_strength_estimate_realization_independent}. Hence,
the mean and variance of cross-covariances are given by
\begin{align}
\left\langle \boldsymbol{C}\right\rangle _{\boldsymbol{W},\boldsymbol{D}} & \approx\left(\boldsymbol{1}-\boldsymbol{M}\right)^{-1}\mathrm{diag}\left(\boldsymbol{a}\right)\left(\boldsymbol{1}-\boldsymbol{M}\right)^{-\T}\,,\label{eq:mean_covs_simplified}\\
\left\langle \delta\boldsymbol{C}^{2}\right\rangle _{\boldsymbol{W,D}} & \approx\left(\boldsymbol{1}-\boldsymbol{S}\right)^{-1}\mathrm{diag}\left(\boldsymbol{a}^{2}\right)\left(\boldsymbol{1}-\boldsymbol{S}\right)^{-\T}\,.\label{eq:var_covs_simplified}
\end{align}
Note that any inverse matrix can be written as $\boldsymbol{A}^{-1}=\det\left(\boldsymbol{A}\right)^{-1}\mathrm{adj}\left(\boldsymbol{A}\right)$,
where $\mathrm{adj}\left(\boldsymbol{A}\right)$ denotes the adjugate
matrix. As a result, the elements of an inverse matrix $\boldsymbol{A}^{-1}$
diverge if the determinant of the matrix $\boldsymbol{A}$ vanishes,
which occurs when at least one eigenvalue of $\boldsymbol{A}$ is
zero. Therefore, the divergence behavior of the mean and variance
of covariances is determined by the eigenvalues of $\boldsymbol{M}$
and $\boldsymbol{S}$ with real parts close to $1$\@.

\prettyref{eq:mean_covs_simplified} reveals that mean cross-covariances
are determined by the mean connectivity $\boldsymbol{M}$. By choosing
$\boldsymbol{D}$ to match the autocovariances of the spiking model,
they are, in particular, unaffected by network heterogeneity, represented
by $\boldsymbol{S}$. A range of important network properties, such
as population structure determining E-I balance \citep{Vreeswijk96_1724,Renart10_587,Tetzlaff12_e1002596,Helias13_023002,Helias14},
spatial structure like distance-dependent connection probabilities
\citep{Rosenbaum14,Rosenbaum17_018103,Darshan18_031072,Smith18_1600,Huang19,dahmen22_e68422},
or low-rank structures \citep{Mastrogiuseppe18_609}, can be encoded
in $\boldsymbol{M}$. Divergences in mean covariances, caused by eigenvalues
of $\boldsymbol{M}$ close to $1$, can thus be indicative of phenomena
like loss of E-I balance with excessive excitation (cf. Fig. 8D of
Ref. \citealp{Helias14}) or instability of the homogeneously active
state in spatially organized networks \citep{Kriener14}.

Variances of cross-covariances are determined by network heterogeneity
{[}\prettyref{eq:var_covs_simplified}{]}, encoded in the connectivity
variance $\boldsymbol{S}$ and are to leading order independent of
the mean connectivity $\boldsymbol{M}$. Note that subleading terms
nevertheless can become sizable if eigenvalues of $\boldsymbol{M}$
are close to the instability line at $1$. As demonstrated by \citet{Aljadeff16_022302},
if $\boldsymbol{S}$ is a block structured matrix encoding different
populations, its eigenvalue spectrum is circular, with a spectral
radius $r$ that is determined by the square root of the maximum eigenvalue
of $\boldsymbol{S}$.

For the E-I network studied here, the matrices $\boldsymbol{M}$
and $\boldsymbol{S}$ have nontrivial block structure with one excitatory
and one inhibitory block (see Appendix \prettyref{appendix:Inference-of-connectivity}).
In this case, the spectral radius is given by \citep{Rajan06}
\[
r^{2}=N_{\mathrm{E}}\sigma_{\mathrm{E}}^{2}+N_{\mathrm{I}}\sigma_{\mathrm{I}}^{2}\,.
\]
The spectral radius, a measure of network heterogeneity, increases
when the variance of synaptic strength grows, which is controlled
by an interplay between the connection probabilities of different
populations and the variances of the associated synaptic weights.
Intuitively, as explained in \citet{dahmen22_e68422}, multisynaptic
signal transmission is very efficient in a network with a large spectral
radius, such that pairs of neurons influence each other via a large
number of neuronal pathways, possibly including differing numbers
of excitatory and inhibitory neurons. The effects of these various
pathways add up, and the large variety of potential pathways results
in a broad distribution of covariances.

We see that the effects of $\boldsymbol{M}$ and $\boldsymbol{S}$
are mostly independent of one another, allowing the mean and variance
of covariances to vary separately. This, however, applies only to
synaptic weights that are identically and independently distributed.
If the weights are correlated, such as through chain structures in
the connectivity, the respective eigenvalues cannot be changed independently.
A more detailed analysis of this behavior is to be published elsewhere.
As a final remark, it is worth noting that the independence of the
mean covariances of $\boldsymbol{S}$ confirms that previously employed
population models \citep{Renart10_587,Tetzlaff12_e1002596,Helias13_023002,Grytskyy13_131,Helias14},
which neglect the variance of connectivity, are valid for computing
mean covariances.

To illustrate how the mean and variance of covariances change as functions
of the network heterogeneity, we plot \prettyref{eq:mean_cov-1} and
\prettyref{eq:var_of_cov} with \prettyref{eq:noise_strength_estimate_realization_independent}
for spectral radii between $0$ and $1$ in \prettyref{fig:Covariance-statistics-at}
(predicted linear). We kept the working point roughly constant for
the different spectral radii by maintaining the mean $\mu$ and variance
$\sigma^{2}$ of the total input to each neuron while modifying the
synaptic efficacy. To compensate for the increased intrinsic input
and fluctuations at larger spectral radii, we reduced the mean and
fluctuations of the external input.

Confirming the discussion of \prettyref{eq:mean_covs_simplified}
and \prettyref{eq:var_covs_simplified}, when the spectral radius
is modified, the variances of covariances vary by several orders of
magnitude, whereas mean covariances remain in the same order of magnitude.
A range of prior research \citep{Tetzlaff12_e1002596,Grytskyy13_131,Helias14}
has shown that a divergence of mean covariances would be observed
as a function of E-I balance, e.g. by altering $g$. Here we focus
on network scenarios away from the excitatory instability (fixed $g=-6$)
and therefore do not see a divergence of mean covariances. Nevertheless,
we observe a change of mean covariances when changing the spectral
radius. This is because in the sparse random network chosen here,
the variance of the synaptic weights is not independent from the mean
of the weights. Adjusting the spectral radius requires modifying the
weights, resulting in the residual change in the mean covariances
visible in \prettyref{fig:Covariance-statistics-at}(a-c). Note that
by keeping the working point of the network constant across spectral
radii, we also keep the noise strength factor in \prettyref{eq:mean_cov-1}
and \prettyref{eq:var_of_cov} constant {[}cf. \prettyref{eq:mean_covs_simplified}
and \prettyref{eq:var_covs_simplified}{]}. If the external noise
strength was instead determined by a fixed external process, i.e $\boldsymbol{D}$
independent of $\boldsymbol{W}$, then mean covariances would also
diverge as a function of the spectral radius due to the factor $\left(\boldsymbol{1}-\boldsymbol{S}\right)^{-1}$,
which enters the noise strength term via $\boldsymbol{Q}^{*}$.

\subsection{Comparison of prediction and measurement of covariance statistics}

\begin{figure*}
\begin{centering}
\includegraphics{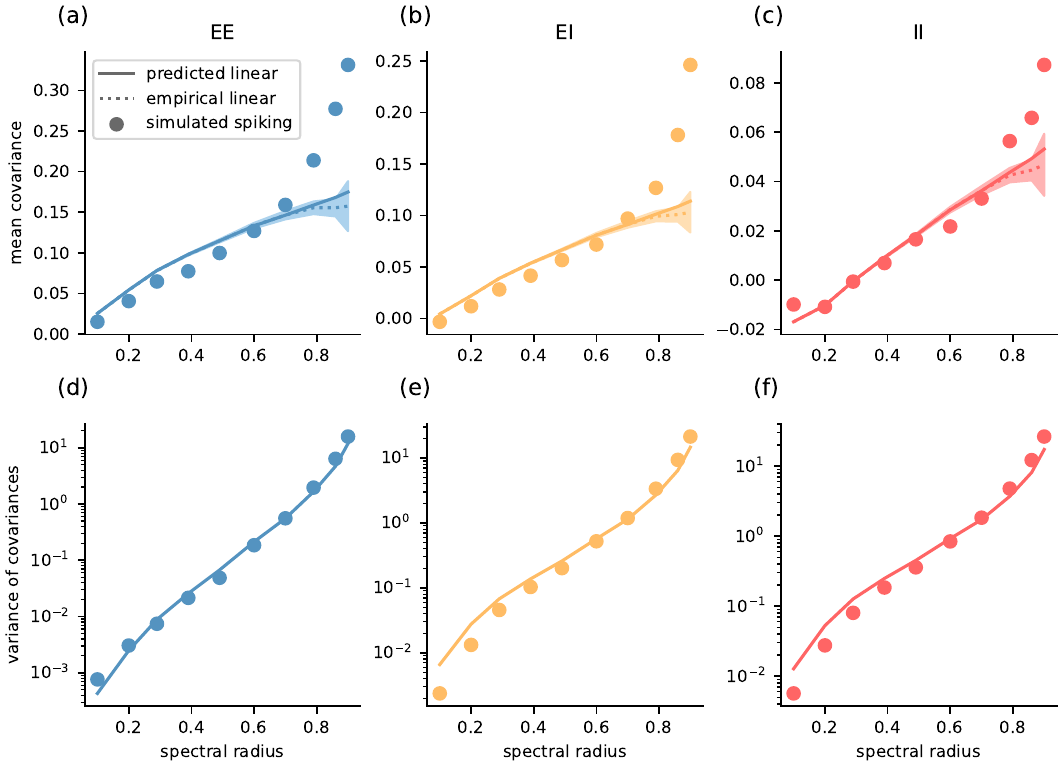}
\par\end{centering}
\caption{Statistics of covariances at different spectral radii of the effective
connectivity matrix $\boldsymbol{W}$. Covariance statistics are resolved
according to population membership of neuron pairs (EE, both neurons
excitatory; EI, one neuron excitatory, one neuron inhibitory; II,
both neurons inhibitory). Dots show the population-resolved {[}(a)-(c){]}
mean or {[}(d)-(f){]} variance of the spike count cross-covariances
{[}\prettyref{eq:spike_count_covariances}{]} measured in network
simulations of spiking E-I networks. Solid lines show the theoretical
predictions of the mean {[}\prettyref{eq:mean_cov-1}{]} and variance
{[}\prettyref{eq:var_of_cov}{]} of cross-covariances, respectively,
using the noise strength estimate {[}\prettyref{eq:noise_strength_estimate_realization_independent}{]}.
The dashed lines show the population-resolved empirical mean and variance
of cross-covariances from \prettyref{eq:covs_standard_eq} averaged
over 20 realizations of $W$, and the shaded area depicts a two-standard-deviation
range around the mean. The variances computed from the simulation
results have been corrected for bias due to finite simulation time
(see Appendix \prettyref{appendix:Bias-correction-of}). Same excitatory-inhibitory
network model as in previous figures. For model details and simulation
parameters see Appendix \prettyref{appendix:nest_simulation}. \label{fig:Covariance-statistics-at}}
\end{figure*}
To check how closely the predictions match the outcomes of spiking
network simulations, we ran ten simulations for different spectral
radii of the effective connectivity matrix $\boldsymbol{W}$ similar
to the one shown in \prettyref{fig:simulation} using the parameters
specified in Appendix \prettyref{appendix:nest_simulation}. The spectral
radius was modified by changing the synaptic weight scale $j$ (see
\prettyref{tab:parameters-1}). In the sparse E-I networks that we
consider here, this affects both the mean connectivity $\boldsymbol{M}$
and the variance of connections $\boldsymbol{S}$ {[}cf. \prettyref{eq:var_Bernoulli}{]}.
We ensured that the spiking networks have roughly similar working
points for the different spectral radii by adjusting the external
input as explained in Appendix \prettyref{appendix:validity_of_theoretical_predictions}.
The theory of \prettyref{eq:mean_cov-1} and \prettyref{eq:var_of_cov}
yields, for any pair of neurons in the network, predictions for the
mean and variance of covariances measured across different network
realizations. In brain circuits, one finds statistical similarities
between different neurons, for example, given by their population
membership. These similarities result in symmetries in the matrices\textbf{
$\boldsymbol{M}$} and $\boldsymbol{S}$. In the current example of
the excitatory-inhibitory network, the statistical similarity among
all excitatory neurons and the similarity among all inhibitory neurons
is reflected in the block structure of the matrices $\boldsymbol{M}$
and $\boldsymbol{S}$. This block structure results in a similar block
structure for the matrices for the mean and variance of covariances.
Any off-diagonal element of one of the blocks (EE, EI, or II) is representative
of the statistics of cross-covariances for this type of neuron pair.
If networks and groups of statistically similar neurons are sufficiently
large, then --- by the self-averaging property \citep{Fischer91,Hertz16_033001,Dahmen19_13051,Helias20_970}
--- an empirical average across these statistically similar neurons
is insensitive to the particular network realization and can be compared
to the results for the statistics across network realizations from
the theory. We thus computed the empirical mean and variance of the
measured covariances for each type of neuron pair (EE, EI, and II),
corrected the variances for bias due to finite simulation time (see
Appendix \prettyref{appendix:Bias-correction-of} for details), and
compared the results to the predictions by \prettyref{eq:mean_cov-1}
and \prettyref{eq:var_of_cov}. The results are displayed in \prettyref{fig:Covariance-statistics-at}:
The top row shows changes in mean covariances with spectral radius
of $\boldsymbol{W}$ {[}via according changes in $\boldsymbol{M}$,
see \prettyref{eq:mean_cov-1}{]}, and the bottom row shows changes
in the variance of covariances with spectral radius of $\boldsymbol{W}$
{[}via according changes in $\boldsymbol{S}$; see \prettyref{eq:var_of_cov}{]}.

We observe that the order of magnitude of mean and variance are well
predicted by \prettyref{eq:mean_cov-1} and \prettyref{eq:var_of_cov},
which is especially evident for the variances {[}\prettyref{fig:Covariance-statistics-at}(d-f){]},
which span several orders of magnitude. However, there is some quantitative
discrepancy between the predictions of the presented linear theory
and the results of the simulated spiking network, which is visible
in \prettyref{fig:Covariance-statistics-at}(a-c), indicating that
a linear theory cannot fully capture the nonlinear spiking dynamics
at high spectral radii, where potential nonrenewal effects of spiking
arise \citep{Ostojic14}. To verify that the discrepancy originates
mostly from the linear-response approximation rather than our disorder-average
approximations, we plotted the predictions of the linear theory {[}\prettyref{eq:covs_standard_eq}{]}
for 20 different network realizations: At small spectral radii, the
predicted disorder-average-based mean is equal to the empirical mean
of the linear networks, and for large spectral radii, the predicted
mean appears to be within the range of two standard deviations around
the empirical mean. This shows that the deviations to the spiking
network results mostly stem from the linear-response approximation.
The remaining difference between the predicted and the empirical mean
in linear networks could be explained by the fact that, for high spectral
radii, the effective connectivity matrix contributes much more strongly
to the noise strength, such that we can no longer disregard its contribution
to the noise strength (cf. \prettyref{fig:noise_strength_autocorr})
and averaging over $\boldsymbol{W}$ and $\boldsymbol{D}$ separately
is no longer feasible.

\section{Discussion\label{sec:Discussion}}

In this study, we introduce theoretical tools based on statistical
physics of disordered systems to investigate the role of heterogeneous
network connectivity in shaping the coordination structure in neural
networks. While the presented methods are applicable to arbitrary
independent connectivity statistics, for illustration we focus our
analysis on the prototypical network model for cortical dynamics by
\citet{Brunel00_183}, which is a spiking network of randomly connected
excitatory and inhibitory leaky integrate-and-fire neurons receiving
uncorrelated external Poisson input. This model has been extensively
studied before using mean-field and linear-response methods to understand
neuronal spiking statistics such as average firing rates and CVs \citep{Amit97,Trousdale12_e1002408}
as well as average cross-covariances between populations of neurons
\citep{Lindner05_061919,Tetzlaff12_e1002596,Helias13_023002,Helias14,Rosenbaum14}.
In this study, we go beyond the population level and introduce tools
from field theory of disordered systems to study the heterogeneity
of activity across individual neurons. We show how to turn a linear-response
result on the link between covariances and connectivity \citep{Lindner05_061919,Pernice11_e1002059,Pernice12_031916,Trousdale12_e1002408,Grytskyy13_131}
into a field-theoretic problem using moment-generating functions.
Then we apply disorder averages, replica theory, and beyond-mean-field
approximations to obtain quantitative predictions for the mean and
variance of cross-covariances that take into account the statistics
of connectivity, but are independent of individual network realizations.
We show that this theory can faithfully predict the statistics of
cross-covariances of spiking leaky integrate-and-fire networks across
the whole linearly stable regime. In doing this, we fixed the statistics
of individual neurons according to their theoretical prediction and
showed that this one working point, defined by the firing rates of
all neurons in the network, can correspond to very distinct correlations
structures. Furthermore, we demonstrate that while the heterogeneity
in single-neuron activities directly impacts the statistics of neuronal
autocovariances, it does not have a sizable impact on the heterogeneity
in cross-covariances (cf. \prettyref{fig:cov_stats_dependence_on_D_heterogeneity}).
The latter heterogeneity is determined by the heterogeneity in neuronal
couplings, quantified by the spectral radius of effective connectivity
bulk eigenvalues.

Technically, by employing linear-response theory, we study two systems:
the spiking leaky integrate-and-fire network and a network of linear
rate neurons. We derive a procedure to set the external input noise
of the linear model in such a way that the covariance statistics of
the spiking network and the linear network match quantitatively. This
way, the autocovariances are fixed to values determined by single-neuron
firing rates and CVs, as predicted by renewal theory for spike trains.
Consequently, autocovariances remain finite in the matched rate network
even when approaching the point of linear instability. This is achieved
by reducing external input fluctuations to account for the increased
intrinsically generated fluctuations when increasing the heterogeneity
in network connectivity. As a result, also neuronal cross-covariances
remain finite close to linear instability. The variance of cross-covariances
nevertheless displays a residual divergence, which is why, within
the linear regime, mean cross-covariances only vary mildly, while
the variance of cross-covariances spans many orders of magnitude when
changing the spectral radius of bulk connectivity eigenvalues.

The methods presented here are restricted to the linearly stable network
regime, usually referred to as the asynchronous irregular state of
the Brunel model \citep{Brunel00_183}. We show that, while mean covariances
are low in this state \citep{Vreeswijk96_1724,Renart10_587,Ecker10},
individual cross-covariances between pairs of neurons can still be
large, reflected by the large variance of cross-covariances in strongly
heterogeneous network settings. Linear stability can, for example,
be realized in excitatory-inhibitory networks if the overall recurrent
feedback in the network is inhibition dominated or only marginally
positive \citep{Tetzlaff12_e1002596} and if synaptic amplitudes are
not too strong. Previous work \citep{Ostojic14,Kriener2014_136} has
shown that the here-considered model transitions to a different asynchronous
activity state if synaptic amplitudes become larger. This state, however,
is not well described by linear-response theory, as slow network fluctuations
and nontrivial spike-train autocorrelations emerge, causing deviations
from the renewal assumptions on spike trains used here. Note that
such slow network fluctuations have not been observed in previous
studies on spontaneous activity in macaque motor cortex \citep{Dahmen19_13051,dahmen22_e68422}
and mouse visual cortex \citep{Dahmen22_365072v3} that employed first
results of the more general theoretical approach presented here to
explain experimentally observed features, such as the large dispersion
of covariances, long-range neuronal coordination, a rich repertoire
of time-scales and low-dimensional activity. These studies relied
on Wick's theorem to calculate the variance of covariances, which
is, however, restricted to linear systems. Here we instead employ
a more general replica approach that can be straightforwardly applied
to nonlinear rate models \citep{Sompolinsky88_259}, as extensively
studied in the recent theoretical neuroscience literature \citep{Stern14_062710,Aljadeff15_088101,Aljadeff16_022302,Marti18_062314,Schuecker18_041029,Crisanti18_062120,Muscinelli19_e1007122,Beiran19_e1006893}.
Importantly, the replica theory reveals in a systematic manner that
the variance of covariances is an observable that is $\mathcal{O}(1/N)$
in the network size and requires beyond-mean-field methods to be computed.
In mean-field or saddle-point approximation, the replica coupling
term that yields the nontrivial variance of covariances vanishes.
We here calculate the next-to-leading-order Gaussian fluctuations
around saddle points that yield good quantitative results across the
whole linear regime. The fact that the linear rate model captures
the covariance statistics of the spiking leaky integrate-and-fire
model further shows that the presented results on the link between
connectivity and covariances do not depend on model details and are
generally valid in the linear regime, which enables applications to
experimental data \citep{Dahmen19_13051,dahmen22_e68422,Dahmen22_365072v3}.

In this paper, we focus on intrinsic mechanisms for heterogeneity
and study the first- and second-order statistics of network connectivity.
The formalism can be applied to any network topology, as arbitrary
connectivity structures can be encoded in the mean and variance matrices
that are the central objects of the theory. The results of the formalism
are particularly useful for comparing covariance statistics within
a single network with groups of statistically equivalent neurons sharing
the same connectivity statistics, such as different cell types within
neural circuits. Such statistical equivalence imposes symmetries on
the structure of the matrices $\boldsymbol{M}$ and $\boldsymbol{S}$
that encode the first- and second-order connectivity statistics. These
symmetries allow a dimensionality reduction of the problem and, by
the self-averaging property \citep{Fischer91,Hertz16_033001,Dahmen19_13051,Helias20_970},
the comparison of theoretical results to empirical averages over covariances
of different pairs of neurons within a single circuit. Here we study
a network with two populations, excitatory and inhibitory, leading
to a block structure in both $\boldsymbol{M}$ and $\boldsymbol{S}$.
This example has been chosen as the simplest but relevant setting
that goes beyond the fully homogeneous random network in \citet{Dahmen19_13051}
with a correspondingly simpler homogeneous theory. Extensions to more
populations or neuron clusters as well as more complex population-specific
connectivity statistics are straight forward by adding more blocks
to $\boldsymbol{M}$ and $\boldsymbol{S}$. Another application of
the here derived theory to a more complex scenario, including interneuron
distance dependence of connection statistics, has been studied (without
derivation) by \citet{dahmen22_e68422}. Notably, in our theory we
assume that connection weights are independently drawn across different
neuron pairs from an arbitrarily complex probability distribution
with finite cumulants. The focus on mean $M_{ij}$ and variance $S_{ij}$
of this distribution is justified as long as connection weights scale
at least as $\mathcal{O}\left(1/\sqrt{N}\right)$, a scaling that
is often employed to preserve fluctuations in the limit of large networks
\citep{Vreeswijk96_1724,Vreeswijk98_1321}. In this scaling, effects
of higher-order-connectivity cumulants are suppressed by the typically
large network size, and the Gaussian approximation of the connectivity
yields accurate results for network dynamics, as here demonstrated
on the example of excitatory-inhibitory networks with population-specific
connectivity statistics comprising sparseness (Bernoulli distribution)
in addition to distributed (Gaussian) synaptic amplitudes of existing
connections. Generalization of dynamic mean-field methods to heavy-tailed
connectivity, which cannot be expanded in cumulants, has been proposed
for studying single-neuron activity statistics \citep{Kusmierz_20_028101,Wardak22}.
A similar approach may potentially be combined with the methods presented
here to investigate cross-covariances. Furthermore, extensions to
correlated connection weights, reflecting an over- or under-representation
of reciprocal, convergent, divergent, and chain motifs, have been
proposed in Ref. \citep{Dahmen22_365072v3}.

In addition to network connectivity, external inputs can be correlated
and heterogeneous and thereby cause heterogeneity in covariances of
local circuits. Previous works have shown that external inputs can
have a strong impact on local covariances, especially in the limit
of infinite network size \citep{Renart10_587,Rosenbaum14,Rosenbaum16_107,Baker19_052414}.
For biologically realistic network sizes of local circuits, intrinsically
generated covariances via local recurrent activity reverberations,
however, make up a substantial contribution to cross-neuronal coordination
\citep{Helias14,Dahmen19_13051}. This contribution is explainable
with the here-presented methods. In general, more research is required
to decipher the precise interplay between intrinsic heterogeneity
and external inputs to arrive at a complete picture for the mechanistic
origin of heterogeneous covariance structures in local circuits.

Many previous studies have linked connectivity and dynamics on an
average level, taking into account particular connection pathways
between neural populations \citep{vanAlbada22_201}, clustering \citep{LitwinKumar12_1498,Brinkman22},
or the spatial dependence of connections \citep{Bressloff12,Zeraati23_1858,Shi23_013005}.
In contrast, we here focus on heterogeneity as a key feature of neural
network connectivity and show that it yields a wealth of complex coordination
patterns that are progressively becoming experimentally accessible
via recent advances in measurement techniques \citep{Campagnola2022_eabj5861}.
Our theoretical framework to systematically incorporate structural
heterogeneity and predict dynamical heterogeneity in biologically
plausible neural network models enables the use of this experimental
knowledge about neural systems. Likewise, the current framework can
be used for the inverse problem of inferring network properties from
measured covariances, as we demonstrated in Ref. \citep{Dahmen19_13051}
for the spectral radius in homogeneous random networks (see Appendix
\prettyref{appendix:Inference-of-connectivity} for an extension to
E-I networks) and in Ref. \citep{dahmen22_e68422} for long-range,
multisynaptic interactions in networks with spatially organized connectivity.
Our work thus opens new avenues for the interpretation of data on
network structure and dynamics and proposes a change of focus from
population-averaged observables to higher-order s\textcolor{black}{tatistics
that uncover the central role of heterogeneity in biological networks.}

All code and data to reproduce the simulations and figures of this
study are publicly available under the Ref. \citep{Layer24_zenodo}.

\subsection*{Acknowledgments}

This work was partially supported by the European Union\textquoteright s
Horizon 2020 research and innovation program under Grant Agreement
No. 945539 (Human Brain Project SGA3) and by the Deutsche Forschungsgemeinschaft
(DFG, German Research Foundation) Grant No. 368482240/GRK2416. Open-access
publication is funded by the Deutsche Forschungsgemeinschaft (DFG,
German Research Foundation), Grant No. 491111487. We are grateful
to our colleagues in the NEST developer community for continuous collaboration.
All network simulations were carried out with NEST \citep[commit dd5b61342]{Gewaltig_07_11204}.
We thank Hannah Bos for the initial numerical implementation of the
CVs.

\noindent 

\newpage{}

\noindent \onecolumngrid

\section*{Appendix}

\subsection{Network model and NEST simulation\label{appendix:nest_simulation}}

We simulate networks of leaky integrate-and-fire neuron models, where
the subthreshold dynamics of the membrane potential $V_{i}$ of neuron
$i$ is given by
\begin{equation}
\tau_{\mathrm{m}}\frac{\mathrm{d}V_{i}(t)}{\mathrm{d}t}=-V_{i}(t)+RI_{i}(t)\,,
\end{equation}
with total input current $I_{i}(t)$ that consists of recurrent input
via connections with strength $J_{ij}$ and delay $d$ as well as
external input: 
\begin{equation}
RI_{i}(t)=\tau_{\mathrm{m}}\left(\sum_{j}J_{ij}s_{j}(t-d)+js_{\mathrm{ext},\mathrm{E}}(t)+gjs_{\mathrm{ext},\mathrm{I}}(t)+\frac{I_{\mathrm{ext}}}{C}\right)\,.
\end{equation}
Synaptic currents are instantaneous without synaptic filtering \citep{Stein67a,Tuckwell88a}.
The external input is decomposed into a constant current $I_{\mathrm{ext}}$
and Poisson spike trains $s_{\mathrm{ext},\mathrm{E}}(t)$ of rate
$\nu_{\mathrm{ext},\mathrm{E}}$ and $s_{\mathrm{ext},\mathrm{I}}(t)$
of rate $\nu_{\mathrm{ext},\mathrm{I}}$ that affect neurons with
excitatory weight $j$ and inhibitory weight $gj$, respectively.
$R$ and $C$ denote the membrane resistance and capacitance, respectively.
More information on the model parameters and their values can be found
in \prettyref{tab:simulation_parameters} and \prettyref{tab:parameters-1}.

We consider random sparse connectivity $\boldsymbol{J}$, with connection
probability $10\,\%$. Sparse connections are realized with a fixed
excitatory indegree $K_{\mathrm{\mathrm{E}}}=800$ and inhibitory
indegree $K_{\mathrm{I}}=200$, with potential self-connections and
prohibiting multiple connections between the same pair of neurons.
To compensate for the imbalance in excitatory and inhibitory neuron
count, we scale the strengths of existing inhibitory connections with
respect to excitatory ones by a factor $g=-6$ to obtain an asynchronous
irregular dynamic regime \citep{Brunel00_183}. We distribute synaptic
amplitudes of existing connections according to population-specific
normal distributions $j_{\mathrm{E}}\propto\mathcal{N}\left(j,0.2j\right)$
and $j_{\mathrm{I}}\propto\mathcal{N}\left(gj,0.2j\right)$. The parameter
$j$ determining the overall scale of connection strength is varied
to modify the overall heterogeneity of connections as measured by
the variance
\begin{equation}
\mathrm{Var}(J_{ij})\propto j^{2}\label{eq:var_Bernoulli}
\end{equation}
and thereby the spectral radius of bulk connectivity eigenvalues (\prettyref{tab:parameters-1}).

\begin{table}[h]
\centering{}%
\begin{tabular}{lll}
\hline 
\multicolumn{3}{l}{\textbf{Network parameters}}\tabularnewline
\hline 
Neuron type & \texttt{iaf\_psc\_delta} & \tabularnewline
Synapse type & \texttt{static\_synapse} & \tabularnewline
Connection rule & \texttt{fixed\_indegree} & \tabularnewline
autapses & \texttt{True} & Connections of a neuron to itself\tabularnewline
multapses & \texttt{False} & Multiple connections between a pair of neurons\tabularnewline
$N_{\mathrm{E}}$ & $8000$ & Number of excitatory neurons\tabularnewline
$N_{\mathrm{I}}$ & $2000$ & Number of inhibitory neurons\tabularnewline
$K_{\mathrm{E}}$ & $800$ & Number of excitatory inputs\tabularnewline
$K_{\mathrm{I}}$ & $200$ & Number of inhibitory inputs\tabularnewline
$C$ & $1\,\mathrm{pF}$ & Membrane capacitance\tabularnewline
$\tau_{\mathrm{m}}$ & $20\,\mathrm{ms}$ & Membrane time constant\tabularnewline
$\tau_{\mathrm{r}}$ & $2\,\mathrm{ms}$ & Refractory period\tabularnewline
$V_{\mathrm{th}}$ & $15\,\mathrm{mV}$ & Relative threshold voltage\tabularnewline
$d$ & $1\,\mathrm{ms}$ & Synaptic delay\tabularnewline
$j$ & $\left[0.04,0.38\right]\,\mathrm{mV}$ & Excitatory synaptic weight\tabularnewline
$g$ & $-6$ & Ratio of inhibitory to excitatory weight\tabularnewline
$\sigma_{j}$ & $20\,\%$ of $j_{\mathrm{E}}$ & Std of Gaussian distribution of E and I weights\tabularnewline
$I_{\mathrm{ext}}$ & $\left[5,125\right]\,\mathrm{pA}$ & External DC current\tabularnewline
$\nu_{\mathrm{ext},\mathrm{E}}$ & $\left[800.73,315049.84\right]\,\mathrm{Hz}$ & Rate of external excitatory Poisson noise\tabularnewline
$\nu_{\mathrm{ext},\mathrm{I}}$ & $\left[640.42,572214.84\right]\,\mathrm{Hz}$ & Rate of external inhibitory Poisson noise\tabularnewline
\hline 
\multicolumn{3}{l}{\textbf{Simulation parameters}}\tabularnewline
\hline 
$\mathrm{d}t$ & $0.1\,\mathrm{ms}$ & Simulation step size\tabularnewline
$t_{\mathrm{sim}}$ & $10,000,000\,\mathrm{ms}$ & Simulation time\tabularnewline
\hline 
\multicolumn{3}{l}{\textbf{Analysis parameters}}\tabularnewline
$T$ & $1000\,\mathrm{ms}$ & Bin width for calculating spike-count correlations\tabularnewline
$T_{\mathrm{init}}$ & $1000\,\mathrm{ms}$ & Initialization time\tabularnewline
\hline 
\end{tabular}\caption{Parameters used for NEST simulations and subsequent analysis.\label{tab:simulation_parameters}}
\end{table}
\begin{table}[h]
\centering{}%
\begin{tabular}{|c||c|c|c|c|c|c|c|c|c|c|}
\hline 
$r$ & $0.10$ & $0.20$ & $0.29$ & $0.39$ & $0.49$ & $0.60$ & $0.70$ & $0.79$ & $0.86$ & $0.90$\tabularnewline
$j$~(mV) & $0.04$ & $0.08$ & $0.12$ & $0.16$ & $0.2$ & $0.25$ & $0.29$ & $0.33$ & $0.36$ & $0.38$\tabularnewline
$I_{\mathrm{ext}}$~(pA) & $125.0$ & $65.0$ & $40.0$ & $25.0$ & $20.0$ & $15.0$ & $10.0$ & $8.0$ & $6.0$ & $5.0$\tabularnewline
$\nu_{\mathrm{ext},\text{E}}$~(Hz) & $315049.84$ & $35406.98$ & $27510.16$ & $32862.34$ & $13335.56$ & $4292.70$ & $6393.05$ & $2149.08$ & $1593.05$ & $800.73$\tabularnewline
$\nu_{\mathrm{ext},\mathrm{I}}$~(Hz) & $572214.84$ & $139878.53$ & $58597.12$ & $29923.17$ & $17262.46$ & $9063.65$ & $5147.04$ & $2722.54$ & $1360.93$ & $640.42$\tabularnewline
\hline 
\end{tabular}\caption{Parameters adjusted for setting different spectral radii while keeping
the firing rate constant: The spectral radius of the connectivity
is set by choosing different synaptic strengths $j$ of connections.
The synaptic strength not only affects the mean connectivity but also
the variance of connections, i.e., the heterogeneity of the network,
that determines the spectral radius {[}see \prettyref{eq:var_Bernoulli}{]}.
The parameters of the external inputs, which model the total excitatory
and inhibitory input from external populations of neurons, are furthermore
adjusted to maintain a constant firing rate across different spectral
radii. This is done to isolate effects of changing spectral radii
on correlations from effects of changing firing rates (see Appendix
\prettyref{appendix:Derivation-of-moment} and \prettyref{fig:r_set_vs_sim}).
For small spectral radii and thus weak recurrent input, strong external
input is needed to drive the network to moderate firing rates, while
for large spectral radii and strong recurrent input, only weak external
input is needed for moderate firing rates. \label{tab:parameters-1}}
\end{table}

\subsection{Time-lag integrated covariances\label{appendix:covariances_definitions}}

The cross-covariance function of two stochastic zero-mean processes
$x_{i}(t)$ and $x_{j}(t)$ is defined as
\[
C_{ij}\left(s,t\right)=\left\langle x_{i}(s)x_{j}(t)\right\rangle \,,
\]
where the average is over the ensemble of realizations of the processes.
If the stochastic processes are stationary, the cross-covariance function
solely depends on the time-lag $\tau=t-s$
\[
C_{ij}\left(\tau\right)=\left\langle x_{i}(s)x_{j}(s+\tau)\right\rangle \,.
\]
Here we are considering the time-lag integrated covariances, as they
can be linked to the experimentally accessible spike-count covariances
\citep{Cohen11_811,Dahmen19_13051}, 
\begin{eqnarray*}
C_{ij} & \coloneqq & \int_{-\infty}^{\infty}C_{ij}\left(\tau\right)\mathrm{d}\tau\\
 & = & \lim_{T\rightarrow\infty}\frac{1}{T}\left(\left\langle n_{i}n_{j}\right\rangle -\left\langle n_{i}\right\rangle \left\langle n_{j}\right\rangle \right)\,,
\end{eqnarray*}
which can be interpreted as a zero-frequency Fourier transform. The
Wiener--Khinchin theorem (\prettyref{appendix:Wiener=002013Khinchin-theorem})
allows expressing the time-lag integrated covariances in terms of
the time series's Fourier components $X_{i}\left(\omega\right)$ at
frequency zero 
\begin{equation}
C_{ij}=\left\langle X_{i}\left(0\right)X_{j}\left(0\right)\right\rangle \,.\label{eq:covs_khinchin}
\end{equation}

\subsection{Wiener--Khinchin theorem\label{appendix:Wiener=002013Khinchin-theorem}}

Here in parts we follow the book by \citet{Gardiner85}. Let $x(t)$
and $y(t)$ be stochastic, stationary processes. Stationary means
that for any $n$-tuple $\left(t_{1},t_{2},\ldots,t_{n}\right)$ of
time points and any real number $u$ the samples $x\left(t_{1}\right),\ldots,x\left(t_{n}\right)$
follow the same distribution as the samples $x\left(t_{1}+u\right),\ldots,x\left(t_{n}+u\right)$
\citep{Khintchine34_604}. Consequently, we may define a raw correlation
function as 
\begin{eqnarray*}
c\left(\tau\right) & = & \left\langle x\left(t\right)y\left(t+\tau\right)\right\rangle \,,
\end{eqnarray*}
which, due to the assumption of stationarity, does not depend on the
time $t$. The average is over the ensemble of realizations of the
processes. If the Fourier transforms of $x$ and $y$ exist, we may
calculate the ensemble average over $X\left(\omega\right)$ and $Y\left(\omega\right)$
as
\begin{eqnarray}
\left\langle X\left(\omega\right)Y\left(\omega^{\prime}\right)\right\rangle  & = & \int\mathrm{d}t\,e^{-\mathrm{i}\omega t}\int\mathrm{d}t^{\prime}\,e^{-\mathrm{i}\omega^{\prime}t^{\prime}}\left\langle x\left(t\right)y\left(t^{\prime}\right)\right\rangle \nonumber \\
 & \stackrel{\text{subst. }t^{\prime}=t+\tau}{=} & \int\mathrm{d}t\,e^{-\mathrm{i}\omega t}\int\mathrm{d}\tau\,e^{-\mathrm{i}\omega^{\prime}(t+\tau)}\left\langle x(t)y(t+\tau)\right\rangle \nonumber \\
 & = & \int\mathrm{d}t\,e^{-\mathrm{i}\left(\omega+\omega^{\prime}\right)t}\int\mathrm{d}\tau\,e^{-\mathrm{i}\omega^{\prime}\tau}\left\langle x(t)y(t+\tau)\right\rangle \nonumber \\
 & = & 2\pi\delta\left(\omega+\omega^{\prime}\right)\int\mathrm{d}\tau\,e^{-\mathrm{i}\omega^{\prime}\tau}\,c(\tau)\nonumber \\
 & = & 2\pi\delta\left(\omega+\omega^{\prime}\right)C\left(\omega^{\prime}\right)\,,\label{eq:cross_spectrum_stationary}
\end{eqnarray}
where we used the identity $\int\mathrm{d}t\,e^{-\mathrm{i}\omega t}=2\pi\delta(\omega)$
which follows from $\frac{1}{2\pi}\int\mathrm{d}\omega\,e^{\mathrm{i}\omega t}2\pi\delta(\omega)=1$,
so that $2\pi\delta(\omega)$ is the Fourier transform of the constant
function and vice versa. Equation \eqref{eq:cross_spectrum_stationary}
states that the cross spectrum between two stationary processes vanishes
except at those frequencies $\omega=-\omega^{\prime}$, where it is
proportional to a $\delta$-distribution times the Fourier transform
of the autocorrelation function.

\subsection{Validity of theoretical predictions\label{appendix:validity_of_theoretical_predictions}}

In this section, we discuss the conditions under which the theory
and simulation described in this paper yield the same results. There
are several factors to consider: the limits of the theory we built
upon, the limitations of the newly presented theory, and the simulation's
constraints.

\begin{figure*}
\begin{centering}
\includegraphics{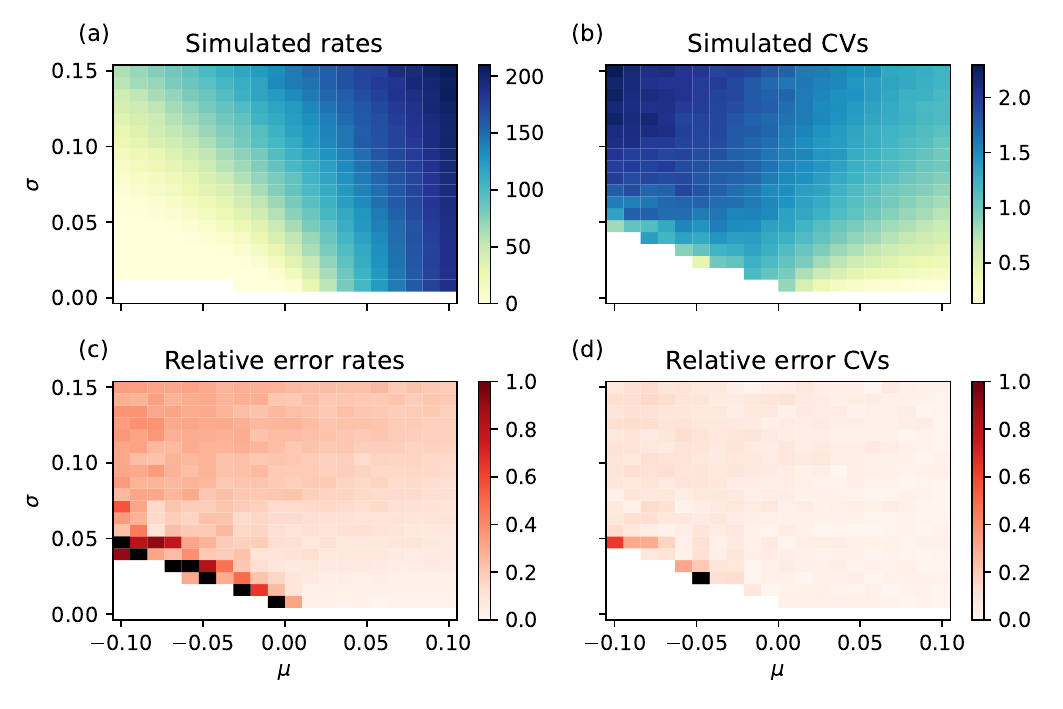}
\par\end{centering}
\caption{Validity of firing rate and CV prediction for single LIF neuron with
instantaneous synapses provided with two independent Poisson inputs.
(a) Simulated rates and (b) simulated CVs at given mean input $\mu$
and input variance $\sigma^{2}$. (c) Relative error $\epsilon=\left|\nu_{\mathrm{sim}}-\nu_{\mathrm{thy}}\right|/\nu_{\mathrm{sim}}$
of rate and (d) relative error of CV prediction using \prettyref{eq:siegert}
and \prettyref{eq:cvs}. Black pixels denote error values larger than
$1.00$.\label{fig:rate_and_cv_error}}
\end{figure*}
\begin{figure}
\begin{centering}
\includegraphics{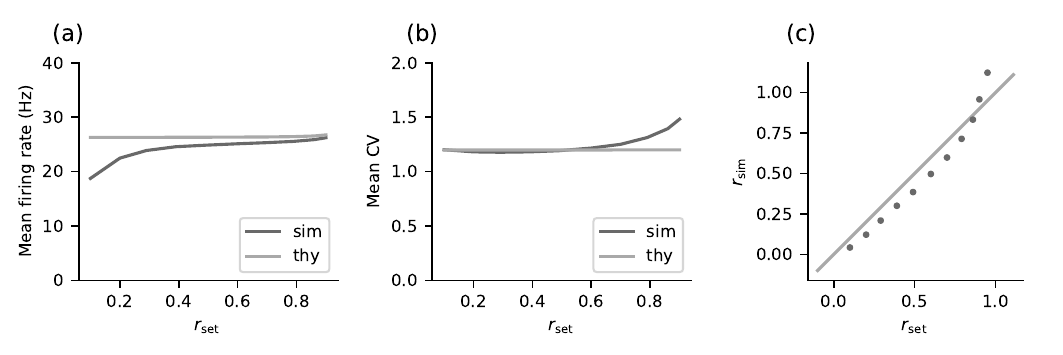}
\par\end{centering}
\caption{Prediction accuracy of spiking network simulation properties. Predicted
and measured (a) mean firing rates and (b) mean CVs at different spectral
radii. (c) Set spectral radius $r_{\mathrm{set}}$ vs. measured spectral
radius $r_{\mathrm{sim}}$. Same excitatory-inhibitory network model
as in previous figures. For model details and simulation parameters
see Appendix \prettyref{appendix:nest_simulation}. \label{fig:r_set_vs_sim}}
\end{figure}
\begin{figure*}
\begin{centering}
\includegraphics{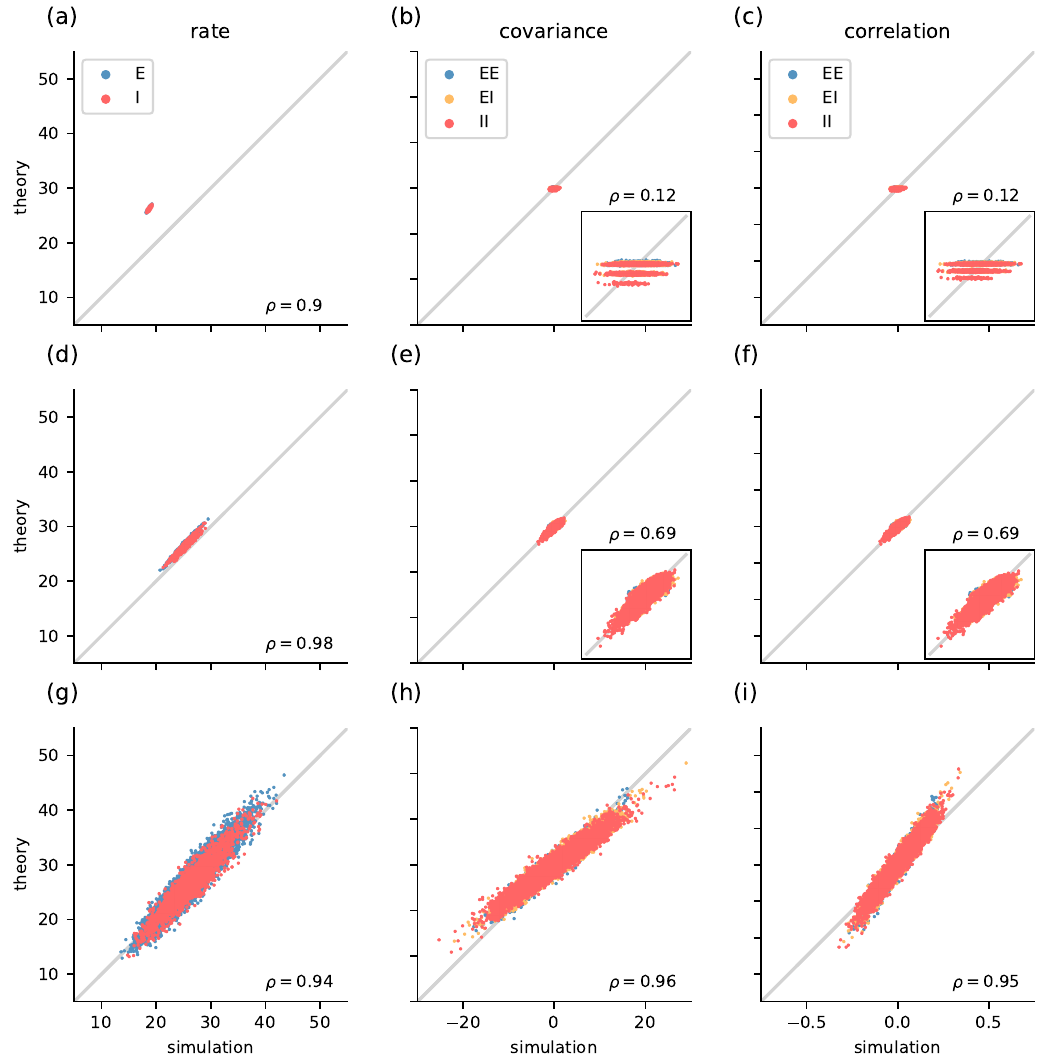}
\par\end{centering}
\caption{Simulation results vs\@. theoretical estimates for E-I networks with
three different spectral radii {[}(a)-(c){]} $r=0.10$, {[}(d)-(f){]}
$r=0.49$, and {[}(g)-(i){]} $r=0.90$ (g, h, i). $\rho$ denotes
the Pearson correlation coefficient. {[}(a), (d), (g){]} Firing rates.
{[}(b), (e), (h){]} Covariances. {[}(c), (f), (i){]} Correlation coefficients.
The insets show a closer look at the data points. Same excitatory-inhibitory
network model as in previous figures. For model details and simulation
parameters see Appendix \prettyref{appendix:nest_simulation}. \label{fig:thy_vs_sim_different_spectral_radii}}
\end{figure*}

The estimation of covariances presented in this paper relies on the
proper estimation of firing rates and CVs, for which we employ \prettyref{eq:siegert}
and \prettyref{eq:cvs} \citep{Brunel00_183}, which rely on a diffusion
approximation that substitutes spiking input by uncorrelated white
noise input. Due to this approximation, these formulas have their
own limitations, and they do not yield good estimates in all parameter
regimes, as shown in \prettyref{fig:rate_and_cv_error} and \prettyref{fig:r_set_vs_sim}(a)
and (b) for a single neuron receiving Poisson input. Note that a correction
due to the finite amplitude of spiking input could in principle be
accounted for \citep{Helias10_1000929}; the diffusion approximation
was shown to typically overestimate firing rates, as can be seen,
for example, in \prettyref{fig:thy_vs_sim}(a). In a network context,
the diffusion approximation furthermore neglects any nontrivial structure
of the autocorrelation of inputs, as well as their cross-correlation
structure. Apart from an offset, \prettyref{fig:thy_vs_sim}(a) also
shows a clustering of inhibitory firing rates in theory. The difference
between the two clusters can be traced back to the presence or absence
of self-connections: Firing rates of neurons with self-connections
are more in line with simulated rates. This fact, however, does not
seem to generalize as networks with other spectral radii do not show
any apparent clustering in firing rate predictions (\prettyref{fig:thy_vs_sim_different_spectral_radii}).
Because the estimates for firing rates and CVs are used to calculate
the effective connectivity matrix and noise strength, a poor estimate
has a direct impact on the covariance estimation. 
Furthermore, the quality of the rate estimates affects how closely
the simulated network matches its analytical counterpart due to the
way we set the parameters for the simulation: We fix the mean and
variance of the single neuron input, and therefore their firing rates
$\boldsymbol{\nu}_{\mathrm{set}}$, and adjust the external input
to set the spectral radius $r$, which we estimate using the result
of \citet{Rajan06} for random Bernoulli E-I networks
\[
r_{\mathrm{set}}=\sqrt{w_{\mathrm{eff,E}}^{2}\left(\boldsymbol{v}_{\mathrm{set}}\right)p\left(1-p\right)N_{\mathrm{E}}+w_{\mathrm{eff,I}}^{2}\left(\boldsymbol{v}_{\mathrm{set}}\right)p\left(1-p\right)N_{\mathrm{I}}}\,,
\]
with connection probability $p$. The effective weights $w_{\mathrm{eff,E}}\left(\boldsymbol{\nu}\right)$,
$w_{\mathrm{eff,I}}\left(\boldsymbol{\nu}\right)$ are computed using
\prettyref{eq:effective_weights}. Once we simulate the network, we
can measure the firing rates, extract the connectivity matrix, and
compute the effective connectivity matrix realized in the simulation.
Its largest eigenvalue determines the spectral radius $r_{\mathrm{sim}}$.
A comparison of $r_{\mathrm{set}}$ and $r_{\mathrm{sim}}$ is shown
in \prettyref{fig:r_set_vs_sim}(c). They do not coincide perfectly,
which is a direct result of the unreliable estimation of the firing
rates, which are slightly overestimated by the theory {[}see \prettyref{fig:thy_vs_sim}(a),
\prettyref{fig:r_set_vs_sim}(a), and \prettyref{fig:thy_vs_sim_different_spectral_radii}(a),
(d), and (g){]}. To make sure the simulated network is always in a
linearly stable regime, we restrict our analysis to spectral radii
$r_{\mathrm{set}}\leq0.90$, for which $r_{\mathrm{sim}}\leq0.94$.

\begin{figure}
\begin{centering}
\includegraphics{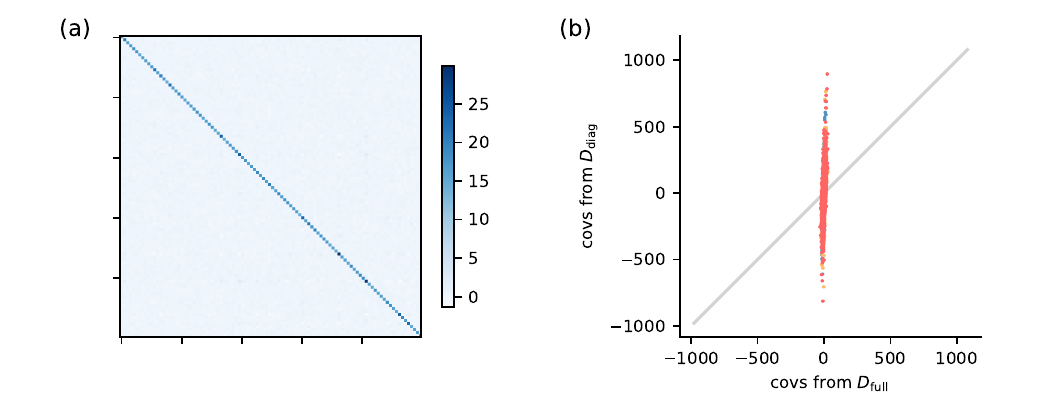}
\par\end{centering}
\caption{\label{fig:negative_D}Noise strength properties. (a) First 100 entries
of $\boldsymbol{D}_{\mathrm{full}}$ computed from simulated covariances.
(b) Comparison of covariances computed using $\boldsymbol{D}_{\mathrm{full}}$
and using $\boldsymbol{D}_{\mathrm{diag}}=\mathrm{diag}\left(\boldsymbol{D}_{\mathrm{full}}\right)$.
Same excitatory-inhibitory network model as in previous figures. For
model details and simulation parameters see Appendix \prettyref{appendix:nest_simulation}
for spectral radius $r=0.9$.}
\end{figure}

We estimate the noise strength $\boldsymbol{D}$ by computing the
variances using \prettyref{eq:variances}, assuming that $\boldsymbol{D}$
is diagonal, and inverting \prettyref{eq:covs_standard_eq} which
yields \prettyref{eq:noise_strength_estimate}. First of all, the
equation for the variances \prettyref{eq:variances} relies on the
assumption that the spike trains are well described by renewal processes
\citep{Cox66}. Therefore, the noise strength estimate is reliable
only if the spike trains are not too bursty. However, even for networks
with $\mathrm{CV}\approx1$ we observed that for large spectral radii
this approach of estimating the noise strength can yield negative
values for $\boldsymbol{D}$, which has no physical interpretation.
Measuring the covariances in a simulation and inverting \prettyref{eq:covs_standard_eq}
without restricting $\boldsymbol{D}$ to be diagonal, yields a matrix
that seems to be almost diagonal, shown in \prettyref{fig:negative_D}(a).
Setting the off-diagonal elements to zero and using the result to
compute the covariances via \prettyref{eq:covs_standard_eq}, however,
reveals that the off-diagonal contribution cannot be neglected {[}\prettyref{fig:negative_D}(b){]},
which means that the external noise sources do have to be correlated
to explain the observed covariance. In cases in which the lowest eigenvalue
of $\boldsymbol{D}_{\mathrm{full}}$ is negative, we conclude that
it is not possible to find a physical linear system (positive-definite
$\boldsymbol{D}$) that explains the individual pair-wise covariances
observed in the spiking network simulation with a large spectral radius.
Our theoretical predictions for the mean and variance of cross-covariances,
\prettyref{eq:mean_cov-1} and \prettyref{eq:var_of_cov}, based on
$\boldsymbol{D}$ computed with \prettyref{eq:noise_strength_estimate}
and its averaged analog, \prettyref{eq:noise_strength_estimate_realization_independent},
nevertheless yield quantitatively matching results with respect to
the spiking network simulations also in this regime (\prettyref{fig:Covariance-statistics-at}),
because, as we show in \prettyref{fig:cov_stats_dependence_on_D_heterogeneity},
the results only depend on the average of $\boldsymbol{D}$. The theory
based on the statistics of connections is therefore found to be more
robust than the theory based on individual connectivity realizations. 

Finally, simulations have one major limitation: their finite simulation
time, which results in a biased estimation of the covariances at the
single-neuron level. As seen in \prettyref{fig:thy_vs_sim_different_spectral_radii}(b),
(c), (e), (f), (h), and (i), there is some variance in the simulations
that is not explained by the theory. This variance is caused by the
finite simulation time and vanishes for longer simulations. The relative
unexplained variance is larger for small spectral radii, since the
firing rates of the neurons are slightly smaller in these networks
leading to poorer estimation of covariances, and overall the covariances
are smaller for small spectral radii.

\subsection{Derivation of moment generating function\label{appendix:Derivation-of-moment}}

As discussed in \prettyref{sec:Background}, in absence of correlated
external input and in the regime of low average covariances, covariances
can be understood in linear response theory \citep{Grytskyy13_131},
where the dynamical equation of LIF neurons describes a model network
of Ornstein-Uhlenbeck processes {[}\prettyref{eq:OUP}{]}. \citet{Grytskyy13_131}
further showed that relation \prettyref{eq:covs_standard_eq} between
time-lag-integrated covariances and effective connections is independent
of the particular filter kernel $h(t)$ and whether noise is injected
in the input or output of neurons. Therefore, we here for simplicity
choose Gaussian white noise in the input and $h(t)$ to be an exponential
kernel with unit time constant. The stochastic differential equation
becomes
\begin{equation}
\mathrm{d}\boldsymbol{x}(t)=-\boldsymbol{x}(t)\mathrm{d}t+\boldsymbol{W}\boldsymbol{x}(t)\mathrm{d}t+\mathrm{d}\boldsymbol{\xi}(t)\,,\label{eq:OUP_functional}
\end{equation}
with generating functional \citep{Dahmen19_13051}
\begin{align*}
Z\left(\boldsymbol{j}\right)= & \int\mathcal{D}\boldsymbol{x}\int\mathcal{D}\widetilde{\boldsymbol{x}}\,\exp\left[\widetilde{\boldsymbol{x}}^{\T}\left(\partial_{t}+1-\boldsymbol{W}\right)\boldsymbol{x}+\frac{D}{2}\widetilde{\boldsymbol{x}}^{\T}\widetilde{\boldsymbol{x}}+\boldsymbol{j}^{\T}\boldsymbol{x}\right]\,.
\end{align*}
The latter can easily be interpreted in the Fourier domain due to
the linearity of \prettyref{eq:OUP_functional} and the invariance
of scalar products under unitary transforms,
\begin{align*}
Z\left(\boldsymbol{J}\right)= & \int\mathcal{D}\boldsymbol{X}\int\mathcal{D}\widetilde{\boldsymbol{X}}\,\exp\left[\widetilde{\boldsymbol{X}}^{\T}\left(\mathrm{i}\omega+1-\boldsymbol{W}\right)\boldsymbol{X}+\frac{D}{2}\widetilde{\boldsymbol{X}}^{\T}\widetilde{\boldsymbol{X}}+\boldsymbol{J}^{\T}\boldsymbol{X}\right]\,,
\end{align*}
with Fourier-transformed variables denoted by capital letters. The
scalar product in the frequency domain reads $\widetilde{\boldsymbol{X}}^{\T}\widetilde{\boldsymbol{X}}=\sum_{j}\int\mathrm{d}\omega\,\widetilde{X}_{j}(-\omega)X_{j}(\omega)$.
The generating functional factorizes into generating functions for
each frequency $\omega$. As we use \prettyref{eq:covs_wiener_khinchin}
to calculate the time-lag integrated covariances, we only require
the zero-frequency components $\boldsymbol{X}\left(0\right)$. In
the following, we therefore only discuss zero frequency and omit the
frequency argument, i.e. we write $\boldsymbol{X}\equiv\boldsymbol{X}\left(0\right)$
and correspondingly for sources $\boldsymbol{J}$. After integrating
over all nonzero frequencies, we obtain the generating function for
zero frequency,
\begin{align}
Z\left(\boldsymbol{J}\right)= & \lambda\int\mathcal{D}\boldsymbol{X}\int\mathcal{D}\widetilde{\boldsymbol{X}}\,\exp\left[\widetilde{\boldsymbol{X}}^{\T}\left(\boldsymbol{1}-\boldsymbol{W}\right)\boldsymbol{X}+\frac{D}{2}\widetilde{\boldsymbol{X}}^{\T}\widetilde{\boldsymbol{X}}+\boldsymbol{J}^{\T}\boldsymbol{X}\right]\,,\label{eq:ZJ}
\end{align}
with the single-frequency scalar product defined as $\widetilde{\boldsymbol{X}}^{\T}\widetilde{\boldsymbol{X}}=\sum_{j}\widetilde{X}_{j}\widetilde{X}_{j}$,
integration measures $\int\mathcal{D}\boldsymbol{X}=\prod_{j}\int_{-\infty}^{\infty}\mathrm{d}X_{j}$
and $\int\mathcal{D}\widetilde{\boldsymbol{X}}=\prod_{j}\frac{1}{2\pi\mathrm{i}}\int_{-\mathrm{i}\infty}^{\mathrm{i}\infty}\mathrm{d}\widetilde{X}_{j}$,
and normalization prefactor $\lambda$. We introduce another source
variable $\widetilde{\boldsymbol{J}}$ so that later we can also compute
correlators that include $\widetilde{\boldsymbol{X}}$
\begin{align*}
Z\left(\boldsymbol{J},\widetilde{\boldsymbol{J}}\right)= & \lambda\int\mathcal{D}\boldsymbol{X}\int\mathcal{D}\widetilde{\boldsymbol{X}}\,\exp\left[\widetilde{\boldsymbol{X}}^{\T}\left(\boldsymbol{1}-\boldsymbol{W}\right)\boldsymbol{X}+\frac{D}{2}\widetilde{\boldsymbol{X}}^{\T}\widetilde{\boldsymbol{X}}+\boldsymbol{J}^{\T}\boldsymbol{X}+\widetilde{\boldsymbol{J}}^{\T}\widetilde{\boldsymbol{X}}\right]\,.
\end{align*}
The Gaussian integrals are solved as follows

\begin{align*}
Z\left(\boldsymbol{J},\widetilde{\boldsymbol{J}}\right) & =\lambda\prod_{i}\int_{-\infty}^{\infty}\mathrm{d}X_{i}\prod_{j}\frac{1}{2\pi\mathrm{i}}\int_{-\mathrm{i}\infty}^{\mathrm{i}\infty}\mathrm{d}\widetilde{X}_{j}\exp\left[\widetilde{\boldsymbol{X}}^{\T}\left(\boldsymbol{1}-\boldsymbol{W}\right)\boldsymbol{X}+\frac{D}{2}\widetilde{\boldsymbol{X}}^{\T}\widetilde{\boldsymbol{X}}+\boldsymbol{J}^{\T}\boldsymbol{X}+\widetilde{\boldsymbol{J}}^{\T}\widetilde{\boldsymbol{X}}\right]\\
 & =\lambda\prod_{i}\int_{-\infty}^{\infty}\mathrm{d}X_{i}\prod_{j}\frac{1}{2\pi}\int_{-\infty}^{\infty}\mathrm{d}\widetilde{X}_{j}\exp\left[\mathrm{i}\widetilde{\boldsymbol{X}}^{\T}\left(\boldsymbol{1}-\boldsymbol{W}\right)\boldsymbol{X}-\frac{D}{2}\widetilde{\boldsymbol{X}}^{\T}\widetilde{\boldsymbol{X}}+\boldsymbol{J}^{\T}\boldsymbol{X}+\mathrm{i}\widetilde{\boldsymbol{J}}^{\T}\widetilde{\boldsymbol{X}}\right]\\
 & =\lambda\left(\frac{1}{2\pi}\right)^{N}\prod_{i}\int_{-\infty}^{\infty}\mathrm{d}X_{i}\prod_{j}\int_{-\infty}^{\infty}\mathrm{d}\widetilde{X}_{j}\exp\left[-\frac{1}{2}\left(\boldsymbol{X}^{\mathrm{T}},\widetilde{\boldsymbol{X}}^{\mathrm{T}}\right)\left(\begin{array}{cc}
0 & -\mathrm{i}\left(\boldsymbol{1}-\boldsymbol{W}^{\T}\right)\\
-\mathrm{i}\left(\boldsymbol{1}-\boldsymbol{W}\right) & D
\end{array}\right)\left(\begin{array}{c}
\boldsymbol{X}\\
\widetilde{\boldsymbol{X}}
\end{array}\right)+\left(\boldsymbol{J}^{\mathrm{T}},\mathrm{i}\widetilde{\boldsymbol{J}}^{\mathrm{T}}\right)\left(\begin{array}{c}
\boldsymbol{X}\\
\widetilde{\boldsymbol{X}}
\end{array}\right)\right]\\
 & =\lambda\left(\frac{1}{2\pi}\right)^{N}\sqrt{\frac{\left(2\pi\right)^{2N}}{\det\left(\begin{array}{cc}
0 & -\mathrm{i}\left(\boldsymbol{1}-\boldsymbol{W}^{\mathrm{T}}\right)\\
-\mathrm{i}\left(\boldsymbol{1}-\boldsymbol{W}\right) & D
\end{array}\right)}}\exp\left[\frac{1}{2}\left(\boldsymbol{J}^{\mathrm{T}},\mathrm{i}\widetilde{\boldsymbol{J}}^{\mathrm{T}}\right)\left(\begin{array}{cc}
0 & -\mathrm{i}\left(\boldsymbol{1}-\boldsymbol{W}^{\T}\right)\\
-\mathrm{i}\left(\boldsymbol{1}-\boldsymbol{W}\right) & D
\end{array}\right)^{-1}\left(\begin{array}{c}
\boldsymbol{J}\\
\mathrm{i}\widetilde{\boldsymbol{J}}
\end{array}\right)\right]\\
 & =\frac{\lambda}{\sqrt{\det\left(\begin{array}{cc}
0 & -\mathrm{i}\left(\boldsymbol{1}-\boldsymbol{W}^{\mathrm{T}}\right)\\
-\mathrm{i}\left(\boldsymbol{1}-\boldsymbol{W}\right) & D
\end{array}\right)}}\exp\left[\frac{1}{2}\left(\boldsymbol{J}^{\mathrm{T}},\mathrm{i}\widetilde{\boldsymbol{J}}^{\mathrm{T}}\right)\left(\begin{array}{cc}
\left(\boldsymbol{1}-\boldsymbol{W}\right)^{-1}D\left(\boldsymbol{1}-\boldsymbol{W}^{\mathrm{T}}\right)^{-1} & \mathrm{i}\left(\boldsymbol{1}-\boldsymbol{W}\right)^{-1}\\
\mathrm{i}\left(\boldsymbol{1}-\boldsymbol{W}^{\T}\right)^{-1} & 0
\end{array}\right)\left(\begin{array}{c}
\boldsymbol{J}\\
\mathrm{i}\widetilde{\boldsymbol{J}}
\end{array}\right)\right]\\
 & =\frac{\lambda}{\sqrt{\det\left(\begin{array}{cc}
0 & -\mathrm{i}\left(\boldsymbol{1}-\boldsymbol{W}^{\mathrm{T}}\right)\\
-\mathrm{i}\left(\boldsymbol{1}-\boldsymbol{W}\right) & D
\end{array}\right)}}\exp\left[\frac{1}{2}\left(\boldsymbol{J}^{\mathrm{T}},\widetilde{\boldsymbol{J}}^{\mathrm{T}}\right)\left(\begin{array}{cc}
\left(\boldsymbol{1}-\boldsymbol{W}\right)^{-1}D\left(\boldsymbol{1}-\boldsymbol{W}^{\mathrm{T}}\right)^{-1} & -\left(\boldsymbol{1}-\boldsymbol{W}\right)^{-1}\\
-\left(\boldsymbol{1}-\boldsymbol{W}^{\T}\right)^{-1} & 0
\end{array}\right)\left(\begin{array}{c}
\boldsymbol{J}\\
\widetilde{\boldsymbol{J}}
\end{array}\right)\right]\,.
\end{align*}
The identity matrix and the matrix of ones commute, therefore we can
use $\det\left(\begin{array}{cc}
A & B\\
C & D
\end{array}\right)=\det(AD-BC)$, and we get
\begin{align*}
\det\left(\begin{array}{cc}
0 & -\mathrm{i}\left(\boldsymbol{1}-\boldsymbol{W}^{\mathrm{T}}\right)\\
-\mathrm{i}\left(\boldsymbol{1}-\boldsymbol{W}\right) & D
\end{array}\right) & =\det\left[-\mathrm{i}^{2}\left(\boldsymbol{1}-\boldsymbol{W}^{\T}\right)\left(\boldsymbol{1}-\boldsymbol{W}\right)\right]\\
 & =\left[\det\left(\boldsymbol{1}-\boldsymbol{W}\right)\right]^{2}\,.
\end{align*}
The normalization condition $Z\left(\boldsymbol{J}=0\right)=1$ yields
$\lambda=\left|\det\left(1-\boldsymbol{W}\right)\right|$, and the
generating function becomes

\begin{align*}
Z\left(\boldsymbol{J},\widetilde{\boldsymbol{J}}\right) & =\exp\left[\frac{1}{2}\left(\boldsymbol{J}^{\mathrm{T}},\widetilde{\boldsymbol{J}}^{\mathrm{T}}\right)\left(\begin{array}{cc}
\left(\boldsymbol{1}-\boldsymbol{W}\right)^{-1}D\left(\boldsymbol{1}-\boldsymbol{W}^{\mathrm{T}}\right)^{-1} & -\left(\boldsymbol{1}-\boldsymbol{W}\right)^{-1}\\
-\left(\boldsymbol{1}-\boldsymbol{W}\right)^{-\T} & 0
\end{array}\right)\left(\begin{array}{c}
\boldsymbol{J}\\
\widetilde{\boldsymbol{J}}
\end{array}\right)\right]\,,
\end{align*}
or 
\begin{align}
Z\left(\boldsymbol{J},\widetilde{\boldsymbol{J}}=0\right) & =\left|\det\left(1-\boldsymbol{W}\right)\right|\int\mathcal{D}\boldsymbol{X}\int\mathcal{D}\widetilde{\boldsymbol{X}}\,\exp\left[\widetilde{\boldsymbol{X}}^{\T}\left(\boldsymbol{1}-\boldsymbol{W}\right)\boldsymbol{X}+\frac{D}{2}\widetilde{\boldsymbol{X}}^{\T}\widetilde{\boldsymbol{X}}+\boldsymbol{J}^{\T}\boldsymbol{X}\right]\label{eq:ZJ-1}\\
 & =\exp\left(\frac{1}{2}\boldsymbol{J}^{\mathrm{T}}\left(\boldsymbol{1}-\boldsymbol{W}\right)^{-1}\boldsymbol{D}\left(\boldsymbol{1}-\boldsymbol{W}\right)^{-\T}\boldsymbol{J}\right)\,,\nonumber 
\end{align}
respectively. We obtain the time-lag-integrated covariances

\begin{alignat*}{1}
C_{ij} & =\left\langle X_{i}X_{j}\right\rangle \\
 & =\left.\frac{\mathrm{\partial}}{\partial J_{i}}\frac{\mathrm{\partial}}{\partial J_{j}}Z\left(\boldsymbol{J},\widetilde{\boldsymbol{J}}\right)\right|_{\boldsymbol{J},\widetilde{\boldsymbol{J}}=0}\\
 & =\left[\left(\boldsymbol{1}-\boldsymbol{W}\right)^{-1}\boldsymbol{D}\left(\boldsymbol{1}-\boldsymbol{W}\right)^{-\T}\right]_{ij}\,.
\end{alignat*}

\subsection{Saddle points and correlators of activity fields\label{appendix:Saddle-points-and}}

The saddle points $\boldsymbol{Q}^{*},\widetilde{\boldsymbol{Q}}^{*}$
are given by $\left.\frac{\partial}{\partial Q_{i}}S\left(\boldsymbol{Q},\widetilde{\boldsymbol{Q}}\right)\right|_{\boldsymbol{Q}^{*},\widetilde{\boldsymbol{Q}}^{*}}=0$
and $\left.\frac{\partial}{\partial\widetilde{Q}_{i}}S\left(\boldsymbol{Q},\widetilde{\boldsymbol{Q}}\right)\right|_{\boldsymbol{Q}^{*},\widetilde{\boldsymbol{Q}}^{*}}=0$,
which yield

\begin{eqnarray*}
Q_{i}^{*} & = & \frac{1}{N}\sum_{j}\Delta_{ij}\left\langle X_{j}X_{j}\right\rangle _{\boldsymbol{Q}^{*},\widetilde{\boldsymbol{Q}}^{*}}\,,\\
\widetilde{Q}_{i}^{*} & = & \frac{1}{2N}\left\langle \widetilde{X}_{i}\widetilde{X}_{i}\right\rangle _{\boldsymbol{Q}^{*},\widetilde{\boldsymbol{Q}}^{*}}\,,
\end{eqnarray*}
including correlators evaluated at the saddle point $\boldsymbol{Q}^{*},\widetilde{\boldsymbol{Q}}^{*}$.
To evaluate them, we need to solve the Gaussian integral in \prettyref{eq:interaction_with_aux_fields}
\begin{eqnarray*}
\widetilde{Z}\left(\boldsymbol{J},\widetilde{\boldsymbol{J}}\right) & = & \int\mathcal{D}\boldsymbol{X}\int\mathcal{D}\widetilde{\boldsymbol{X}}\,\exp\left[\widetilde{\boldsymbol{X}}^{\mathrm{T}}\left(\boldsymbol{1}-\boldsymbol{M}\right)\boldsymbol{X}+\frac{1}{2}\widetilde{\boldsymbol{X}}^{\mathrm{T}}\left[\boldsymbol{D}+\mathrm{diag}\left(\boldsymbol{Q}\right)\right]\widetilde{\boldsymbol{X}}+\boldsymbol{X}^{\T}\mathrm{diag}\left(\widetilde{\boldsymbol{Q}}^{\T}\boldsymbol{\Delta}\right)\boldsymbol{X}+\boldsymbol{J}^{\mathrm{T}}\boldsymbol{X}+\widetilde{\boldsymbol{J}}^{\T}\widetilde{\boldsymbol{X}}\right]\,,
\end{eqnarray*}
where we added the additional source term $\widetilde{\boldsymbol{J}}^{\T}\widetilde{\boldsymbol{X}}$
to allow for the calculation of correlators including $\widetilde{\boldsymbol{X}}$.
We can rewrite the equation as
\begin{align*}
\widetilde{Z}\left(\boldsymbol{J}\right) & =\int\mathcal{D}\boldsymbol{Z}\,\exp\left[-\frac{1}{2}\boldsymbol{Z}^{\T}\boldsymbol{A}\boldsymbol{Z}+\boldsymbol{B}^{\T}\boldsymbol{Z}\right]\\
 & =\left(2\pi\right)^{-N}\sqrt{\frac{\left(2\pi\right)^{2n}}{\mathrm{det}\boldsymbol{A}}}\exp\left(\frac{1}{2}\boldsymbol{B}^{T}\boldsymbol{A}^{-1}\boldsymbol{B}\right)\\
 & =\frac{1}{\sqrt{\mathrm{det}\boldsymbol{A}}}\exp\left(\frac{1}{2}\boldsymbol{B}^{T}\boldsymbol{A}^{-1}\boldsymbol{B}\right)\,,
\end{align*}
using
\begin{eqnarray*}
\boldsymbol{Z} & = & \left(\begin{array}{c}
\boldsymbol{X}\\
\widetilde{\boldsymbol{X}}
\end{array}\right)\,,\quad\boldsymbol{B}=\left(\begin{array}{c}
\boldsymbol{J}\\
\mathrm{i}\widetilde{\boldsymbol{J}}
\end{array}\right)\,,\\
\boldsymbol{A} & = & \left(\begin{array}{cc}
-2\mathrm{diag}\left(\widetilde{\boldsymbol{Q}}^{\T}\boldsymbol{\Delta}\right) & -\mathrm{i}\left(\boldsymbol{1}-\boldsymbol{M}\right)^{\T}\\
-\mathrm{i}\left(\boldsymbol{1}-\boldsymbol{M}\right) & \left[\boldsymbol{D}+\mathrm{diag}\left(\boldsymbol{Q}\right)\right]
\end{array}\right)\,,
\end{eqnarray*}
where the prefactor $\left(2\pi\right)^{-N}$ comes from the integration
measure $\mathcal{D}\widetilde{\boldsymbol{X}}$ , such that 
\begin{eqnarray*}
\boldsymbol{A}_{11}^{-1} & = & \left\{ -2\mathrm{diag}\left(\widetilde{\boldsymbol{Q}}^{\T}\boldsymbol{\Delta}\right)+\left(\boldsymbol{1}-\boldsymbol{M}\right)^{\T}\left[\boldsymbol{D}+\mathrm{diag}\left(\boldsymbol{Q}\right)\right]^{-1}\left(\boldsymbol{1}-\boldsymbol{M}\right)\right\} ^{-1}\,,\\
\boldsymbol{A}_{12}^{-1} & = & \mathrm{i}\left\{ -2\mathrm{diag}\left(\widetilde{\boldsymbol{Q}}^{\T}\boldsymbol{\Delta}\right)+\left(\boldsymbol{1}-\boldsymbol{M}\right)^{\T}\left[\boldsymbol{D}+\mathrm{diag}\left(\boldsymbol{Q}\right)\right]^{-1}\left(\boldsymbol{1}-\boldsymbol{M}\right)\right\} ^{-1}\left(\boldsymbol{1}-\boldsymbol{M}\right)^{\T}\left[\boldsymbol{D}+\mathrm{diag}\left(\boldsymbol{Q}\right)\right]^{-1}\,,\\
\boldsymbol{A}_{21}^{-1} & = & \mathrm{i}\left[\boldsymbol{D}+\mathrm{diag}\left(\boldsymbol{Q}\right)\right]^{-1}\left(\boldsymbol{1}-\boldsymbol{M}\right)\left\{ -2\mathrm{diag}\left(\widetilde{\boldsymbol{Q}}^{\T}\boldsymbol{\Delta}\right)+\left(\boldsymbol{1}-\boldsymbol{M}\right)^{\T}\left[\boldsymbol{D}+\mathrm{diag}\left(\boldsymbol{Q}\right)\right]^{-1}\left(\boldsymbol{1}-\boldsymbol{M}\right)\right\} ^{-1}\,,\\
\boldsymbol{A}_{22}^{-1} & = & \left[\boldsymbol{D}+\mathrm{diag}\left(\boldsymbol{Q}\right)\right]^{-1}\\
 &  & -\left[\boldsymbol{D}+\mathrm{diag}\left(\boldsymbol{Q}\right)\right]^{-1}\left(\boldsymbol{1}-\boldsymbol{M}\right)\left\{ -2\mathrm{diag}\left(\widetilde{\boldsymbol{Q}}^{\T}\boldsymbol{\Delta}\right)+\left(\boldsymbol{1}-\boldsymbol{M}\right)^{\T}\left[\boldsymbol{D}+\mathrm{diag}\left(\boldsymbol{Q}\right)\right]^{-1}\left(\boldsymbol{1}-\boldsymbol{M}\right)\right\} ^{-1}\left(\boldsymbol{1}-\boldsymbol{M}\right)^{\T}\left[\boldsymbol{D}+\mathrm{diag}\left(\boldsymbol{Q}\right)\right]^{-1}\,.
\end{eqnarray*}
Deriving the normalized moment-generating function $Z\left(\boldsymbol{J},\widetilde{\boldsymbol{J}}\right)=\widetilde{Z}\left(\boldsymbol{J},\widetilde{\boldsymbol{J}}\right)/\widetilde{Z}\left(0,0\right)$
twice with respect to $\widetilde{\boldsymbol{J}}$ yields
\begin{eqnarray*}
\widetilde{\boldsymbol{Q}}^{*} & = & \frac{1}{2N}\left\langle \widetilde{\boldsymbol{X}}\widetilde{\boldsymbol{X}}^{\T}\right\rangle _{\boldsymbol{Q}^{*},\widetilde{\boldsymbol{Q}}^{*}}\\
 & = & \frac{1}{2N}A_{22}^{-1}\\
 & = & \frac{1}{2N}\biggl(\left[\boldsymbol{D}+\mathrm{diag}\left(\boldsymbol{Q}\right)\right]^{-1}\\
 &  & -\left[\boldsymbol{D}+\mathrm{diag}\left(\boldsymbol{Q}\right)\right]^{-1}\left(\boldsymbol{1}-\boldsymbol{M}\right)\left\{ -2\mathrm{diag}\left(\widetilde{\boldsymbol{Q}}^{\T}\boldsymbol{\Delta}\right)+\left(\boldsymbol{1}-\boldsymbol{M}\right)^{\T}\left[\boldsymbol{D}+\mathrm{diag}\left(\boldsymbol{Q}\right)\right]^{-1}\left(\boldsymbol{1}-\boldsymbol{M}\right)\right\} ^{-1}\left(\boldsymbol{1}-\boldsymbol{M}\right)^{\T}\left[\boldsymbol{D}+\mathrm{diag}\left(\boldsymbol{Q}\right)\right]^{-1}\biggr)\,,
\end{eqnarray*}
which is solved by $\widetilde{\boldsymbol{Q}}^{*}=0$. Inserting
this result, we find 
\begin{eqnarray*}
\left\langle \boldsymbol{X}\boldsymbol{X}^{\T}\right\rangle _{\boldsymbol{Q}^{*},\widetilde{\boldsymbol{Q}}^{*}} & = & \left(\boldsymbol{1}-\boldsymbol{M}\right)^{-1}\left[\boldsymbol{D}+\mathrm{diag}\left(\boldsymbol{Q}\right)\right]\left(\boldsymbol{1}-\boldsymbol{M}\right)^{-\T}\,,\\
\left\langle \widetilde{\boldsymbol{X}}\boldsymbol{X}^{\T}\right\rangle _{\boldsymbol{Q}^{*},\widetilde{\boldsymbol{Q}}^{*}} & = & -\left(\boldsymbol{1}-\boldsymbol{M}\right)^{-\T}\,,\\
\left\langle \widetilde{\boldsymbol{X}}\widetilde{\boldsymbol{X}}^{\T}\right\rangle _{\boldsymbol{Q}^{*},\widetilde{\boldsymbol{Q}}^{*}} & = & 0\,.
\end{eqnarray*}
Inserting the correlators into the saddle-point equations and solving
for $\boldsymbol{Q}^{*}$ yields
\begin{eqnarray}
Q_{i}^{*} & = & \frac{1}{N}\sum_{j,k,l,m}\left(\boldsymbol{1}-\frac{1}{N}\Delta\cdot\boldsymbol{R}^{\circ2}\right)_{ij}^{-1}\Delta_{jk}R_{kl}D_{lm}R_{km}\,,\label{eq:saddle_point_solution}\\
\widetilde{Q}_{i}^{*} & = & 0\,.\nonumber 
\end{eqnarray}
with $\boldsymbol{R}=\left(\boldsymbol{1}-\boldsymbol{M}\right)^{-1}$.

The saddle point of the auxiliary fields $\boldsymbol{Q}_{XX}$, $\boldsymbol{Q}_{XY}$,
$\boldsymbol{Q}_{YY}$ in the replica theory are determined by finding
the zeros of the first derivative of the action {[}\prettyref{eq:replica_action}{]}.
This yields
\begin{eqnarray*}
Q_{XX,i}^{*} & = & \frac{1}{N}\sum_{j}\Delta_{ij}\left\langle X_{j}X_{j}\right\rangle _{\boldsymbol{Q}^{*},\widetilde{\boldsymbol{Q}}^{*}}=Q_{i}^{*}\,,\\
Q_{YY,i}^{*} & = & \frac{1}{N}\sum_{j}\Delta_{ij}\left\langle Y_{j}Y_{j}\right\rangle _{\boldsymbol{Q}^{*},\widetilde{\boldsymbol{Q}}^{*}}=Q_{i}^{*}\,,\\
Q_{XY,i}^{*} & = & \frac{1}{N}\sum_{j}\Delta_{ij}\left\langle X_{j}Y_{j}\right\rangle _{\boldsymbol{Q}^{*},\widetilde{\boldsymbol{Q}}^{*}}\,,\\
\widetilde{Q}_{XX,i}^{*} & = & \frac{1}{2N}\left\langle \widetilde{X}_{i}\widetilde{X}_{i}\right\rangle _{\boldsymbol{Q}^{*},\widetilde{\boldsymbol{Q}}^{*}}=0\,,\\
\widetilde{Q}_{YY,i}^{*} & = & \frac{1}{2N}\left\langle \widetilde{Y}_{i}\widetilde{Y}_{i}\right\rangle _{\boldsymbol{Q}^{*},\widetilde{\boldsymbol{Q}}^{*}}=0\,,\\
\widetilde{Q}_{XY,i}^{*} & = & \frac{1}{N}\left\langle \widetilde{X}_{i}\widetilde{Y}_{i}\right\rangle _{\boldsymbol{Q}^{*},\widetilde{\boldsymbol{Q}}^{*}}=0\,,
\end{eqnarray*}
and in a fashion analogous to the derivation above, we find
\[
\left\langle \boldsymbol{X}\boldsymbol{Y}^{\T}\right\rangle _{\boldsymbol{Q}^{*},\widetilde{\boldsymbol{Q}}^{*}}=\left(\boldsymbol{1}-\boldsymbol{M}\right)^{-1}\mathrm{diag}\left(\boldsymbol{Q}_{XY}^{*}\right)\left(\boldsymbol{1}-\boldsymbol{M}\right)^{-\T}\,.
\]
Inserting the latter solution into the saddle-point equations again
yields a linear self-consistency equation for $\boldsymbol{Q}_{XY}^{*}$
with the solution
\[
Q_{XY,i}^{*}=0\,,
\]
such that
\[
\left\langle \boldsymbol{X}\boldsymbol{Y}^{\T}\right\rangle _{\boldsymbol{Q}^{*},\widetilde{\boldsymbol{Q}}^{*}}=0\,.
\]

\subsection{Fluctuations around saddle-points\label{appendix:fluctuations_around_saddle-points}}

Here we showcase how to perform a fluctuation expansion of the correlator
$\left\langle X_{i}Y_{i}\right\rangle _{\boldsymbol{Q},\widetilde{\boldsymbol{Q}}}$
around $\boldsymbol{Q}_{XY}^{*}$ and $\widetilde{\boldsymbol{Q}}_{XY}^{*}$.
Other correlators follow analogously. Following the definition in
\prettyref{eq:disorder_average_mapped_to_q_average_replica}, the
correlator is given by
\begin{align*}
\left\langle X_{i}Y_{i}\right\rangle _{\boldsymbol{Q},\widetilde{\boldsymbol{Q}}} & =\left.\frac{\partial}{\partial J_{i}}\frac{\partial}{\partial K_{i}}Z_{\boldsymbol{Q},\widetilde{\boldsymbol{Q}}}\left(\boldsymbol{J},\boldsymbol{K}\right)\right|_{\boldsymbol{J},\boldsymbol{K}=0}\,.
\end{align*}
Now, we expand $Z_{\boldsymbol{Q},\widetilde{\boldsymbol{Q}}}$ around
the saddle points
\begin{align*}
Z_{\boldsymbol{Q},\widetilde{\boldsymbol{Q}}}\left(\boldsymbol{J},\boldsymbol{K}\right)\approx & Z_{\boldsymbol{Q}^{*},\widetilde{\boldsymbol{Q}}^{*}}\left(\boldsymbol{J},\boldsymbol{K}\right)\\
 & +\sum_{k}\left.\frac{\partial}{\partial Q_{XY,k}}Z_{\boldsymbol{Q},\widetilde{\boldsymbol{Q}}}\left(\boldsymbol{J},\boldsymbol{K}\right)\right|_{\boldsymbol{Q}^{*},\widetilde{\boldsymbol{Q}}^{*}}\left(Q_{XY,k}-Q_{XY,k}^{*}\right)\\
 & +\sum_{k}\left.\frac{\partial}{\partial\widetilde{Q}_{XY,k}}Z_{\boldsymbol{Q},\widetilde{\boldsymbol{Q}}}\left(\boldsymbol{J},\boldsymbol{K}\right)\right|_{\boldsymbol{Q}^{*},\widetilde{\boldsymbol{Q}}^{*}}\left(\widetilde{Q}_{XY,k}-\widetilde{Q}_{XY,k}^{*}\right)\,,
\end{align*}
and use $Z_{\boldsymbol{Q},\widetilde{\boldsymbol{Q}}}\left(\boldsymbol{J},\boldsymbol{K}\right)=\widetilde{Z}_{\boldsymbol{Q},\widetilde{\boldsymbol{Q}}}\left(\boldsymbol{J},\boldsymbol{K}\right)/\widetilde{Z}_{\boldsymbol{Q},\widetilde{\boldsymbol{Q}}}\left(\boldsymbol{0},\boldsymbol{0}\right)$
to obtain
\begin{align}
\frac{\partial}{\partial Q_{XY,k}}Z_{\boldsymbol{Q},\widetilde{\boldsymbol{Q}}}\left(\boldsymbol{J},\boldsymbol{K}\right) & =\frac{1}{\widetilde{Z}_{\boldsymbol{Q},\widetilde{\boldsymbol{Q}}}\left(\boldsymbol{0},\boldsymbol{0}\right)}\frac{\partial}{\partial Q_{XY,k}}\widetilde{Z}_{\boldsymbol{Q},\widetilde{\boldsymbol{Q}}}\left(\boldsymbol{J},\boldsymbol{K}\right)-\frac{\widetilde{Z}_{\boldsymbol{Q},\widetilde{\boldsymbol{Q}}}\left(\boldsymbol{J},\boldsymbol{K}\right)}{\widetilde{Z}_{\boldsymbol{Q},\widetilde{\boldsymbol{Q}}}\left(\boldsymbol{0},\boldsymbol{0}\right)^{2}}\frac{\partial}{\partial Q_{XY,k}}\widetilde{Z}_{\boldsymbol{Q},\widetilde{\boldsymbol{Q}}}\left(0,0\right)\,.\label{eq:supp_eq1}
\end{align}
Using the definition of $\widetilde{Z}_{\boldsymbol{Q},\widetilde{\boldsymbol{Q}}}\left(\boldsymbol{J},\boldsymbol{K}\right)$
in \prettyref{eq:replica_action}, its derivative is given by

\begin{align*}
\frac{\partial}{\partial Q_{XY,k}}\widetilde{Z}_{\boldsymbol{Q},\widetilde{\boldsymbol{Q}}}\left(\boldsymbol{J},\boldsymbol{K}\right)= & \int\mathcal{D}\boldsymbol{X}\int\mathcal{D}\widetilde{\boldsymbol{X}}\int\mathcal{D}\boldsymbol{Y}\int\mathcal{D}\widetilde{\boldsymbol{Y}}\,\widetilde{X}_{k}\widetilde{Y}_{k}\\
\times\exp\Big[ & S_{\boldsymbol{Q}_{XX},\widetilde{\boldsymbol{Q}}_{XX}}\left(\boldsymbol{X},\widetilde{\boldsymbol{X}}\right)+S_{\boldsymbol{Q}_{YY},\widetilde{\boldsymbol{Q}}_{YY}}\left(\boldsymbol{Y},\widetilde{\boldsymbol{Y}}\right)+\widetilde{\boldsymbol{X}}^{\T}\mathrm{diag}\left(\boldsymbol{Q}_{XY}\right)\widetilde{\boldsymbol{Y}}+\boldsymbol{X}^{\T}\mathrm{diag}\left(\widetilde{\boldsymbol{Q}}_{XY}^{\T}\boldsymbol{\Delta}\right)\boldsymbol{Y}+\boldsymbol{J}^{\mathrm{T}}\boldsymbol{X}+\boldsymbol{K}^{\T}\boldsymbol{Y}\Big]\,,
\end{align*}
such that normalizing and evaluating at the saddle point and for zero
sources yields
\begin{align*}
\left.\frac{\left.\frac{\partial}{\partial J_{i}}\frac{\partial}{\partial K_{i}}\frac{\partial}{\partial Q_{XY,k}}\widetilde{Z}_{\boldsymbol{Q},\widetilde{\boldsymbol{Q}}}\left(\boldsymbol{J},\boldsymbol{K}\right)\right|_{\boldsymbol{J},\boldsymbol{K}=0}}{\widetilde{Z}_{\boldsymbol{Q},\widetilde{\boldsymbol{Q}}}\left(\boldsymbol{0},\boldsymbol{0}\right)}\right|_{\boldsymbol{Q}^{*},\widetilde{\boldsymbol{Q}}^{*}} & =\left\langle X_{i}Y_{i}\widetilde{X}_{k}\widetilde{Y}_{k}\right\rangle _{\boldsymbol{Q}^{*},\widetilde{\boldsymbol{Q}}^{*}}\,.
\end{align*}
The second term on the right-hand side of \prettyref{eq:supp_eq1}
vanishes at the saddle point and for $\boldsymbol{J}=\boldsymbol{K}=0$
\[
\left.\left.\frac{\partial}{\partial J_{i}}\frac{\partial}{\partial K_{i}}\frac{\widetilde{Z}_{\boldsymbol{Q},\widetilde{\boldsymbol{Q}}}\left(\boldsymbol{J},\boldsymbol{K}\right)}{\widetilde{Z}_{\boldsymbol{Q},\widetilde{\boldsymbol{Q}}}\left(\boldsymbol{0},\boldsymbol{0}\right)^{2}}\frac{\partial}{\partial Q_{XY,k}}\widetilde{Z}_{\boldsymbol{Q},\widetilde{\boldsymbol{Q}}}\left(0,0\right)\right|_{\boldsymbol{J},\boldsymbol{K}=0}\right|_{\boldsymbol{Q}^{*},\widetilde{\boldsymbol{Q}}^{*}}=\left\langle X_{i}Y_{i}\right\rangle _{\boldsymbol{Q}^{*},\widetilde{\boldsymbol{Q}}^{*}}\left\langle \widetilde{X}_{k}\widetilde{Y}_{k}\right\rangle _{\boldsymbol{Q}^{*},\widetilde{\boldsymbol{Q}}^{*}}=0\,.
\]
The derivative with respect to $\widetilde{Q}_{XY,k}$ can be computed
analogously, with $\widetilde{X}_{k}\widetilde{Y}_{k}$ replaced by
$\sum_{l}\Delta_{kl}X_{l}Y_{l}$. Therefore, the first-order expansion
in the replica coupling term reads
\begin{equation}
\left\langle X_{i}Y_{i}\right\rangle _{\boldsymbol{Q},\widetilde{\boldsymbol{Q}}}=\underbrace{\left\langle X_{i}Y_{i}\right\rangle _{\boldsymbol{Q}^{*},\widetilde{\boldsymbol{Q}}^{*}}}_{=0}+\sum_{k}\left\langle X_{i}Y_{i}\widetilde{X}_{k}\widetilde{Y}_{k}\right\rangle _{\boldsymbol{Q}^{*},\widetilde{\boldsymbol{Q}}^{*}}\left(Q_{XY,k}-Q_{XY,k}^{*}\right)+\sum_{k,l}\Delta_{kl}\left\langle X_{i}Y_{i}X_{l}Y_{l}\right\rangle _{\boldsymbol{Q}^{*},\widetilde{\boldsymbol{Q}}^{*}}\left(\widetilde{Q}_{XY,k}-\widetilde{Q}_{XY,k}^{*}\right)\,,
\end{equation}
which we use in \prettyref{eq:expanded_correlator}.

\subsection{Correlators of auxiliary fields\label{appendix:Correlators-of-auxiliary}}

We consider the Gaussian approximation of $p\left(\boldsymbol{Q},\widetilde{\boldsymbol{Q}}\right)$
with

\[
\mathcal{S}\left(\boldsymbol{Q},\widetilde{\boldsymbol{Q}}\right)=\mathcal{S}\left(\boldsymbol{Q}^{*},\widetilde{\boldsymbol{Q}}^{*}\right)+\frac{1}{2}\left(\delta\boldsymbol{Q}_{XY},\delta\widetilde{\boldsymbol{Q}}_{XY}\right)\boldsymbol{\mathcal{S}}^{(2)}\left(\begin{array}{c}
\delta\boldsymbol{Q}_{XY}\\
\delta\widetilde{\boldsymbol{Q}}_{XY}
\end{array}\right)\,,
\]
where $\boldsymbol{\mathcal{S}}^{(2)}$ contains the second derivatives
with respect to the auxiliary fields
\[
\boldsymbol{\mathcal{S}}^{(2)}=\left(\begin{array}{cc}
\left.\frac{\partial\mathcal{S}\left(\boldsymbol{Q},\widetilde{\boldsymbol{Q}}\right)}{\partial\boldsymbol{Q}_{XY}\partial\boldsymbol{Q}_{XY}}\right|_{\boldsymbol{Q}^{*},\widetilde{\boldsymbol{Q}}^{*}} & \left.\frac{\partial\mathcal{S}\left(\boldsymbol{Q},\widetilde{\boldsymbol{Q}}\right)}{\partial\boldsymbol{Q}_{XY}\partial\widetilde{\boldsymbol{Q}}_{XY}}\right|_{\boldsymbol{Q}^{*},\widetilde{\boldsymbol{Q}}^{*}}\\
\left.\frac{\partial\mathcal{S}\left(\boldsymbol{Q},\widetilde{\boldsymbol{Q}}\right)}{\partial\widetilde{\boldsymbol{Q}}_{XY}\partial\boldsymbol{Q}_{XY}}\right|_{\boldsymbol{Q}^{*},\widetilde{\boldsymbol{Q}}^{*}} & \left.\frac{\partial\mathcal{S}\left(\boldsymbol{Q},\widetilde{\boldsymbol{Q}}\right)}{\partial\widetilde{\boldsymbol{Q}}_{XY}\partial\widetilde{\boldsymbol{Q}}_{XY}}\right|_{\boldsymbol{Q}^{*},\widetilde{\boldsymbol{Q}}^{*}}
\end{array}\right)\,,
\]
with
\begin{eqnarray*}
\mathcal{S}_{11,ij}^{(2)} & = & 0\\
\mathcal{S}_{12,ij}^{(2)} & = & N\delta_{ij}-\sum_{k}\Delta_{jk}\left\langle \widetilde{X}_{i}\widetilde{Y}_{i}X_{k}Y_{k}\right\rangle _{\boldsymbol{Q}^{*},\widetilde{\boldsymbol{Q}}^{*}}=N\delta_{ij}-\sum_{k}\Delta_{jk}R_{ki}^{2}\\
\mathcal{S}_{21,ij}^{(2)} & = & N\delta_{ij}-\sum_{k}\Delta_{ik}\left\langle \widetilde{X}_{j}\widetilde{Y}_{j}X_{k}Y_{k}\right\rangle _{\boldsymbol{Q}^{*},\widetilde{\boldsymbol{Q}}^{*}}=N\delta_{ij}-\sum_{k}\Delta_{ik}R_{kj}^{2}\\
\mathcal{S}_{22,ij}^{(2)} & = & -\sum_{k,l}\Delta_{ik}\Delta_{jl}\left\langle X_{k}Y_{k}X_{l}Y_{l}\right\rangle _{\boldsymbol{Q}^{*},\widetilde{\boldsymbol{Q}}^{*}}=-\sum_{k,l}\Delta_{ik}\Delta_{jl}\left\langle X_{k}X_{l}\right\rangle _{\boldsymbol{Q}^{*},\widetilde{\boldsymbol{Q}}^{*}}\left\langle Y_{k}Y_{l}\right\rangle _{\boldsymbol{Q}^{*},\widetilde{\boldsymbol{Q}}^{*}}\,.
\end{eqnarray*}
Using

\[
\boldsymbol{\mathcal{S}}^{(2)}=\left(\begin{array}{cc}
0 & \boldsymbol{\mathcal{S}}_{21}^{(2)\T}\\
\boldsymbol{\mathcal{S}}_{21}^{(2)} & \boldsymbol{\mathcal{S}}_{22}^{(2)}
\end{array}\right)\,,\quad\left(\boldsymbol{\mathcal{S}}^{(2)}\right)^{-1}=\left(\begin{array}{cc}
-\boldsymbol{\mathcal{S}}_{21}^{(2)-1}\boldsymbol{\mathcal{S}}_{22}^{(2)}\boldsymbol{\mathcal{S}}_{21}^{(2)-\mathrm{T}} & \boldsymbol{\mathcal{S}}_{21}^{(2)-1}\\
\boldsymbol{\mathcal{S}}_{21}^{(2)-\T} & 0
\end{array}\right)\,,
\]
we find
\begin{eqnarray*}
\left\langle \delta\boldsymbol{Q}_{XY}\delta\boldsymbol{Q}_{XY}^{\T}\right\rangle  & = & -\boldsymbol{\mathcal{S}}_{21}^{(2)-1}\boldsymbol{\mathcal{S}}_{22}^{(2)}\boldsymbol{\mathcal{S}}_{21}^{(2)-\mathrm{T}}\\
 & = & \frac{1}{N^{2}}\left[\boldsymbol{1}-\frac{1}{N}\boldsymbol{\Delta}\cdot\boldsymbol{R}^{\circ2}\right]^{-1}\boldsymbol{\Delta}\left\langle \boldsymbol{X}\boldsymbol{X}^{\T}\right\rangle _{\boldsymbol{Q}^{*},\widetilde{\boldsymbol{Q}}^{*}}^{\circ2}\boldsymbol{\Delta}^{\T}\left[\boldsymbol{1}-\frac{1}{N}\boldsymbol{\Delta}\cdot\boldsymbol{R}^{\circ2}\right]^{-\T}\\
\left\langle \delta\boldsymbol{Q}_{XY}\delta\widetilde{\boldsymbol{Q}}_{XY}^{\T}\right\rangle  & = & \boldsymbol{\mathcal{S}}_{21}^{(2)-1}=\frac{1}{N}\left[\boldsymbol{1}-\frac{1}{N}\boldsymbol{\Delta}\cdot\boldsymbol{R}^{\circ2}\right]^{-1}\\
\left\langle \delta\widetilde{\boldsymbol{Q}}_{XY}\delta\widetilde{\boldsymbol{Q}}_{XY}^{\T}\right\rangle  & = & 0\,.
\end{eqnarray*}

\subsection{Disorder-averaged variance of covariances\label{appendix:Disorder-averaged-variance-of}}

Starting with \prettyref{eq:covariance_after_wick}, we find
\begin{eqnarray*}
\left\langle C_{ij}^{2}\right\rangle _{\boldsymbol{W}} & = & \left\langle \left\langle X_{i}X_{j}\right\rangle ^{2}\right\rangle _{\boldsymbol{W}}\\
 & = & \int D\boldsymbol{Q}\int D\widetilde{\boldsymbol{Q}}\,p\left(\boldsymbol{Q},\widetilde{\boldsymbol{Q}}\right)\left\langle X_{i}X_{j}Y_{i}Y_{j}\right\rangle _{\boldsymbol{Q},\widetilde{\boldsymbol{Q}}}\\
 & = & \int D\boldsymbol{Q}\int D\widetilde{\boldsymbol{Q}}\,p\left(\boldsymbol{Q},\widetilde{\boldsymbol{Q}}\right)\left(\left\langle X_{i}X_{j}\right\rangle _{\boldsymbol{Q},\widetilde{\boldsymbol{Q}}}\left\langle Y_{i}Y_{j}\right\rangle _{\boldsymbol{Q},\widetilde{\boldsymbol{Q}}}+\left\langle X_{i}Y_{i}\right\rangle _{\boldsymbol{Q},\widetilde{\boldsymbol{Q}}}\left\langle X_{j}Y_{j}\right\rangle _{\boldsymbol{Q},\widetilde{\boldsymbol{Q}}}+\left\langle X_{i}Y_{j}\right\rangle _{\boldsymbol{Q},\widetilde{\boldsymbol{Q}}}\left\langle X_{j}Y_{i}\right\rangle _{\boldsymbol{Q},\widetilde{\boldsymbol{Q}}}\right)\\
 & \approx & \left\langle \left\langle X_{i}X_{j}\right\rangle \right\rangle _{\boldsymbol{W}}^{2}+\left(1+\delta_{ij}\right)\int D\boldsymbol{Q}\int D\widetilde{\boldsymbol{Q}}\,p\left(\boldsymbol{Q},\widetilde{\boldsymbol{Q}}\right)\left\langle X_{i}Y_{i}\right\rangle _{\boldsymbol{Q},\widetilde{\boldsymbol{Q}}}\left\langle X_{j}Y_{j}\right\rangle _{\boldsymbol{Q},\widetilde{\boldsymbol{Q}}}\\
 & = & \left\langle C_{ij}\right\rangle _{\boldsymbol{W}}^{2}+\left(1+\delta_{ij}\right)\int D\boldsymbol{Q}\int D\widetilde{\boldsymbol{Q}}\,p\left(\boldsymbol{Q},\widetilde{\boldsymbol{Q}}\right)\left\langle X_{i}Y_{i}\right\rangle _{\boldsymbol{Q},\widetilde{\boldsymbol{Q}}}\left\langle X_{j}Y_{j}\right\rangle _{\boldsymbol{Q},\widetilde{\boldsymbol{Q}}}\,.
\end{eqnarray*}
\newpage Inserting the derived expressions for the correlators into
\prettyref{eq:covs_second_term_gaussian_expanded} yields
\begin{eqnarray*}
\int D\boldsymbol{Q} & \int & D\widetilde{\boldsymbol{Q}}\,p\left(\boldsymbol{Q},\widetilde{\boldsymbol{Q}}\right)\left\langle X_{i}Y_{i}\right\rangle _{\boldsymbol{Q},\widetilde{\boldsymbol{Q}}}\left\langle X_{j}Y_{j}\right\rangle _{\boldsymbol{Q},\widetilde{\boldsymbol{Q}}}\\
 & = & \sum_{k,l}\left\langle X_{i}\widetilde{X}_{k}\right\rangle _{\boldsymbol{Q}^{*},\widetilde{\boldsymbol{Q}}^{*}}^{2}\left\langle X_{j}\widetilde{X}_{l}\right\rangle _{\boldsymbol{Q}^{*},\widetilde{\boldsymbol{Q}}^{*}}^{2}\left\langle \delta Q_{XY,k}\delta Q_{XY,l}\right\rangle _{\boldsymbol{Q},\widetilde{\boldsymbol{Q}}}\\
 &  & +\sum_{k,l,m}\left\langle X_{i}\widetilde{X}_{k}\right\rangle _{\boldsymbol{Q}^{*},\widetilde{\boldsymbol{Q}}^{*}}^{2}\Delta_{lm}\left\langle X_{j}X_{m}\right\rangle _{\boldsymbol{Q}^{*},\widetilde{\boldsymbol{Q}}^{*}}^{2}\left\langle \delta Q_{XY,k}\delta\widetilde{Q}_{XY,l}\right\rangle _{\boldsymbol{Q},\widetilde{\boldsymbol{Q}}}\\
 &  & +\sum_{k,l,m}\left\langle X_{j}\widetilde{X}_{k}\right\rangle _{\boldsymbol{Q}^{*},\widetilde{\boldsymbol{Q}}^{*}}^{2}\Delta_{lm}\left\langle X_{i}X_{m}\right\rangle _{\boldsymbol{Q}^{*},\widetilde{\boldsymbol{Q}}^{*}}^{2}\left\langle \delta Q_{XY,k}\delta\widetilde{Q}_{XY,l}\right\rangle _{\boldsymbol{Q},\widetilde{\boldsymbol{Q}}}\\
 &  & +\sum_{k,l,m,n}\left\langle X_{i}X_{m}\right\rangle _{\boldsymbol{Q}^{*},\widetilde{\boldsymbol{Q}}^{*}}^{2}\left\langle X_{j}X_{n}\right\rangle _{\boldsymbol{Q}^{*},\widetilde{\boldsymbol{Q}}^{*}}^{2}\Delta_{km}\Delta_{ln}\left\langle \delta\widetilde{Q}_{XY,k}\delta\widetilde{Q}_{XY,l}\right\rangle _{\boldsymbol{Q},\widetilde{\boldsymbol{Q}}}\\
 & = & \sum_{k,l}R_{ik}^{2}\left\{ \frac{1}{N^{2}}\left[\boldsymbol{1}-\frac{1}{N}\boldsymbol{\Delta}\cdot\boldsymbol{R}^{\circ2}\right]^{-1}\boldsymbol{\Delta}\left\langle \boldsymbol{X}\boldsymbol{X}^{\T}\right\rangle _{\boldsymbol{Q}^{*},\widetilde{\boldsymbol{Q}}^{*}}^{\circ2}\boldsymbol{\Delta}^{\T}\left[\boldsymbol{1}-\frac{1}{N}\boldsymbol{\Delta}\cdot\boldsymbol{R}^{\circ2}\right]^{-\T}\right\} _{kl}R_{jl}^{2}\\
 &  & +\sum_{k,l,m}R_{ik}^{2}\Delta_{lm}\left\langle X_{j}X_{m}\right\rangle _{\boldsymbol{Q}^{*},\widetilde{\boldsymbol{Q}}^{*}}^{2}\left\{ \frac{1}{N}\left[\boldsymbol{1}-\frac{1}{N}\boldsymbol{\Delta}\cdot\boldsymbol{R}^{\circ2}\right]^{-1}\right\} _{kl}\\
 &  & +\sum_{k,l,m}R_{jk}^{2}\Delta_{lm}\left\langle X_{i}X_{m}\right\rangle _{\boldsymbol{Q}^{*},\widetilde{\boldsymbol{Q}}^{*}}^{2}\left\{ \frac{1}{N}\left[\boldsymbol{1}-\frac{1}{N}\boldsymbol{\Delta}\cdot\boldsymbol{R}^{\circ2}\right]^{-1}\right\} _{kl}\\
 & = & \left\{ \text{\ensuremath{\boldsymbol{R}^{\circ2}}}\frac{1}{N^{2}}\left[\boldsymbol{1}-\frac{1}{N}\boldsymbol{\Delta}\cdot\boldsymbol{R}^{\circ2}\right]^{-1}\boldsymbol{\Delta}\left\langle \boldsymbol{X}\boldsymbol{X}^{\T}\right\rangle _{\boldsymbol{Q}^{*},\widetilde{\boldsymbol{Q}}^{*}}^{\circ2}\boldsymbol{\Delta}^{\T}\left[\boldsymbol{1}-\frac{1}{N}\boldsymbol{\Delta}\cdot\boldsymbol{R}^{\circ2}\right]^{-\T}\text{\ensuremath{\boldsymbol{R}^{\circ2\T}}}\right\} _{ij}\\
 &  & +\left\{ \text{\ensuremath{\boldsymbol{R}^{\circ2}\frac{1}{N}\left[\boldsymbol{1}-\frac{1}{N}\boldsymbol{\Delta}\cdot\boldsymbol{R}^{\circ2}\right]^{-1}} }\boldsymbol{\Delta}\left\langle \boldsymbol{X}\boldsymbol{X}^{\T}\right\rangle _{\boldsymbol{Q}^{*},\widetilde{\boldsymbol{Q}}^{*}}^{\circ2}\right\} _{ij}\\
 &  & +\left\{ \left\langle \boldsymbol{X}\boldsymbol{X}^{\T}\right\rangle _{\boldsymbol{Q}^{*},\widetilde{\boldsymbol{Q}}^{*}}^{\circ2}\boldsymbol{\Delta}^{\T}\frac{1}{N}\left[\boldsymbol{1}-\frac{1}{N}\boldsymbol{\Delta}\cdot\boldsymbol{R}^{\circ2}\right]^{-\T}\boldsymbol{R}^{\circ2\T}\right\} _{ij}\\
 & = & \left\{ \boldsymbol{R}^{\circ2}\boldsymbol{\Delta}\frac{1}{N^{2}}\left[\boldsymbol{1}-\frac{1}{N}\boldsymbol{R}^{\circ2}\cdot\boldsymbol{\Delta}\right]^{-1}\left\langle \boldsymbol{X}\boldsymbol{X}^{\T}\right\rangle _{\boldsymbol{Q}^{*},\widetilde{\boldsymbol{Q}}^{*}}^{\circ2}\boldsymbol{\Delta}^{\T}\left[\boldsymbol{1}-\frac{1}{N}\boldsymbol{\Delta}\cdot\boldsymbol{R}^{\circ2}\right]^{-\T}\text{\ensuremath{\boldsymbol{R}^{\circ2\T}}}\right\} _{ij}\\
 &  & +\left\{ \text{\ensuremath{\boldsymbol{R}^{\circ2}\boldsymbol{\Delta}\frac{1}{N}\left[\boldsymbol{1}-\frac{1}{N}\boldsymbol{R}^{\circ2}\cdot\boldsymbol{\Delta}\right]^{-1}} }\left\langle \boldsymbol{X}\boldsymbol{X}^{\T}\right\rangle _{\boldsymbol{Q}^{*},\widetilde{\boldsymbol{Q}}^{*}}^{\circ2}\right\} _{ij}\\
 &  & +\left\{ \left\langle \boldsymbol{X}\boldsymbol{X}^{\T}\right\rangle _{\boldsymbol{Q}^{*},\widetilde{\boldsymbol{Q}}^{*}}^{\circ2}\frac{1}{N}\left[\boldsymbol{1}-\frac{1}{N}\boldsymbol{R}^{\circ2}\cdot\boldsymbol{\Delta}\right]^{-\T}\boldsymbol{\Delta}^{\T}\boldsymbol{R}^{\circ2\T}\right\} _{ij}\\
 & = & \Bigg\{\left(\boldsymbol{1}+\frac{1}{N}\text{\ensuremath{\boldsymbol{R}^{\circ2}}}\boldsymbol{\Delta}\left[\boldsymbol{1}-\frac{1}{N}\boldsymbol{R}^{\circ2}\cdot\boldsymbol{\Delta}\right]^{-1}\right)\left\langle \boldsymbol{X}\boldsymbol{X}^{\T}\right\rangle _{\boldsymbol{Q}^{*},\widetilde{\boldsymbol{Q}}^{*}}^{\circ2}\left(\boldsymbol{1}+\left[\boldsymbol{1}-\frac{1}{N}\boldsymbol{R}^{\circ2}\cdot\boldsymbol{\Delta}\right]^{-\T}\frac{1}{N}\boldsymbol{\Delta}^{\T}\text{\ensuremath{\boldsymbol{R}^{\circ2\T}}}\right)\Bigg\}_{ij}-\left[\left\langle \boldsymbol{X}\boldsymbol{X}^{\T}\right\rangle _{\boldsymbol{Q}^{*},\widetilde{\boldsymbol{Q}}^{*}}^{\circ2}\right]_{ij}\\
 & = & \left\{ \left[\boldsymbol{1}-\frac{1}{N}\boldsymbol{R}^{\circ2}\cdot\boldsymbol{\Delta}\right]^{-1}\left\langle \boldsymbol{X}\boldsymbol{X}^{\T}\right\rangle _{\boldsymbol{Q}^{*},\widetilde{\boldsymbol{Q}}^{*}}^{\circ2}\left[\boldsymbol{1}-\frac{1}{N}\boldsymbol{R}^{\circ2}\cdot\boldsymbol{\Delta}\right]^{-\T}\right\} _{ij}-\left[\left\langle \boldsymbol{X}\boldsymbol{X}^{\T}\right\rangle _{\boldsymbol{Q}^{*},\widetilde{\boldsymbol{Q}}^{*}}^{\circ2}\right]_{ij}\\
 & = & \left\{ \left[\boldsymbol{1}-\frac{1}{N}\boldsymbol{R}^{\circ2}\cdot\boldsymbol{\Delta}\right]^{-1}\left\langle \boldsymbol{X}\boldsymbol{X}^{\T}\right\rangle _{\boldsymbol{Q}^{*},\widetilde{\boldsymbol{Q}}^{*}}^{\circ2}\left[\boldsymbol{1}-\frac{1}{N}\boldsymbol{R}^{\circ2}\cdot\boldsymbol{\Delta}\right]^{-\T}\right\} _{ij}-\left\langle C_{ij}\right\rangle _{\boldsymbol{W}}^{2}
\end{eqnarray*}
where we used
\begin{eqnarray*}
\boldsymbol{R}^{\circ2}\left[\boldsymbol{1}-\frac{1}{N}\boldsymbol{\Delta}\boldsymbol{R}^{\circ2}\right]^{-1}\boldsymbol{\Delta} & = & \boldsymbol{R}^{\circ2}\sum_{k}\left[\frac{1}{N}\boldsymbol{\Delta}\boldsymbol{R}^{\circ2}\right]^{k}\boldsymbol{\Delta}\\
 & = & \boldsymbol{R}^{\circ2}\left[\boldsymbol{1}+\frac{1}{N}\boldsymbol{\Delta}\boldsymbol{R}^{\circ2}+\frac{1}{N^{2}}\boldsymbol{\Delta}\boldsymbol{R}^{\circ2}\boldsymbol{\Delta}\boldsymbol{R}^{\circ2}+...\right]\boldsymbol{\Delta}\\
 & = & \boldsymbol{R}^{\circ2}\boldsymbol{\Delta}\left[\boldsymbol{1}+\frac{1}{N}\boldsymbol{R}^{\circ2}\boldsymbol{\Delta}+\frac{1}{N^{2}}\boldsymbol{R}^{\circ2}\boldsymbol{\Delta}\boldsymbol{R}^{\circ2}\boldsymbol{\Delta}+...\right]\\
 & = & \boldsymbol{R}^{\circ2}\boldsymbol{\Delta}\left[\boldsymbol{1}-\frac{1}{N}\boldsymbol{R}^{\circ2}\boldsymbol{\Delta}\right]^{-1}\,.
\end{eqnarray*}
Finally, we obtain the second moment to leading order
\begin{align*}
\left\langle C_{ij}^{2}\right\rangle _{\boldsymbol{W}} & =\left(1+\delta_{ij}\right)\left\{ \left[\boldsymbol{1}-\frac{1}{N}\boldsymbol{R}^{\circ2}\cdot\boldsymbol{\Delta}\right]^{-1}\left\langle \boldsymbol{X}\boldsymbol{X}^{\T}\right\rangle _{\boldsymbol{Q}^{*},\widetilde{\boldsymbol{Q}}^{*}}^{\circ2}\left[\boldsymbol{1}-\frac{1}{N}\boldsymbol{R}^{\circ2}\cdot\boldsymbol{\Delta}\right]^{-\T}\right\} _{ij}-\delta_{ij}\left\langle C_{ij}\right\rangle _{\boldsymbol{W}}^{2}\,,
\end{align*}
and for the covariance we find
\begin{eqnarray*}
\left\langle \delta C_{ij}^{2}\right\rangle _{\boldsymbol{W}} & = & \left\langle C_{ij}^{2}\right\rangle _{\boldsymbol{W}}-\left\langle C_{ij}\right\rangle _{\boldsymbol{W}}^{2}\\
 & = & \left(1+\delta_{ij}\right)\left\{ \left[\boldsymbol{1}-\frac{1}{N}\boldsymbol{R}^{\circ2}\cdot\boldsymbol{\Delta}\right]^{-1}\left\langle \boldsymbol{X}\boldsymbol{X}^{\T}\right\rangle _{\boldsymbol{Q}^{*},\widetilde{\boldsymbol{Q}}^{*}}^{\circ2}\left[\boldsymbol{1}-\frac{1}{N}\boldsymbol{R}^{\circ2}\cdot\boldsymbol{\Delta}\right]^{-\T}\right\} _{ij}-\left(1+\delta_{ij}\right)\left\langle C_{ij}\right\rangle _{\boldsymbol{W}}^{2}\\
 & \approx & \left(1+\delta_{ij}\right)\left[\left(\boldsymbol{1}-\boldsymbol{S}\right)^{-1}\left[\boldsymbol{D}+\mathrm{diag}\left(\boldsymbol{Q}\right)\right]\left(\boldsymbol{1}-\boldsymbol{S}\right)^{-\T}\right]_{ij}-\left(1+\delta_{ij}\right)\left\langle C_{ij}\right\rangle _{\boldsymbol{W}}^{2}\,.
\end{eqnarray*}

\newpage{}

\subsection{Dependence of population-resolved covariance statistics on heterogeneity
in noise strength $\boldsymbol{D}$}

\begin{figure}[H]
\begin{centering}
\includegraphics{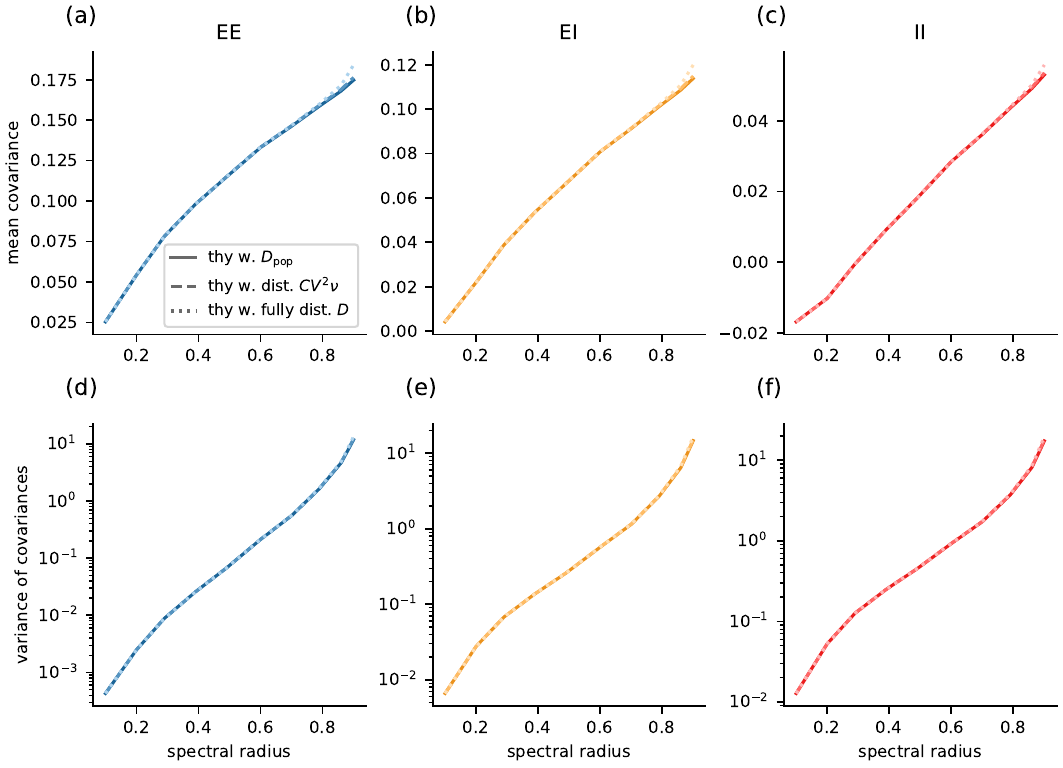}
\par\end{centering}
\caption{Dependence of population-resolved covariance statistics on heterogeneity
in noise strength $\boldsymbol{D}$. The continuous lines show the
results using the realization independent estimate of $\boldsymbol{D}$
\prettyref{eq:noise_strength_estimate_realization_independent}. For
the dashed lines, \prettyref{eq:noise_strength_estimate_realization_independent}
with the single-neuron resolved estimates of $\mathrm{CV}_{i}^{2}\nu_{i}$
introduced in \prettyref{sec:Background} was used, whereas the dotted
lines show the results using the full single-neuron resolved estimate
of $\boldsymbol{D}$ {[}\prettyref{eq:noise_strength_estimate}{]}.
{[}(a)-(c){]} Mean cross-covariances. {[}(d)-(f){]} Variance of cross-covariances.
Same excitatory-inhibitory network model as in previous figures. For
model details and simulation parameters see Appendix \prettyref{appendix:nest_simulation}.\label{fig:cov_stats_dependence_on_D_heterogeneity}}
\end{figure}

In this Appendix, we study the dependence of population-resolved covariance
statistics on heterogeneity in noise strength. The heterogeneity in
noise strength has two sources: variability in $\mathrm{CV}$s and
firing rates across neurons, and variability indirectly induced by
the connectivity via the matrix $\mathbf{B}$ {[}\prettyref{eq:noise_strength_estimate}{]}.
We first replace the full single-neuron resolved estimate of $\mathbf{D}$
{[}\prettyref{eq:noise_strength_estimate}{]} by an estimate that
is independent of the specific connectivity realization in $\mathbf{B}$
(see \prettyref{eq:noise_strength_estimate_realization_independent}
but with neuron-resolved firing rates and $\mathrm{CV}$s). Subsequently,
we additionally replace the individual neuron firing rates and $\mathrm{CV}$s
by their population mean {[}\prettyref{eq:noise_strength_estimate_realization_independent}{]}.
Both simplifications have hardly any effect on the mean and variance
of covariances (\prettyref{fig:cov_stats_dependence_on_D_heterogeneity}),
confirming that these statistics are determined by the effective connectivity
statistics rather than variability in single-neuron firing statistics.

\subsection{Bias correction of variance of covariances\label{appendix:Bias-correction-of}}

\begin{figure*}
\begin{centering}
\includegraphics{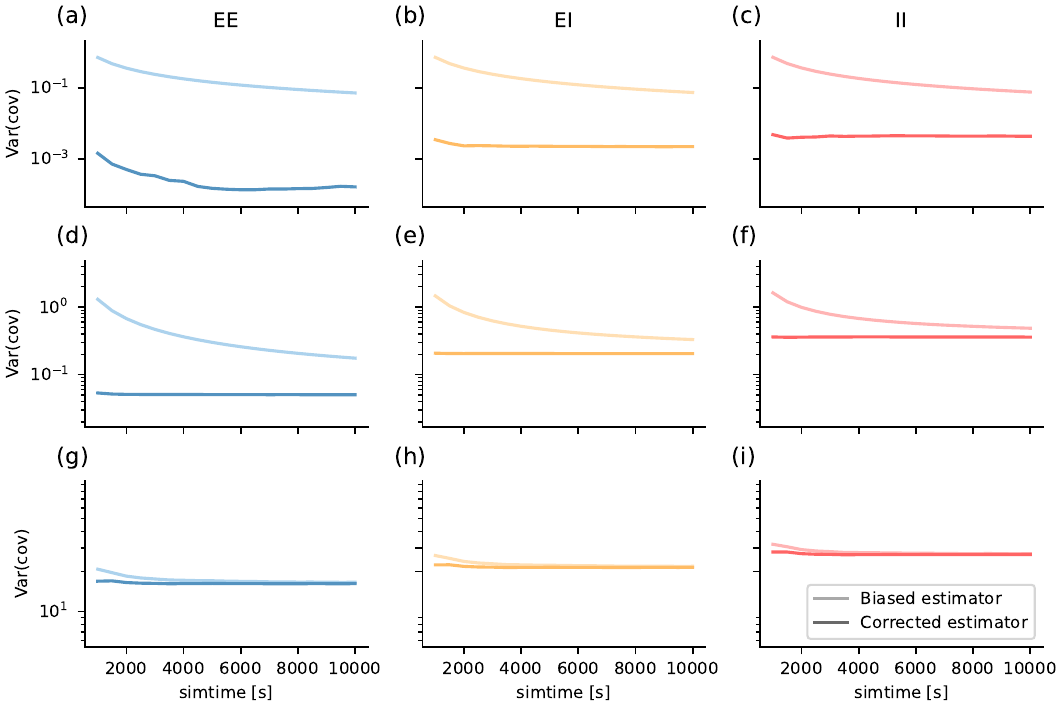}
\par\end{centering}
\caption{Bias correction of variance of covariance estimation for different
simulation lengths at three different spectral radii $r$. The light
color depicts the biased estimator, the dark color the corrected estimator
\prettyref{eq:bias_correction}. {[}(a)-(c){]} $r=0.10$; {[}(d)-(f){]}
$r=0.49$; {[}(g)-(i){]} $r=0.90$. Same excitatory-inhibitory network
model as in previous figures. For model details and simulation parameters
see Appendix \prettyref{appendix:nest_simulation}.\label{fig:Bias-correction-of}}
\end{figure*}
\begin{figure*}
\begin{centering}
\includegraphics{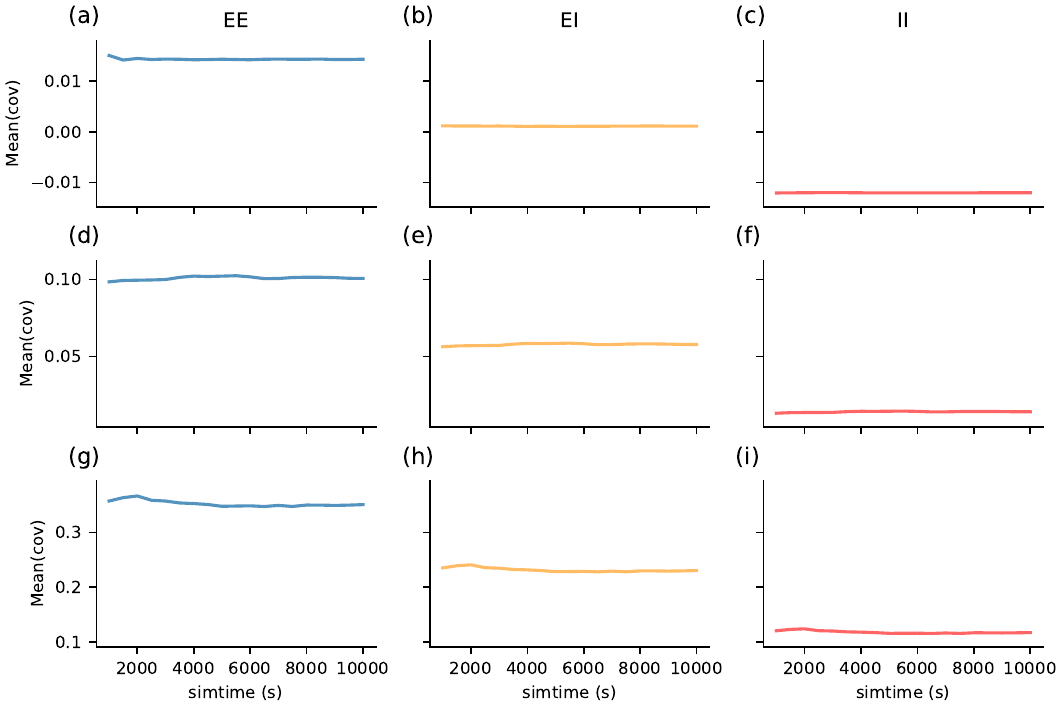}
\par\end{centering}
\caption{Mean of covariance estimation for different simulation lengths at
three different spectral radii $r$. {[}(a)-(c){]} $r=0.10$; {[}(d)-(f){]}
$r=0.49$; {[}(g)-(i){]} $r=0.90$. Same excitatory-inhibitory network
model as in previous figures. For model details and simulation parameters
see Appendix \prettyref{appendix:nest_simulation}.\label{fig:mean_of_simtime}}
\end{figure*}
We utilize Eq. (4) from the supplementary information of Ref. \citep{Dahmen19_13051}
to correct for the bias in the estimation of the variances of covariances
due to the finite simulation time. The analogous correction for two
populations $a$ and $b$ is given by
\begin{equation}
\delta C_{ab}^{2}=\delta\hat{C}_{ab}^{2}-\frac{\left\langle A_{a}\right\rangle \left\langle A_{b}\right\rangle -\left\langle C_{ab}\right\rangle ^{2}}{N+1}\,,\label{eq:bias_correction}
\end{equation}
with the biased estimator of the variance of cross-covariances $\delta\hat{C}_{ab}^{2}$,
mean autocovariance $\left\langle A_{a}\right\rangle $, mean cross-covariance
$\left\langle C_{ab}\right\rangle $, and the number of bins the spike
trains are divided into $N=T_{\mathrm{sim}}/T_{\mathrm{bin}}$. \prettyref{fig:Bias-correction-of}
illustrates that after a simulation time of $10000\,\mathrm{s}$,
the corrected estimator converges to a fixed value while the biased
estimator does not, especially for smaller spectral radii. In contrast,
the mean covariance estimator converges much faster for all spectral
radii, as shown in \prettyref{fig:mean_of_simtime}.

\subsection{Numerical implementation of CVs}

For computing the theoretical prediction of the CVs, we make use of
the equation found in Appendix A.1 in Ref. \citep{Brunel00_183},
which in our units reads
\begin{eqnarray}
\mathrm{CV}^{2} & = & 2\pi\left(\tau_{\mathrm{m}}\nu\right)^{2}\int_{y_{\mathrm{r}}}^{y_{\mathrm{th}}}\mathrm{d}x\,\mathrm{e}^{x^{2}}\int_{-\infty}^{x}\mathrm{d}s\,\mathrm{e}^{s^{2}}\left[1+\mathrm{erf}\left(s\right)\right]^{2}\,.\label{eq:cvs_brunel}
\end{eqnarray}
However, a naive implementation of \prettyref{eq:cvs_brunel} is numerically
unstable due to the diverging integrals. To proceed, we rewrite \prettyref{eq:cvs_brunel}
using the following steps: 
\begin{align*}
\mathrm{CV}^{2} & =2\pi\left(\tau_{\mathrm{m}}\nu\right)^{2}\int_{y_{\mathrm{r}}}^{y_{\mathrm{th}}}\mathrm{d}x\,\mathrm{e}^{x^{2}}\int_{-\infty}^{x}\mathrm{d}s\,\mathrm{e}^{s^{2}}\left(1+\mathrm{erf}\left(s\right)\right)^{2}\\
 & =2\pi\left(\tau_{\mathrm{m}}\nu\right)^{2}\int_{y_{\mathrm{r}}}^{y_{\mathrm{th}}}\mathrm{d}x\,\mathrm{e}^{x^{2}}\int_{-\infty}^{x}\mathrm{d}s\,\mathrm{e}^{s^{2}}\left(\frac{2}{\sqrt{\pi}}\int_{-\infty}^{s}\mathrm{e}^{-w^{2}}\mathrm{d}w\right)^{2}\\
 & =8\left(\tau_{\mathrm{m}}\nu\right)^{2}\int_{y_{\mathrm{r}}}^{y_{\mathrm{th}}}\mathrm{d}x\int_{-\infty}^{x}\mathrm{d}s\int_{-\infty}^{s}\mathrm{d}v\int_{-\infty}^{s}\mathrm{d}w\,\mathrm{e}^{x^{2}+s^{2}-v^{2}-w^{2}}\,.
\end{align*}
We make a change of variables $v^{\prime}=s-v$, $w^{\prime}=s-w$,
where we immediately drop the prime, yielding
\begin{eqnarray*}
\mathrm{CV}^{2} & = & 8\left(\tau_{\mathrm{m}}\nu\right)^{2}\int_{y_{\mathrm{r}}}^{y_{\mathrm{th}}}\mathrm{d}x\int_{0}^{\infty}\mathrm{d}v\int_{0}^{\infty}\mathrm{d}w\int_{-\infty}^{x}\mathrm{d}s\,\mathrm{e}^{x^{2}-s^{2}-v^{2}-w^{2}+2\left(v+w\right)s}\,.
\end{eqnarray*}
Another change of variables $s^{\prime}=x+v+w-s$ gives
\begin{eqnarray*}
\mathrm{CV}^{2} & = & 8\left(\tau_{\mathrm{m}}\nu\right)^{2}\int_{y_{\mathrm{r}}}^{y_{\mathrm{th}}}\mathrm{d}x\int_{0}^{\infty}\mathrm{d}v\int_{0}^{\infty}\mathrm{d}w\int_{v+w}^{\infty}\mathrm{d}s\,\mathrm{e}^{-s^{2}+2vw+2sx}\,.
\end{eqnarray*}
We switch the order of integration using $\int_{0}^{\infty}\mathrm{d}v\int_{0}^{\infty}\mathrm{d}w\int_{v+w}^{\infty}\mathrm{d}s=\int_{0}^{\infty}\mathrm{d}s\int_{0}^{s}\mathrm{d}v\int_{0}^{s-v}\mathrm{d}w$,
which yields the form we used for the numerical implementation:
\begin{align*}
\mathrm{CV}^{2} & =8\left(\tau_{\mathrm{m}}\nu\right)^{2}\int_{0}^{\infty}\mathrm{d}s\int_{0}^{s}\mathrm{d}v\int_{0}^{s-v}\mathrm{d}w\int_{y_{\mathrm{r}}}^{y_{\mathrm{th}}}\mathrm{d}x\,\mathrm{e}^{-s^{2}+2vw+2sx}\\
 & =8\left(\tau_{\mathrm{m}}\nu\right)^{2}\int_{0}^{\infty}\mathrm{d}s\int_{0}^{s}\mathrm{d}v\,\mathrm{e}^{-s^{2}}\int_{0}^{s-v}\mathrm{d}w\,\mathrm{e}^{2vw}\int_{y_{\mathrm{r}}}^{y_{\mathrm{th}}}\mathrm{d}x\,\mathrm{e}^{2sx}\\
 & =2\left(\tau_{\mathrm{m}}\nu\right)^{2}\int_{0}^{\infty}\mathrm{d}s\int_{0}^{s}\mathrm{d}v\,\mathrm{e}^{-s^{2}}\frac{1}{v}\left[\mathrm{e}^{2v\left(s-v\right)}-1\right]\frac{1}{s}\left[\mathrm{e}^{2sy_{\mathrm{th}}}-\mathrm{e}^{2sy_{\mathrm{r}}}\right]\\
 & =2\left(\tau_{\mathrm{m}}\nu\right)^{2}\int_{0}^{\infty}\mathrm{d}s\,\frac{1}{s}\left[\mathrm{e}^{2sy_{\mathrm{th}}}-\mathrm{e}^{2sy_{\mathrm{r}}}\right]\int_{0}^{s}\mathrm{d}v\,\frac{1}{v}\left[\mathrm{e}^{-s^{2}-2v^{2}+2sv}-\mathrm{e}^{-s^{2}}\right]\,.
\end{align*}

\subsection{Inference of connectivity features from covariances\label{appendix:Inference-of-connectivity}}

\begin{figure}
\begin{centering}
\includegraphics{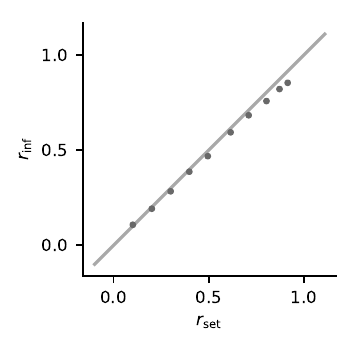}
\par\end{centering}
\caption{Set spectral radius $r_{\mathrm{set}}$ (defined in Appendix \prettyref{appendix:validity_of_theoretical_predictions})
vs inferred spectral radius $r_{\mathrm{inf}}$. Same excitatory-inhibitory
network model as in previous figures. For model details and simulation
parameters see Appendix \prettyref{appendix:nest_simulation}. \label{fig:inference_conn_params}}
\end{figure}
Equations \eqref{eq:mean_covs_simplified} and \eqref{eq:var_covs_simplified}
relate the mean connectivity $\boldsymbol{M}$ and the connectivity's
variance $\boldsymbol{S}$ to the mean covariances $\left\langle \boldsymbol{C}\right\rangle _{\boldsymbol{W},\boldsymbol{D}}$
and the variance of covariances $\left\langle \delta\boldsymbol{C}^{2}\right\rangle _{\boldsymbol{W},\boldsymbol{D}}$,
respectively. These relations can be inverted to infer features of
$\boldsymbol{M}$ and $\boldsymbol{S}$ from the statistics of covariances.
Here, we focus on $\boldsymbol{S}$ and, in particular, the spectral
radius $r=\sqrt{N_{\mathrm{E}}S_{\mathrm{E}}+N_{\mathrm{I}}S_{\mathrm{I}}}$,
which substantially modulates the size of the variance of covariances,
as shown in \prettyref{fig:Covariance-statistics-at}. In the E-I
network considered in this work, $\boldsymbol{S}$ has the structure
\begin{equation}
\boldsymbol{S}=\left(\begin{array}{cc}
S_{\mathrm{E}}\left\{ \boldsymbol{1}\right\} _{\mathrm{EE}} & \,\,S_{\mathrm{I}}\left\{ \boldsymbol{1}\right\} _{\mathrm{EI}}\\
S_{\mathrm{E}}\left\{ \boldsymbol{1}\right\} _{\mathrm{IE}} & \,\,S_{\mathrm{I}}\left\{ \boldsymbol{1}\right\} _{\mathrm{II}}
\end{array}\right)\,,
\end{equation}
with $\{\boldsymbol{1}\}_{XY}$ being the matrix of ones of dimension
$N_{X}\times N_{Y}$. The inverse matrix $\left(\boldsymbol{1}-\boldsymbol{S}\right)^{-1}$
then follows as
\begin{equation}
\left(\boldsymbol{1}-\boldsymbol{S}\right)^{-1}=\left(\begin{array}{cc}
\boldsymbol{1}_{\mathrm{EE}}+\widetilde{S}_{E}\left\{ \boldsymbol{1}\right\} _{\mathrm{EE}} & \,\,\widetilde{S}_{\mathrm{I}}\left\{ \boldsymbol{1}\right\} _{\mathrm{EI}}\\
\widetilde{S}_{\mathrm{E}}\left\{ \boldsymbol{1}\right\} _{\mathrm{IE}} & \,\,\boldsymbol{1}_{\mathrm{II}}+\widetilde{S}_{\mathrm{I}}\left\{ \boldsymbol{1}\right\} _{\mathrm{II}}
\end{array}\right)\,,
\end{equation}
with 
\begin{align}
\widetilde{S}_{\mathrm{E}} & =\frac{S_{\mathrm{E}}}{1-N_{\mathrm{E}}S_{\mathrm{E}}-N_{\mathrm{I}}S_{\mathrm{I}}}\,,\label{eq:tilde_alpha}\\
\widetilde{S}_{\mathrm{I}} & =\frac{S_{\mathrm{I}}}{1-N_{\mathrm{E}}S_{\mathrm{E}}-N_{\mathrm{I}}S_{\mathrm{I}}}\,.\label{eq:tilde_beta}
\end{align}
From \prettyref{eq:var_covs_simplified} the variance of covariances
follows as $\left\langle \delta\boldsymbol{C}^{2}\right\rangle _{\boldsymbol{W},\boldsymbol{D}}=\left(\overline{\mathrm{CV}^{2}}\overline{\nu}\right)^{2}\boldsymbol{F}$
with 
\begin{align*}
\boldsymbol{F} & =\left(\boldsymbol{1}-\boldsymbol{S}\right)^{-1}\left(\boldsymbol{1}-\boldsymbol{S}\right)^{-\T}\\
 & =\boldsymbol{1}+\left(\begin{array}{cc}
2\widetilde{S}_{\mathrm{E}}\left\{ \boldsymbol{1}\right\} _{\mathrm{EE}} & \,\,\left(\widetilde{S}_{\mathrm{E}}+\widetilde{S}_{\mathrm{I}}\right)\left\{ \boldsymbol{1}\right\} _{\mathrm{EI}}\\
\left(\widetilde{S}_{\mathrm{E}}+\widetilde{S}_{\mathrm{I}}\right)\left\{ \boldsymbol{1}\right\} _{\mathrm{IE}} & \,\,2\widetilde{S}_{\mathrm{I}}\left\{ \boldsymbol{1}\right\} _{\mathrm{II}}
\end{array}\right)+\left(\widetilde{S}_{\mathrm{E}}^{2}N_{\mathrm{E}}+\widetilde{S}_{\mathrm{I}}^{2}N_{\mathrm{I}}\right)\left\{ \boldsymbol{1}\right\} \,.
\end{align*}
Focusing on the off-diagonal elements of each block of $\boldsymbol{F}$
yields quadratic equations
\begin{align*}
F_{\mathrm{EE}} & =2\widetilde{S}_{\mathrm{E}}+\left(\widetilde{S}_{\mathrm{E}}^{2}N_{\mathrm{E}}+\widetilde{S}_{\mathrm{I}}^{2}N_{\mathrm{I}}\right)\\
F_{\mathrm{II}} & =2\widetilde{S}_{\mathrm{I}}+\left(\widetilde{S}_{\mathrm{E}}^{2}N_{\mathrm{E}}+\widetilde{S}_{\mathrm{I}}^{2}N_{\mathrm{I}}\right)\\
F_{\mathrm{EI}} & =\left(\widetilde{S}_{\mathrm{E}}+\widetilde{S}_{\mathrm{I}}\right)+\left(\widetilde{S}_{\mathrm{E}}^{2}N_{\mathrm{E}}+\widetilde{S}_{\mathrm{I}}^{2}N_{\mathrm{I}}\right)=\frac{1}{2}\left(F_{\mathrm{EE}}+F_{\mathrm{II}}\right)
\end{align*}
that can be solved for $\widetilde{S}_{\mathrm{E}}$ and $\widetilde{S}_{\mathrm{I}}$
and subsequently, using \prettyref{eq:tilde_alpha} and \prettyref{eq:tilde_beta},
for the connectivity parameters $S_{\mathrm{E}}$ and $S_{\mathrm{I}}$:
\begin{align}
S_{\mathrm{E}} & =\frac{1}{N}\left(1-\frac{\sigma\left(1-\frac{1}{2}\left(F_{\mathrm{EE}}-F_{\mathrm{II}}\right)N_{\mathrm{I}}\right)}{\sqrt{1+N\frac{F_{\mathrm{EE}}-\frac{1}{4}\left(F_{\mathrm{EE}}-F_{\mathrm{II}}\right)^{2}N_{\mathrm{I}}}{\left[1-\frac{1}{2}\left(F_{\mathrm{EE}}-F_{\mathrm{II}}\right)N_{\mathrm{I}}\right]^{2}}}}\right)\,,\\
S_{\mathrm{I}} & =\frac{1}{N}\left(1-\frac{\sigma\left(1+\frac{1}{2}\left(F_{\mathrm{EE}}-F_{\mathrm{II}}\right)N_{\mathrm{E}}\right)}{\sqrt{1+N\frac{F_{\mathrm{II}}-\frac{1}{4}\left(F_{\mathrm{II}}-F_{\mathrm{EE}}\right)^{2}N_{\mathrm{E}}}{\left[1-\frac{1}{2}\left(F_{\mathrm{II}}-F_{\mathrm{EE}}\right)N_{\mathrm{E}}\right]^{2}}}}\right)\,,
\end{align}
with $\sigma(x)=1$ for $x\geq0$ and $\sigma(x)=-1$ for $x<0$,
 $F_{\mathrm{EE}}=\left\langle \delta C_{\mathrm{EE}}^{2}\right\rangle _{\boldsymbol{W},\boldsymbol{D}}/\left(\overline{\mathrm{CV}^{2}}\overline{\nu}\right)^{2}$,
and $F_{\mathrm{II}}=\left\langle \delta C_{\mathrm{II}}^{2}\right\rangle _{\boldsymbol{W},\boldsymbol{D}}/\left(\overline{\mathrm{CV}^{2}}\overline{\nu}\right)^{2}$.
As shown in \prettyref{fig:inference_conn_params}, this in particular
allows the inference of the spectral radius of bulk connectivity eigenvalues,
\begin{equation}
r_{\mathrm{inf}}=\sqrt{N_{\mathrm{E}}S_{\mathrm{E}}+N_{\mathrm{I}}S_{\mathrm{I}}}
\end{equation}
from the variance of covariances. In principle, a similar procedure
could be followed to infer the mean connectivity $M_{\mathrm{E}}$
and $M_{\mathrm{I}}$ from the mean covariances. However, we performed
a number of simplifications $\left(\boldsymbol{1}-\boldsymbol{M}\right)^{-1}\approx\boldsymbol{1}$
in calculating the effective noise {[}see, e.g., \prettyref{eq:approx_D}{]}
to arrive at \prettyref{eq:mean_covs_simplified}. Therefore, the
inference of $M_{\mathrm{E}}$ and $M_{\mathrm{I}}$ based on measured
mean covariances is expected to be less accurate and would potentially
require a more careful mathematical treatment.

\bibliographystyle{apsrev4-1_prx}
\bibliography{manuscript_postprint.bbl}

\end{document}